\documentclass[preprint,12pt]{elsarticle}

\usepackage{filecontents}

\usepackage{amssymb}
\usepackage{bm}
\usepackage{tikz}
\usetikzlibrary{shapes.geometric,arrows.meta,shadows,calc,positioning,backgrounds}
\usepackage[acronym]{glossaries}
\usepackage{float}
\usepackage{subcaption}
\usepackage{chemformula} % Formula subscripts using \ch{}
\usepackage{multicol}
\usepackage[super]{nth}
\usepackage{adjustbox}
\usepackage{braket}
\usepackage{mathrsfs}
\usepackage{amsmath}
\usepackage{tabularx}
\usepackage{array}
\usepackage{braket}
\usepackage{gensymb}
\usepackage{siunitx}

\newcolumntype{Y}{>{\centering\arraybackslash}X} % tabularx X but centered

\def\tensor#1{\underline{\underline{\bm{#1}}}}
\tikzstyle{arrow} = [thick,->,>=stealth]
\tikzstyle{line} = [thick,-]
\tikzstyle{dottedarrow} = [dotted,thickBiconjugate gradient stabilized method ,->,>=stealth]
\usepackage{enumerate}   

\makeatletter
\newcommand*{\addFileDependency}[1]{% argument=file name and extension
\typeout{(#1)}% latexmk will find this if $recorder=0
\@addtofilelist{#1}
\IfFileExists{#1}{}{\typeout{No file #1.}}
}\makeatother

\newacronym{dns}{DNS}{Direct Numerical Simulation}
\newacronym{fdtd}{FDTD}{Finite-Difference Time-Domain}
\newacronym{em}{EM}{Electromagnetic}
\newacronym{bc}{BC}{Boundary Condition}
\newacronym{cpml}{CPML}{Convolutional Perfectly Matched Layer}
\newacronym{cfl}{CFL}{Courant–Friedrichs–Lewy}
\newacronym{bicgstab}{BiCGSTAB}{Biconjugate gradient stabilized method}
\newacronym{pde}{PDE}{Partial Differential Equation}
\newacronym{si}{SI}{Supporting Information}

\makeatletter
\def\ps@pprintTitle{%
  \let\@oddhead\@empty
  \let\@evenhead\@empty
  \def\@oddfoot{\reset@font\hfil\thepage\hfil} % Keeps page number
  \let\@evenfoot\@oddfoot}
\makeatother

\usepackage{xr}
\makeatletter
\@input{si.aux}
\makeatother
\begin{document}

\begin{frontmatter}



\title{A computational approach for the study of electromagnetic interactions in reacting flows}


\author[inst1,inst2]{Efstratios M. Kritikos\corref{cor1}}
\author[inst3]{Stewart Cant}
\author[inst1]{Andrea Giusti}

\affiliation[inst1]{{Department of Mechanical Engineering, Imperial College London, London SW7 2AZ, United Kingdom}}
\affiliation[inst2]{{Department of Applied Physics and Materials Science, California Institute of Technology, Pasadena, 91125, United States}}
\affiliation[inst3]{{Department of Engineering, University of Cambridge, CB2 1PZ, United Kingdom}}

\cortext[cor1]{Corresponding author.\\
E-mail address: emk@caltech.edu.}

\begin{abstract}
A computational fluid dynamics methodology for the simulation of electromagnetic interactions in compressible reacting flows has been formulated. The developed code, named Electromagnetic Integrator (EMI), is based on the SENGA Direct Numerical Simulation (DNS) software. Static electric and magnetic fields are solved using Gauss's laws of Maxwell's equations. Electromagnetic wave propagation is solved by discretizing Ampere's and Faraday's equations using the explicit Finite-Difference Time-Domain (FDTD) method. The equations for the electromagnetic fields are fully coupled with the Navier-Stokes equations, such that interactions between the electromagnetic fields and the fluid are included in the formulation. The interaction terms include the Lorentz, polarization, and magnetization forces. These forces determine volume forces that affect the transport of momentum, as well as the diffusion velocity in the transport of species and energy conservation equations. In addition, the medium's properties affect the propagation of the electromagnetic fields via electrical permittivity and conductivity, charge density, and magnetic permeability. The solution of electromagnetic fields is validated against analytical and numerical solutions. The implementation of the coupling between electromagnetic fields and conservation equations for species, energy, and momentum is validated with laminar reacting flow numerical solutions from the literature. The key capabilities of the proposed formulation are investigated by means of a range of laminar methane-air computations under electrostatic, magnetostatic, and high-frequency electromagnetic waves. The validity of the electrostatic formulation in the presence of currents related to the movement of charged species is also assessed. Results demonstrate that EMI-SENGA can capture the fundamental effects of electromagnetic fields on reacting flows and the dynamics of charged species and their effect on flame shape and reactivity. The formulation proposed in this work provides a comprehensive modeling of multiphysics interactions in reacting flows, and it is of interest to emerging areas of research and development that exploit electromagnetic fields to manipulate species diffusion and reactivity.
\end{abstract}

% \begin{highlights}
% \item Static electric or magnetic fields and time-varying electromagnetic waves in compressible reacting flows.
% \item Effects of Lorentz, polarization, and magnetization forces on the species of the flow.
% \item Impact of the flow's properties on the propagation of electromagnetic fields.
% \item Study of laminar methane-air flames under various electromagnetic fields.
% \end{highlights}


\end{frontmatter}


\section{Introduction}
\label{sec:intro}

The comprehensive study of multiphysics phenomena within multiphase reacting flows, especially the ones arising from the interaction of different species and external electromagnetic fields, is an everlasting pursuit in the fields of engineering, physics, and chemistry. Numerical simulations are a valuable tool for the study of complex multiphysics phenomena that span a wide range of scales, including the small ones, which are difficult or impossible to capture experimentally.

Numerous experiments showed that external electric (e.g., Refs.~\cite{Lawton1969,Noorani1985,Ata2005,Chien2019}) and magnetic (e.g., Refs.~\cite{Ueno1987,Wakayama1993,Jocher2019,Xie2021}) fields can affect the behaviour of reacting flows. On the one hand, external electromagnetic fields result in forces acting on the species forming the medium. The total electromagnetic force acting on a species includes the contributions of the Lorentz, polarization, and magnetization forces. The Lorentz force is applied to the ionic species of the fluid, the polarization force to polar species and species with electrically induced polarization, and the magnetization force to species with an intrinsic spin. The total electromagnetic force, arising from all these contributions, directly affects conservation equations, resulting in a bulk volume force as well as preferential diffusion of some of the species. On the other hand, the constituent species of the reacting flow affect the electromagnetic properties of the medium and therefore the distribution and propagation of electromagnetic waves. Charge density, permeability, and permittivity are the relevant quantities that determine the distribution of electromagnetic fields in the fluid domain. In addition, moving ions and electrons may result in currents, which are sources of electromagnetic waves. As a consequence, both the species and the electromagnetic fields directly interact and therefore influence the behaviour of the reacting flow. 

Numerical studies conducted so far typically make assumptions or neglect some of the physical phenomena either in their mathematical or numerical formulations, mainly to reduce the computational cost. The assumptions met in the literature that the formulation introduced in the present study aims to remove are detailed below. First, studies typically assume static fields, either electrostatic (e.g., Refs.~\cite{Speelman2015,Belhi2013,Esclapez2020}) or magnetostatic (e.g., Refs.~\cite{Yamada2002,Delmaere2010,Sarh2014}), which implies that electromagnetic fields generated by the motion of charged species are neglected. Second, the impact of the species' electric or magnetic moments on the electromagnetic fields' distribution is usually neglected. In addition, in the case of magnetostatic fields, the magnetic field is simply imposed rather than solved based on boundary conditions and the flow's evolution. Third, to the best of the authors' knowledge, there are no numerical studies that investigate the impact of electromagnetic waves on reactive flows. However, solving reacting flows under the action of electromagnetic waves is an important capability to be developed since experimental studies showed that AC currents and pulsed discharges have unique effects on reacting flows compared to static fields~\cite{Criner2006,Drews2012,Duan2015}. Also, studies on the magnetic field effects on reacting flows typically neglect the existence of ionic species and electrons. Fourth, numerical studies mainly focused on one-dimensional or two-dimensional configurations, due to the high computational cost that three-dimensional simulations of these multiphysics phenomena would require. Nonetheless, a three-dimensional formulation is undoubtedly more realistic and necessary, especially when turbulence is present. Only two recent studies investigated the effects of electrostatic fields on the ionic species of laminar~\cite{Belhi2019} and turbulent~\cite{DiRenzo2023} reacting flows.

A comprehensive numerical framework that incorporates all the species' electromagnetic properties and their interactions with the electromagnetic fields has not been developed yet. Additionally, a computational model able to simulate the evolution in space and time of electromagnetic waves must be developed. In this paper, we present a multiphysics solver that can simulate electromagnetic interactions in compressible reacting flows. The proposed continuum model is characterized by a detailed description of the physical phenomena involved, which are the result of interactions across the quantum, atomistic, and macroscopic scales. In addition, the framework aims to achieve an accurate representation of the physics through the direct solution of the governing equations and their coupling.

The objectives of this study are: (i) to present a unified theoretical and numerical framework that provides a comprehensive description of the interactions between electromagnetic fields and reacting flows; (ii) to develop an efficient code that allows for the investigation of electromagnetic field effects on laminar and turbulent reacting flows; (iii) to expand the framework to describe spatially and temporally varying electromagnetic fields, propagating through a flow medium; and (iv) to study the contribution of various physical phenomena to the overall electrodynamics of the system. To achieve the first objective, a new numerical framework is formulated. This new framework describes both electric and magnetic interactions with ions, neutrals, and electrons, and takes into account contributions from classical electrodynamics and quantum mechanics. To achieve the second objective, a computational fluid dynamics code that can simulate electromagnetic interactions in compressible reacting flows has been developed. The code, named Electromagnetic Integrator (EMI), is based on the SENGA~\cite{Jenkins1999,Cant1999} Direct Numerical Simulation (DNS) code. Electrostatic and magnetostatic fields are described by solving Gauss's laws of the Maxwell equations. To address the third objective, the \gls{fdtd} method is employed, which solves explicitly for the propagation of electromagnetic waves. The code is validated with analytical and numerical solutions from the literature. For the fourth objective, the key capabilities of the code are examined in one- and two-dimensional laminar methane-air flames under external electrostatic and magnetostatic fields. Then, time-varying electromagnetic waves are studied in a three-dimensional case. The key differences between time-varying and static fields are examined, together with the response of the reacting flow to these fields. 

The paper is organized as follows: first, the fundamental equations and the developed numerical framework are discussed in detail; then, validation of the code and key findings are presented. Finally, the main conclusions are summarized.

\section{Methodology}
\subsection{Maxwell's equation}

The time-dependent Maxwell's equations in differential form describe electromagnetic wave propagation through a medium and are given by:
    \begin{equation}
        \nabla \times \bm{H}=\dfrac{\partial \bm{D}}{\partial t} + \bm{J},
        \label{eq:Ampere_law}
    \end{equation}
    \begin{equation}
        \nabla \times \bm{\mathcal{E}}=-\dfrac{\partial \bm{B}}{\partial t} - \bm{\mathcal{M}},
        \label{eq:Faraday_law}
    \end{equation}
    \begin{equation}
        \nabla \cdot \bm{D}=\rho_{q},
        \label{eq:Gauss_law_electric_field}
    \end{equation}
    \begin{equation}
        \nabla \cdot \bm{B}=0,
        \label{eq:Gauss_law_magnetic_field}
    \end{equation}
where Eq.~\eqref{eq:Ampere_law} is Ampere's law, Eq.~\eqref{eq:Faraday_law} is Faraday's law, and Eqs.~\eqref{eq:Gauss_law_electric_field} and~\eqref{eq:Gauss_law_magnetic_field} are Gauss' laws for the electric and magnetic field, respectively. In Eqs.~\eqref{eq:Ampere_law}-\eqref{eq:Gauss_law_magnetic_field}, $\bm{\mathcal{E}}$ is the electric field; $\bm{D}$ is the electric flux density; $\bm{H}$ is the magnetic field; $\bm{B}$ is the magnetic flux density; $\bm{J}$ is the electric current density; $\bm{\mathcal{M}}$ the equivalent magnetic current density; and $\rho_{q}$ is the charge density. 

The constitutive relations for the electric and magnetic flux densities are given by~\cite{Taflove2005}:
\begin{equation}
    \bm{D} = \tensor{\epsilon} \cdot \bm{\mathcal{E}}=\epsilon_0\left(\bm{P}+ \bm{\mathcal{E}}\right)=\epsilon_0 \left(\chi_e+1\right)\bm{\mathcal{E}} = \epsilon \bm{\mathcal{E}},
    \label{eq:electric_flux_density}
\end{equation}
\begin{equation}
    \bm{B} = \tensor{\mu} \cdot \bm{H} = \mu_0 \left(\bm{M}+ \bm{H}\right) = \mu_0 \left(\chi_m + 1\right) \bm{H} = \mu \bm{H}.
    \label{eq:magnetic_flux_density}    
\end{equation}
In Eq.~\eqref{eq:electric_flux_density} and Eq.~\eqref{eq:magnetic_flux_density}, $\tensor{\epsilon}$ is the electrical permittivity tensor and $\tensor{\mu}$ is the magnetic permeability tensor of the medium. 

In this study, it is assumed that the medium is a linear, isotropic, and nondispersive material. Thus, the electric and magnetic properties of the reacting flow are independent of the electromagnetic field strength, direction, and frequency. However, it is noted that these assumptions may not always be valid, even if the medium is pure air. For instance, electromagnetic waves of high strength or specific wavelengths, or the presence of particulate matter in the reacting flow, may cause the violation of these assumptions~\cite{Ross1990,Bohren2008,Griffiths2013}. The validity of the aforementioned assumptions in the context of reacting flows must be verified experimentally in the future.

As a consequence of these assumptions, the electric flux density can be related to the polarization density and to the electric field in the medium. Likewise, the magnetic flux density is defined as a combination of the magnetization density and the magnetic field in the medium. The densities of polarization and magnetization can be expressed as functions of the electric and magnetic susceptibilities, i.e., $\chi_e$ and $\chi_m$, respectively. As a result, the polarization density is defined as $\bm{P}= \chi_e \bm{\mathcal{E}}$ and the magnetization density as $\bm{M}= \chi_m \bm{H}$. Combining these relations, the final right-hand sides of Eqs.~\eqref{eq:electric_flux_density} and \eqref{eq:magnetic_flux_density} are obtained. Note that, in Eqs.~\eqref{eq:electric_flux_density} and \eqref{eq:magnetic_flux_density}, the electrical permittivity and magnetic permeability of the material have been simplified to scalar fields, meaning that $\tensor{\epsilon}\equiv \epsilon \tensor{I}$ and $\tensor{\mu}\equiv \mu \tensor{I}$, i.e., the two tensors are the identity tensor $\tensor{I}$ multiplied by a constant.

The electric and magnetic current densities, $\bm{J}$ and $\bm{\mathcal{M}}$, are given by:
\begin{equation}
    \bm{J}=\bm{J}_c+\bm{J}_s=\tensor{\sigma}^e \cdot \bm{\mathcal{E}}+\bm{J}_s,
    \label{eq:J_em}
\end{equation}
\begin{equation}
\bm{\mathcal{M}}=\bm{\mathcal{M}}_c+\bm{\mathcal{M}}_s=\tensor{\sigma}^m \cdot \bm{H}+\bm{\mathcal{M}}_s,
\end{equation}
with $\bm{J}_c$ being the electric current density due to the movement of charged species and $\bm{J}_s$ being an independent current source of electric field energy. Similarly, $\bm{\mathcal{M}}_c$ corresponds to the equivalent magnetic current density and $\bm{\mathcal{M}}_s$ to an independent source of magnetic field energy. The tensor fields $\tensor{\sigma}^e$ and $\tensor{\sigma}^m$ are the electrical conductivity and equivalent magnetic loss, respectively. For the applications studied here, the electrical conductivity is assumed to be a scalar field, $\sigma^e$, corresponding to isotropic materials. The equivalent magnetic loss is assumed to be zero.

\subsubsection{Electrostatic and magnetostatic fields}
\label{subsubsec:electrostatic_magnetostatic_fields}

Maxwell's equations can be simplified in the case of static fields. Electrostatic fields exist when there is no electric current density (i.e., $\bm{J}=\bm{0}$) or time-dependent electric field fluctuations (i.e., the displacement currents are zero in Eq.~\eqref{eq:Ampere_law}). Equivalently, if there are no magnetic current densities (i.e., $\bm{\mathcal{M}}=\bm{0}$) or time-dependent magnetic field fluctuations (i.e., the displacement current is zero in Eqs.~\eqref{eq:Ampere_law} and~\eqref{eq:Faraday_law}), then the magnetostatics assumption holds. 

In the case of static fields, it is necessary to use the potential form of Maxwell's equations to uniquely define the problem. The relation between electric field and electric scalar potential, $\Phi$, is:
\begin{equation}
\begin{aligned}
    \bm{\mathcal{E}} = -\nabla \Phi.
    \label{eq:electric_potential}
\end{aligned}
\end{equation}
Equivalently a magnetic scalar potential, $\Psi$, can be defined such that:
\begin{equation}
\begin{aligned}
    \bm{H} = -\nabla \Psi.
    \label{eq:magnetic_potential}    
\end{aligned}
\end{equation}
When the scalar potentials are used, the irrotational conditions for electrostatics, $\nabla \times \bm{\mathcal{E}}=0$, and magnetostatics, $\nabla \times \bm{H}=0$, are satisfied. Furthermore, the number of unknowns of the \glspl{pde} is reduced to one and thus it is possible to close the problem.

Note here that it is common in the literature to introduce the magnetic vector potential $\bm{\mathcal{A}}$ (i.e., $\bm{H} = \nabla \times \bm{\mathcal{A}}$)~\cite{Biro1989}. In that case, the divergence of $\bm{B}$ is equal to zero, and thus the remaining Maxwell's equations can be used to obtain a solution. The magnetic vector potential is useful in cases where there is considerable current density due to the moving charged species or time-varying electric and magnetic fields. 

In this study, the effect of magnetostatic fields is explored with a chemical mechanism that does not include ionic species. Therefore, a formulation based on Eq.~\eqref{eq:magnetic_potential} is adopted. Note that previous studies exploring the magnetic field effects on reacting flows also neglected the ionic species (e.g.,~Refs.~\cite{Yamada2002,Yamada2003,Yamada2003a}), due to their low concentration in flames. It should also be noted that if the reacting flow contains a low concentration of charged particles with low mobilities, it would be a sufficient approximation to neglect the resulting weak current density and solve for magnetostatics.

The equations for electrostatics can be derived from Eq.~\eqref{eq:Gauss_law_electric_field} by substituting the relevant definitions of the electric flux density, see Eq.~\eqref{eq:electric_flux_density}, and electric potential field,  see Eq.~\eqref{eq:electric_potential}. As a result, the following equation is derived:
\begin{equation}
\begin{aligned}
    &\left(1+\chi_e\right) \nabla^2 \Phi+\nabla \Phi \cdot \nabla \chi_e =-\dfrac{\rho_{q}}{\epsilon_0},
    \label{eq:temp_e_1}
\end{aligned}
\end{equation}
which in Cartesian coordinates reads:
\begin{equation}
    \dfrac{\partial^2 \Phi}{\partial x^2} + \dfrac{\partial^2 \Phi}{\partial y^2} + \dfrac{\partial^2 \Phi}{\partial z^2} + \dfrac{1}{1+\chi_e} \left(\dfrac{\partial \Phi}{\partial x}\dfrac{\partial \chi_e}{\partial x} + \dfrac{\partial \Phi}{\partial y}\dfrac{\partial \chi_e}{\partial y} + \dfrac{\partial \Phi}{\partial z}\dfrac{\partial \chi_e}{\partial z}\right) = -\dfrac{\rho_q}{\epsilon_0 \left(1+\chi_e\right)}.
    \label{eq:electrostatics}
\end{equation}

Equivalently, the equation that describes magnetostatic fields is found by substituting the magnetic flux density, see Eq.~\eqref{eq:magnetic_flux_density}, and magnetic potential field, see Eq.~\eqref{eq:magnetic_potential}, in Eq.~\eqref{eq:Gauss_law_magnetic_field}. The resulting equation is:
\begin{equation}
\begin{aligned}
    &\left(1+\chi_m\right) \nabla^2 \Psi + \nabla \Psi \cdot \nabla \chi_m = 0,
\end{aligned}
\end{equation}
which in Cartesian coordinates is:
\begin{equation}
   \dfrac{\partial^2 \Psi}{\partial x^2} + \dfrac{\partial^2 \Psi}{\partial y^2} + \dfrac{\partial^2 \Psi}{\partial z^2} + \dfrac{1}{1+\chi_m} \left(\dfrac{\partial \Psi}{\partial x}\dfrac{\partial \chi_m}{\partial x} + \dfrac{\partial \Psi}{\partial y}\dfrac{\partial \chi_m}{\partial y} + \dfrac{\partial \Psi}{\partial z}\dfrac{\partial \chi_m}{\partial z}\right) = 0.
\label{eq:magnetostatics}
\end{equation}

It is noted here that in reacting flows, the effects of the electric and magnetic susceptibilities on the overall spatial distribution of the electric and magnetic fields, respectively, are typically neglected. Nevertheless, the respective terms are kept in the formulation used in this study for three main reasons:
\begin{enumerate}
    \item Considering a one-dimensional static and uniform magnetic field in free space, the change in magnetic susceptibility is responsible for the generation of magnetization forces when a flow of particles with magnetic moments passes through the field. Therefore, it is imperative to include in the formulation the terms involving the variation in space of magnetic susceptibility and evaluate their contribution.
    \item The proposed mathematical framework is meant to be of general validity. The framework should enable the inclusion of particles with strong polarization and magnetization properties in the flow without the need for any modifications to the underlying equations. For instance, iron is a ferromagnetic material, and thus it can be easily magnetized and get strongly attracted to an external magnetic field. Iron can reach a permeability that is 350,000 times higher than the permeability of free space~\cite{Haynes2016}.
    \item In the study of dusty plasmas, Chai and Bellan suggested that the particles' agglomeration process is primarily influenced by the electric field gradients around the particles~\cite{Chai2013}. Consequently, the field around these agglomerates and the polarization force induced on nearby polar species can have a significant impact on the dynamics of the system. Thus, the electric susceptibilities must be included in the formulation in order to be able to accurately describe induced electric fields around particles.
\end{enumerate}

\begin{figure}[t]
\centering
\begin{adjustbox}{width=0.5\linewidth,center}
\begin{tikzpicture}[font={\fontsize{10pt}{12}\selectfont}]
\draw (0,0) rectangle (14,7);
\coordinate (BA) at (0,3.5); \node[above=1pt of BA,rotate=90] (BA) {Dirichlet B.C. for potential};
\draw (0.6,0) -- (0.6,7);
\coordinate (BA) at (0.6,3.5); \node[above=1pt of BA,rotate=90] (BA) {\nth{2} order accuracy in $x$ direction};
\draw (1.2,0) -- (1.2,7);
\coordinate (BA) at (1.2,3.5); \node[above=1pt of BA,rotate=90] (BA) {\nth{4} order accuracy in $x$ direction};
\draw (1.8,0) -- (1.8,7);
\coordinate (BA) at (1.8,3.5); \node[above=1pt of BA,rotate=90] (BA) {\nth{6} order accuracy in $x$ direction};
\draw (2.4,0) -- (2.4,7);
\coordinate (BA) at (2.4,3.5); \node[above=1pt of BA,rotate=90] (BA) {\nth{8} order accuracy in $x$ direction};
\coordinate (BA) at (7,1); \node[above=1pt of BA] (BA) {\nth{10} order accuracy in $x$ direction};
\coordinate (BA) at (7,0.5); \node[above=1pt of BA] (BA) {\nth{10} order accuracy in $x$ \& $z$ direction everywhere};
\coordinate (BA) at (14,3.5); \node[above=1pt of BA,rotate=90] (BA) {Dirichlet B.C. for potential};
\draw (13.4,0) -- (13.4,7);
\coordinate (BA) at (13.4,3.5); \node[above=1pt of BA,rotate=90] (BA) {\nth{2} order accuracy in $x$ direction};
\draw (12.8,0) -- (12.8,7);
\coordinate (BA) at (12.8,3.5); \node[above=1pt of BA,rotate=90] (BA) {\nth{4} order accuracy in $x$ direction};
\draw (12.2,0) -- (12.2,7);
\coordinate (BA) at (12.2,3.5); \node[above=1pt of BA,rotate=90] (BA) {\nth{6} order accuracy in $x$ direction};
\draw (11.6,0) -- (11.6,7);
\coordinate (BA) at (11.6,3.5); \node[above=1pt of BA,rotate=90] (BA) {\nth{8} order accuracy in $x$ direction};
\draw (6,3.5) rectangle (8,5.5);
\draw[dashed] (5.5,3) rectangle (8.5,6);
\coordinate (BA) at (7,5.5); \node[above=1pt of BA] (BA) {Dirichlet B.C.};
\coordinate (BA) at (6.2,6.); \node[above=1pt of BA] (BA) {Processor:};
\coordinate (BA) at (7,4.5); \node[above=1pt of BA] (BA) {\nth{10} order in};\node[below=1pt of BA] (BA) {all directions};
\end{tikzpicture}
\end{adjustbox}
\caption{2D schematic representation of domain discretization}
\label{fig:schematic_electrostatics_magnetostatics}
\end{figure}
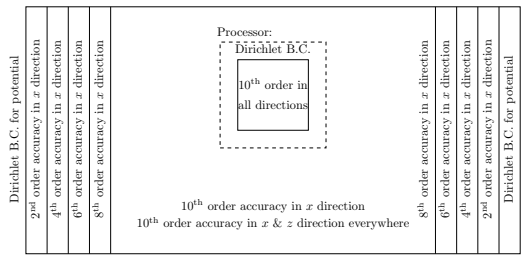

Both Eqs.~\eqref{eq:electrostatics} and~\eqref{eq:magnetostatics} are solved in a similar manner. The discretization strategy is schematically shown in Fig.~\ref{fig:schematic_electrostatics_magnetostatics} and is explained in detail in Section~\ref{subsec:static_solver}. Moreover, Fig.~\ref{fig:flow_chart_static} shows a flow chart of the EMI-static code. The accuracy is \nth{10} order in the direction perpendicular to the external field, whereas in the direction of the external field, it starts from \nth{2} order in the boundaries and increases to \nth{10} order in the middle of the domain. Periodic boundary conditions are applied in the directions perpendicular to the external field, whereas Dirichlet boundary conditions are used for the electric or magnetic potential in the direction of the external field. The BCs can either be fixed in time or time-dependent (such as a sinusoidal time dependency of an electrostatic wave).

The resulting $(A_1+A_2) X = \Omega$ system (see Section~\ref{subsec:static_solver} for more details) is solved via direct or iterative methods. The parallelization strategy of the EMI-static code is based on the domain decomposition approach and is explained in detail in Section~\ref{subsec:static_solver}. Note that the propagation in time in SENGA is performed using a Runge-Kutta algorithm. The static fields are solved in the internal loop of the Runge-Kutta algorithm, as shown in Fig.~\ref{fig:flow_chart_static}.

\subsubsection{Electromagnetic waves}
\label{subsubsec:electromagnetic_waves}

\begin{figure}[t]
\begin{adjustbox}{width=0.5\linewidth,center}
\centering
\def\L{5}
\def\ly{2}
\def\lx{3}
\begin{tikzpicture}
\draw (0,0) rectangle (\L,\L);
\draw (\L,\L) -- (\L+\lx,\L+\ly);
\draw (\L,0) -- (\L+\lx,\ly);
\draw (\lx,\L+\ly) -- (\L+\lx,\L+\ly);
\draw (\L+\lx,\ly) -- (\L+\lx,\L+\ly);
\draw (0,\L) -- (\lx,\L+\ly);
\draw[dashed] (\lx,\ly) -- (\L+\lx,\ly);
\draw[dashed] (\lx,\ly) -- (\lx,\L+\ly);
\draw[dashed] (0,0) -- (\lx,\ly);
\draw[line width=0.2mm,dashed] (\L/2,0) -- (\L/2,\L);
\draw[line width=0.2mm,dashed] (0,\L/2) -- (\L,\L/2);

\draw[line width=0.2mm,dashed] (\L,\L/2) -- (\L+\lx,\L/2+\ly);
\draw[line width=0.2mm,dashed] (\L/2,\L) -- (\L/2+\lx,\L+\ly);
\draw[line width=0.2mm,dashed] (\lx/2,\L+\ly/2) -- (\L+\lx/2,\L+\ly/2);
\draw[line width=0.2mm,dashed] (\L+\lx/2,\ly/2) -- (\L+\lx/2,\L+\ly/2);
\draw[line width=0.2mm,dashed] (\L/2+\lx,\ly) -- (\L/2+\lx,\ly+\L);
\draw[line width=0.2mm,dashed] (\lx,\L/2+\ly) -- (\L+\lx,\ly+\L/2);
\draw[line width=0.2mm,dashed] (0,\L/2) -- (\lx,\L/2+\ly);
\draw[line width=0.2mm,dashed] (\L/2,0) -- (\L/2+\lx,\ly);
\draw[line width=0.2mm,dashed] (\lx/2,\ly/2) -- (\L+\lx/2,\ly/2);
\draw[line width=0.2mm,dashed] (\lx/2,\ly/2) -- (\lx/2,\L+\ly/2);

\draw (0,0) circle (2pt) node[below]{$\left(i,j,k\right)$};
\filldraw[black] (\L/2,0) circle (2pt) node[below]{$\mathcal{E}_x$};
\filldraw[black] (\L/2,\L) circle (2pt) node[above]{$\mathcal{E}_x$};
\filldraw[black] (\L,\L/2) circle (2pt) node[above left]{$\mathcal{E}_z$};
\filldraw[black] (\L+\lx,\L/2+\ly) circle (2pt) node[above left]{$\mathcal{E}_z$};
\filldraw[black] (\lx/2,\L+\ly/2) circle (2pt) node[above left]{$\mathcal{E}_y$};
\filldraw[black] (\L+\lx/2,\L+\ly/2) circle (2pt) node[below right]{$\mathcal{E}_y$};
\filldraw[black] (\L/2+\lx,\ly) circle (2pt) node[above right]{$\mathcal{E}_x$};
\filldraw[black] (\L/2+\lx,\ly+\L) circle (2pt) node[below right]{$\mathcal{E}_x$};
\filldraw[black] (0,\L/2) circle (2pt) node[left]{$\mathcal{E}_z$};
\filldraw[black] (\lx,\L/2+\ly) circle (2pt) node[above right]{$\mathcal{E}_z$};
\filldraw[black] (\lx/2,\ly/2) circle (2pt) node[above left]{$\mathcal{E}_y$};
\filldraw[black] (\L+\lx/2,\ly/2) circle (2pt) node[below right]{$\mathcal{E}_y$};
\filldraw[black] (\L/2,\L/2) circle (2pt) node[above left]{$H_y$};
\filldraw[black] (\L+\lx/2,\L/2+\ly/2) circle (2pt) node[below right]{$H_x$};
\filldraw[black] (\L/2+\lx/2,\L+\ly/2) circle (2pt) node[above]{$H_z$};
\filldraw[black] (\L/2+\lx,\L/2+\ly) circle (2pt) node[above right]{$H_y$};
\filldraw[black] (\lx/2,\L/2+\ly/2) circle (2pt) node[above left]{$H_x$};
\filldraw[black] (\L/2+\lx/2,\ly/2) circle (2pt) node[above]{$H_z$};

\draw[-To,thick] (-2,-1) -- (-0.5,-1);
\draw[-To,thick] (-2,-1) -- (-2,0.5);
\draw[-To,thick] (-2,-1) -- (-1,-0.3333333333333);
\filldraw[black] (-0.5,-1) circle (0) node[right]{$x$};
\filldraw[black] (-2,0.5) circle (0) node[above]{$z$};
\filldraw[black] (-1,-0.3333333333333) circle (0) node[above]{$y$};
\end{tikzpicture}
\end{adjustbox}
\caption{Positions of $\mathcal{E}$ and $H$ field components on the Yee cell~\cite{Yee1966}. The $\mathcal{E}$-components are in the middle of the edges and the $H$-components are in the center of the faces.}
\label{fig:Yee_cell}
\end{figure}
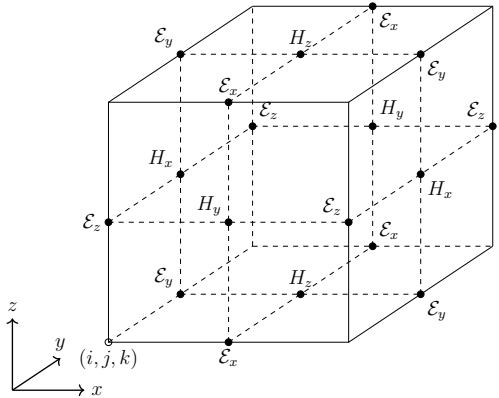

If assumptions cannot be made for the form of the electromagnetic fields, then the time-dependent Maxwell’s curl equations, i.e., Eqs.~\eqref{eq:Ampere_law} and~\eqref{eq:Faraday_law}, must be discretized to solve for arbitrary electromagnetic waves. To do that, the \gls{fdtd} method is implemented~\cite{Yee1966}. According to the \gls{fdtd} method, the electric and magnetic field components are computed at different positions of the cell, known as the Yee cell. The Yee cell, consisting of staggered electric and magnetic grids, is shown in Fig.~\ref{fig:Yee_cell}. Also, the electric and magnetic fields are successively advanced at half-timesteps. In the \gls{fdtd} method, the staggered grids in space and time have \nth{2}-order accuracy. It should be noted that the staggered grid is selected compared to alternative unstaggered grids, due to their increased numerical stability~\cite{Liu1996}. Typically, unstaggered grids require a more refined mesh, which would significantly increase the computational cost.

According to the \gls{fdtd} method, Ampere's and Faraday's laws are solved, while Gauss's laws must be satisfied at any point in time and space. The discretized equations that update each electric and magnetic field component on the Yee cell (see Fig.~\ref{fig:Yee_cell}) of the EMI-FDTD code can be found in Ref.~\cite{Kritikos2023t}. Note that this formulation refers to a Cartesian frame of reference, and it is based on the structured grid implementation of EMI-SENGA.

If the initial conditions satisfy the divergence equations, i.e., Eq.~\eqref{eq:Gauss_law_electric_field} and Eq.~\eqref{eq:Gauss_law_magnetic_field}, then these equations are satisfied for the rest of the simulation~\cite{Villasenor1992,Munz1999}. Specifically, $\nabla \cdot \bm{B}$ remains zero if it was initially satisfied. Furthermore, $\nabla \cdot \bm{\mathcal{E}}$ remains equal to $\rho_q/\epsilon_0$ if the charge conservation equation:
\begin{equation}
\dfrac{\partial \rho_q}{\partial t} = -\nabla \cdot \bm{J}
\label{eq:charge_conservation}
\end{equation}
holds true for the duration of the simulation time. Since the continuity equation is imposed in the \gls{dns} code and the chemical mechanism used in this study (see Section~\ref{subsec:validation}) conserves the charges in the reactions, Eq.~\eqref{eq:charge_conservation} is satisfied. Therefore, it is important to set the correct initial profile of the electric field based on the initial charge distribution of the reacting flow. To achieve that, the electrostatic field solver described in Section~\ref{subsubsec:electrostatic_magnetostatic_fields} is used to initialize the electric field profile. Afterwards, the \gls{fdtd} solver is applied without violating Maxwell's equations.

\glspl{bc} on the domain boundaries must be applied for the EM fields. The EMI-FDTD solver is formulated with the Stretched coordinate Convolutional Perfectly Matched Layer (CPML) absorbing BCs~\cite{Taflove2005,Elsherbeni2016}. According to this method, additional BC cells that absorb the incident electromagnetic waves and suppress any reflections in the computational domain are set. The CPML cells are placed outside the DNS domain, which solves the reacting flow. In other words, the domain of the EMI-FDTD solver corresponds to the DNS domain, expanded by the additional CPML cells. The thickness of the CPML layer can vary. A thicker CPML layer better prevents reflections but at a higher computational cost and memory. It is noted that SENGA already uses 5 ghost cells to achieve a \nth{10}-order discretization scheme. Thus, a 5-cell-thick CPML region does not introduce any additional memory cost. The \gls{cpml} parameters are computed at the \gls{cpml} regions and are superimposed on the field components. The equations that are used for the $\sigma$, $\alpha$, and $\kappa$ parameters of the CPML regions in the EMI-FDTD solver can be found in Section~\ref{sisubsec:cpml}.

Sources of electromagnetic waves are applied at a single point and placed inside the domain. Note that other methods that introduce sources at the boundaries have been formulated~\cite{Rumpf2012,Tan2010}, but are not studied in the present work. The two most common sources that are introduced in the FDTD grid are hard and soft sources~\cite{Costen2009}. In the hard source approach, the electric or magnetic fields are imposed on a node. On the other hand, the soft source approach involves the application of an electric or magnetic current. The hard source can be problematic, since, due to its formulation, it cannot handle incident waves. As a result, the hard source can lead to spurious nonphysical retroreflection~\cite{Taflove2005}. This is expected to also occur in a reacting flow, which is characterized by spatially varying conductivity, permittivity, and permeability, and thus reflected waves can appear from the flame front. One possible solution is to turn off the source after a short time. However, this imposes a serious limitation on the current configuration, since the time scales of electromagnetic wave propagation are significantly lower than the time scales of DNS. Hence, in this study, soft sources are implemented. A common soft source is the Hertzian dipole~\cite{Taflove2005}. The Hertzian dipole is given by:
\begin{equation}
    J_s = \dfrac{\mathcal{I}\Delta l}{\Delta x \Delta y \Delta z},
    \label{eq:J_source}
\end{equation}
where $\mathcal{I}$ is the current of a given waveform and $\Delta l$ is the length of the electric dipole. Depending on the direction of the dipole moment, the $\Delta l$ is set equal to $\Delta x$, $\Delta y$, or $\Delta z$.

Regarding the accuracy of the solution and the numerical stability of the code, two main conditions must be satisfied. For the accurate numerical solution of the electromagnetic field propagation, the Courant–Friedrichs–Lewy (CFL) condition must be satisfied, i.e., $\Delta t \leq 1/c \sqrt{1/\Delta x^2+1/\Delta y^2+1/\Delta z^2}$
where $c$ is the speed of light. In addition, to avoid significant numerical dispersion, the spatial grid should be sufficiently refined to resolve the minimum free-space wavelength $\lambda_{0,\min}=2c\Delta t$. Therefore, the minimum grid size is $\Delta x \approx \dfrac{\lambda_{0,\min}}{N_\lambda}$ where usually $N_\lambda\geq10$. Since electromagnetic waves propagate at a higher speed than ionic species and even electrons, the EMI-FDTD solver operates under a different timestep, $\Delta t_{FDTD}$, than the one that updates the fluid mechanics equations.

The structure of the code is detailed in Section~\ref{sisubsec:fdtd_solver}, and the flow chart of the EMI-FDTD code is shown in Fig.~\ref{fig:flow_chart_fdtd}. Since the FDTD grid is on a staggered grid, the medium properties are interpolated on the Yee grid before the propagation of the electromagnetic fields. Afterwards, the electromagnetic fields are interpolated back into the SENGA grid. The parallelization strategy, which is based on the domain decomposition of the staggered Yee cell, is discussed in Section~\ref{sisubsec:fdtd_solver}.

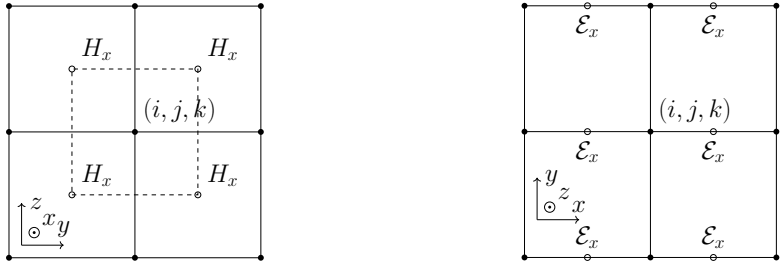
\begin{figure}[t]
\centering
\begin{subfigure}[b]{0.5\textwidth}
\begin{adjustbox}{width=0.5\linewidth,center}
\begin{tikzpicture}
\Large
\draw (0,0) rectangle (6,6);
\draw (0,3) -- (6,3);
\draw (3,0) -- (3,6);
\draw[dashed] (1.5,1.5) rectangle (4.5,4.5);
\coordinate (BA) at (1.5,1.5); \node[above right=1pt of BA] (BA) {$H_{x}$};
\coordinate (BB) at (4.5,1.5); \node[above right=1pt of BB] (BB) {$H_{x}$};
\coordinate (BC) at (4.5,4.5); \node[above right=1pt of BC] (BC) {$H_{x}$};
\coordinate (BD) at (1.5,4.5); \node[above right=1pt of BD] (BD) {$H_{x}$};
\coordinate (BE) at (3,3); \node[above right=0pt of BE] (BE) {${(i,j,k)}$};
\draw (1.5,1.5)  circle[radius=2pt];
\draw (4.5,1.5)  circle[radius=2pt];
\draw (4.5,4.5)  circle[radius=2pt];
\draw (1.5,4.5)  circle[radius=2pt];
\fill (3,3)  circle[radius=2pt];
\fill (0,3)  circle[radius=2pt];
\fill (3,0)  circle[radius=2pt];
\fill (0,0)  circle[radius=2pt];
\fill (6,3)  circle[radius=2pt];
\fill (3,6)  circle[radius=2pt];
\fill (0,6)  circle[radius=2pt];
\fill (6,0)  circle[radius=2pt];
\fill (6,6)  circle[radius=2pt];
\draw[-To] (0.3,0.3) -- (1.3,0.3) node[above] {$y$};
\draw[-To] (0.3,0.3) -- (0.3,1.3) node[right] {$z$};
\draw (0.6,0.6)  circle[radius=3.5pt];
\fill (0.6,0.6)  circle[radius=1pt];
\coordinate[label=above right:$x$] ($x$) at (0.6,0.6);
\end{tikzpicture}
\end{adjustbox}
\end{subfigure}%
\begin{subfigure}[b]{0.5\textwidth}
\begin{adjustbox}{width=0.5\linewidth,center}
\begin{tikzpicture}
\Large
\draw (0,0) rectangle (6,6);
\draw (0,3) -- (6,3);
\draw (3,0) -- (3,6);
\coordinate (EA) at (1.5,0); \node[above=1pt of EA] (EA) {$\mathcal{E}_{x}$};
\coordinate (EB) at (4.5,0); \node[above=1pt of EB] (EB) {$\mathcal{E}_{x}$};
\coordinate (EC) at (1.5,3); \node[below=1pt of EC] (EC) {$\mathcal{E}_{x}$};
\coordinate (ED) at (4.5,3); \node[below=1pt of ED] (ED) {$\mathcal{E}_{x}$};
\coordinate (EE) at (1.5,6); \node[below=1pt of EE] (EE) {$\mathcal{E}_{x}$};
\coordinate (EF) at (4.5,6); \node[below=1pt of EF] (EF) {$\mathcal{E}_{x}$};
\coordinate (EG) at (3,3); \node[above right=0pt of EG] (EG) {${(i,j,k)}$};
\draw (1.5,0)  circle[radius=2pt];
\draw (4.5,0)  circle[radius=2pt];
\draw (1.5,3)  circle[radius=2pt];
\draw (4.5,3)  circle[radius=2pt];
\draw (1.5,6)  circle[radius=2pt];
\draw (4.5,6)  circle[radius=2pt];
\fill (3,3)  circle[radius=2pt];
\fill (0,3)  circle[radius=2pt];
\fill (3,0)  circle[radius=2pt];
\fill (0,0)  circle[radius=2pt];
\fill (6,3)  circle[radius=2pt];
\fill (3,6)  circle[radius=2pt];
\fill (0,6)  circle[radius=2pt];
\fill (6,0)  circle[radius=2pt];
\fill (6,6)  circle[radius=2pt];
\draw[-To] (0.3,0.9) -- (1.3,0.9) node[above] {$x$};
\draw[-To] (0.3,0.9) -- (0.3,1.9) node[right] {$y$};
\draw (0.6,1.2)  circle[radius=3.5pt];
\fill (0.6,1.2)  circle[radius=1pt];
\coordinate[label=above right:$z$] ($z$) at (0.6,1.2);
\end{tikzpicture}
\end{adjustbox}
\end{subfigure}
\caption{Interpolation of $H_{x}$ and $\mathcal{E}_{x}$ on SENGA nodes.}
\label{fig:space_interp}
\end{figure}

Due to the fact that the EMI-FDTD solver is formulated on a staggered grid, interpolation of the medium properties before the FDTD loop and interpolation of the $\bm{\mathcal{E}}$ and $\bm{H}$ fields after the FDTD loop are required. For the spatial interpolation of the electric and magnetic field components from the Yee grid to the SENGA grid, it is expected that a linear interpolation for the electric field components and a bilinear interpolation for the magnetic field components give sufficient accuracy for a sufficiently refined mesh. The interpolation strategy for the electric and magnetic fields is shown in Fig.~\ref{fig:space_interp}. 

These interpolation methods are implemented in EMI-SENGA. For $H_{x}$, interpolation must be done on the $y$-$z$ plane between the 4 adjacent points, since $H^{n+\frac{1}{2}}_{x}\left(i,j+\frac{1}{2},k+\frac{1}{2}\right)$ is already in the appropriate node $i$ of the \gls{dns} grid. Similar considerations apply for $H^{n+\frac{1}{2}}_{y}\left(i+\frac{1}{2},j,k+\frac{1}{2}\right)$ on $x$-$z$ plane and $H^{n+\frac{1}{2}}_{z}\left(i+\frac{1}{2},j+\frac{1}{2},k\right)$ on $x$-$y$ plane. For $\mathcal{E}_{x}$, interpolation must be done on the $x$ axis between 2 adjacent points, since $\mathcal{E}^{n+1}_{x}\left(i+\frac{1}{2},j,k\right)$ is already in the appropriate $j$ and $k$ nodes of the \gls{dns} grid. Similar considerations apply $\mathcal{E}^{n+1}_{y}\left(i,j+\frac{1}{2},k\right)$ on $y$ axis and $\mathcal{E}^{n+1}_{z}\left(i,j,k+\frac{1}{2}\right)$ on $z$ axis. For higher accuracy, other methods, such as cubic or bicubic spline interpolation~\cite{Bhattacharyya1969}, should be implemented. However, the computational cost will increase. Implementation of other spatial interpolation methods is left for future work. 

Note that the magnetic field components are half a timestep ahead of the electric field components. Therefore, at the end of the EMI-FDTD loop and using the semi-implicit approximation, the magnetic field at time $n$ is computed using the equation $\bm{H} = (\bm{H}^{n-\frac{1}{2}}+\bm{H}^{n+\frac{1}{2}})/2$.

\subsection{Electromagnetic force}

The electromagnetic force that charged species (or particles) experience is given by the Lorentz force $F_L$. Species with a charge polarity in spatially inhomogeneous (i.e., non-uniform in space) electric fields also experience a polarization force $F_P$. Similarly, for inhomogeneous magnetic fields, a magnetization force $F_M$ is applied to each species with a net magnetic moment (spin). The conservation equations in Section~\ref{subsec:conservation_eq} are formulated with respect to volume forces (i.e., in units of force per mass of substance or N/kg). Therefore, the total electromagnetic volume force is:
\begin{equation}
\begin{aligned}
    \bm{f}^s &= \bm{f}^s_L + \bm{f}^s_P + \bm{f}^s_M.
\end{aligned}
\end{equation}

Each term is equal to:
\begin{equation}
\begin{aligned}
    \bm{f}^s_L &= \dfrac{q_s N_A}{M_s}\left[\bm{\mathcal{E}}+\mu_0\left(1+\chi_m^s\right) \bm{u}\times\bm{H}\right],
    \label{eq:lorentz_force}
\end{aligned}
\end{equation}

\begin{equation}
\begin{aligned}
    \bm{f}^s_P &= \dfrac{\epsilon_0}{\rho Y_s} \chi_e^s\left(\bm{\mathcal{E}}\cdot\nabla\right) \bm{\mathcal{E}},
    \label{eq:polarization_force}    
\end{aligned}
\end{equation}

and
\begin{equation}
\begin{aligned}
    \bm{f}^s_M &= \dfrac{\mu_0}{\rho Y_s} \chi_m^s\bm{H}(\bm{H} \cdot \nabla \chi_m) + \dfrac{\mu_0}{\rho Y_s}\chi_m^s\left(1+\chi_m\right)\left(\bm{H}\cdot\nabla\right) \bm{H},
    \label{eq:magnetization_force}    
\end{aligned}
\end{equation}
which is derived from Gilbert's model, or 
\begin{equation}
\begin{aligned}
    \bm{f}^s_M &= \dfrac{\mu_0}{\rho Y_s} \left(\dfrac{1+\chi_m}{1+\chi_m^s}\right)^2 \bm{H}^2 \nabla \chi_m^s + \dfrac{2\mu_0\chi_m^s (1+\chi_m)}{\rho Y_s(1+\chi_m^s)} \bm{H}^2 \nabla \chi_m \\
              &+ \dfrac{\mu_0\chi_m^s \left(1+\chi_m\right)^2}{\rho Y_s(1+\chi_m^s)} \nabla\bm{H}^2,
    \label{eq:magnetization_force_2}    
\end{aligned}
\end{equation}
which is derived from Ampere's model.

In the above equations, $M_s$ is the molecular weight of species $s$ and $\bm{u}$ is the velocity vector of the fluid. In addition, $Y_s$ is the mass fraction of species $s$ and $N_A$ is the Avogadro constant.
The derivation of Eqs.~\eqref{eq:polarization_force}, \eqref{eq:magnetization_force}, and \eqref{eq:magnetization_force_2} is given in~\ref{sec:polarization_magnetization_forces}. It is noted that for the remainder of this paper, the magnetization force according to Gilbert's model, i.e., Eq.~\eqref{eq:magnetization_force}, is used, due to the reasons discussed in~\ref{sec:polarization_magnetization_forces}. Note also that the molecular transport properties, i.e., the polynomial fitting method and the transport model for ions, are discussed in Sections~\ref{sisubsubsec:polynomial_fitting_method} and~\ref{sisubsubsec:ion_transport_model}.

\subsection{Conservation equations}
\label{subsec:conservation_eq}

The conservation equations modified to include the electromagnetic forces are presented below~\cite{Poinsot2005}. The momentum equations are:
\begin{equation}
\begin{aligned}
    \dfrac{\partial \rho u_i}{\partial t} + \dfrac{\partial \rho u_i u_j}{\partial x_i} =
    - \dfrac{p}{\partial x_i} + \dfrac{\partial \tau_{ij}}{\partial x_i} + \rho \sum_s Y_s f^s_i,
\end{aligned}
\end{equation}
with the stress tensor $\tau_{ij}$:
\begin{equation}
\begin{aligned}
    \tau_{ij} = -\dfrac{2}{3} \mu \dfrac{\partial u_m}{\partial x_m} \delta_{ij} + \mu \left(\dfrac{\partial u_i}{\partial x_j} +\dfrac{\partial u_j}{\partial x_i} \right).
\end{aligned}
\end{equation}
The transport equation for the mass fraction of a generic species $s$ is:
\begin{equation}
\begin{aligned}
    \dfrac{\partial \rho Y_s}{\partial t} + \dfrac{\partial \rho u_i Y_s}{\partial x_i} =
    - \dfrac{\partial \rho Y_s V^s_i}{\partial x_i} + \Dot{\omega}^s.
\end{aligned}
\end{equation}
The total energy equation is:
\begin{equation}
\begin{aligned}
    \dfrac{\partial \rho E}{\partial t} + \dfrac{\partial \rho u_i E}{\partial x_i} = 
    \dfrac{\partial p u_i}{\partial x_i} - \dfrac{\partial q_i}{\partial x_i} + \dfrac{\partial \tau_{im} u_m}{\partial x_i} + \Dot{q}_R + \rho \sum_s Y_s f^s_i \left(u_i+V^s_i\right),
\end{aligned}
\end{equation}
where the heat flux vector $q_i$ is given by:
\begin{equation}
\begin{aligned}
    q_i = -\lambda \dfrac{\partial T}{\partial x_i} + \sum_{s} \rho h_s Y_s V^s_i.
\end{aligned}
\end{equation}

In the above equations, $\rho$ and $u$ correspond to the density and velocity of the fluid, respectively; $p$ is the pressure of the fluid; $Y_s$ the mass fraction of species $s$; $\delta_{ij}$ is the Kronecker delta; $\Dot{\omega}^s$ is the reaction rate of species $s$, and $E$ is the total energy.

The diffusion velocity due to the external electromagnetic forces based on the Hirschfelder and Curtiss approximation~\cite{Hirschfelder1964} is derived in~\ref{sec:diffusion_drift_velocity}. The equation that is numerically solved is given by:
\begin{equation}
\begin{aligned}
    V_i^s=&-\dfrac{D_s}{Y_s} \dfrac{\partial Y_s}{\partial x_i} + \dfrac{\rho D_s M_{s}}{M_{m} P} \sum_{k=1}^N Y_k\left(f^s_i-f^k_i\right) \\
          &+\sum_{\lambda=1}^N Y_\lambda \left[\dfrac{D_\lambda}{Y_\lambda} \dfrac{\partial Y_\lambda}{\partial x_i} - \dfrac{\rho D_\lambda M_{\lambda}}{M_{m} P} \sum_{k=1}^N Y_k\left(f^\lambda_i-f^k_i\right)\right].
    \label{eq:diffusion_velocity}
\end{aligned}
\end{equation}
where $M_{s}$ is the molecular weight of species $s$ and $M_{m}$ is the average molecular weight of the mixture. 

\subsection{Medium properties}
\label{subsec:medium_properties}

\subsubsection{Electrical permittivity}

The permittivity $\epsilon$ of a medium is computed according to the relation:
\begin{equation}
\begin{aligned}
    \epsilon = \left(1+\chi_e \right) \epsilon_0,
\end{aligned}
\end{equation}
where $\chi_e$ is the electric susceptibility and $\epsilon_0$ is the electrical permittivity of vacuum. The electric susceptibility of a species $s$, is computed from the Clausius-Massotti relation, given by~\cite{Talebian2013}:
\begin{equation}
\begin{aligned}
    &\chi^s_e = \dfrac{3 N_s \alpha_s}{3- N_s \alpha_s},
    \end{aligned}
\label{eq:Clausius_Massotti_species}
\end{equation}
while the total electric susceptibility of a medium consisting of many species is given by~\cite{Talebian2013}:
\begin{equation}
\begin{aligned}
     \chi_e = \dfrac{3\sum_{s} N_s \alpha_s}{3-\sum_{s} N_s \alpha_s},
\end{aligned}
\label{eq:Clausius_Massotti}
\end{equation}
In the above equations, $\alpha_s$ is the molecular polarizability of species $s$. Values of molecular polarizabilities for the species under investigation are reported in Table~\ref{Table:elec_prop}. 

\subsubsection{Electrical conductivity}
\label{sec:electrical_conductivity}

The electrical conductivity, $\sigma_e$, is given by~\cite{Weinberg1986}:
\begin{equation}
\begin{aligned}
    \sigma_e = \sum_s \upsilon_s N_s |q_s|,
    \label{eq:sigma_e}
\end{aligned}
\end{equation}
where $\upsilon$ is the mobility, $N$ is the number density, and $q$ is the charge of species $s$. The mixture-averaged mobility $\upsilon_s$ of a species $s$ is computed as~\cite{McDaniel1973}:
\begin{equation}\label{eq:mobility}
\begin{aligned}
    \upsilon_s = \left(\sum_\nu \dfrac{X_\nu}{\upsilon_{s\nu}} \right)^{-1} = \left(\sum_\nu \dfrac{M_{m} Y_\nu}{M_{\nu}\upsilon_{s\nu}} \right)^{-1}.
\end{aligned}
\end{equation}
In Eq.~(\ref{eq:mobility}), $\upsilon_{s\nu}$ are the binary mobilities of the heavy ions obtained from the Einstein relation~\cite{McDaniel1973}:
\begin{equation}
\begin{aligned}
    \upsilon_{s\nu} = \dfrac{D_{s\nu} |q_s|}{k_B T},
\end{aligned}
\end{equation}
with $k_B$ being the Boltzmann constant. Note that if a constant Lewis number is used, the mobilities are simply computed from $\upsilon_{s} = D |q_s|/k_B T$.

Various values for the electron mobility, $\upsilon_{e^-}$, have been reported in the literature. The determination of an exact electron mobility value is still an active field of research. The electron mobility directly affects the computational cost of the whole simulation. Specifically, if a high $\upsilon_{e^-}$ value is used, electron diffusion is fast and thus the timestep of the simulation must be sufficiently low. This introduces a significant bottleneck in the numerical solution. In this study, $\upsilon_{e^-}=0.2$~m$^2$/(sV)is used, which is consistent with the study of Di Renzo and Cuenot~\cite{DiRenzo2022}. The diffusion coefficient of electrons is computed as $D_{e^-} = \dfrac{\upsilon_{e^-} k_B T}{q_e}$~\cite{McDaniel1973}.

\subsubsection{Magnetic permeability}

The permeability $\mu$ of a medium is given by the relation:
\begin{equation}
\begin{aligned}
    \mu = \left(1+\chi_m \right) \mu_0,
\end{aligned}
\end{equation}
where $\chi_m$ is the volume magnetic susceptibility and $\mu_0$ is the magnetic permeability of vacuum. 

For paramagnetic molecules, the volume magnetic susceptibility is computed according to Curie's law, which is appropriate for most paramagnetic materials at high temperatures. According to Curie's law, the volume magnetic susceptibility of a species $s$ is given by~\cite{Kittel2005}:
\begin{equation}
\begin{aligned}
    \chi_m^s = \dfrac{\mathcal{C}}{T}=\dfrac{N_s \mu_0 \mu_B^2 g_L^2 \mathscr{J}_s \left(\mathscr{J}_s+1\right)}{3k_B T},
    \label{eq:Curie_law}
\end{aligned}
\end{equation}
where $\mathcal{C}$ is the Curie constant, $N$ is the number of magnetic molecules per unit volume, $g_L$ is the Land\'e g-factor, $\mu_B$ is the Bohr magneton constant, and $\mathscr{J}$ is the total angular momentum quantum number. Typically, the contribution of the orbital angular momentum and spin of electrons is many orders of magnitude larger than the nuclear spin contribution~\cite{Levitt2013}. Moreover, in molecules and especially in organic molecules, the orbital angular momentum is typically quenched and thus the spin-only contribution is considered~\cite{Gerson2003}.

Therefore, the approximation $\mathscr{J}\approx \mathcal{S}$, where $\mathcal{S}$ is the total electron spin, is adopted. The Bohr magneton constant is given by $\mu_B=e\hbar/(2m_e)$, where $e$ is the elementary charge, $\hbar$ is the reduced Planck constant, and $m_e$ is the electron rest mass. In SI units, $\mu_B=9.27401\times 10^{-24}$ J/T. For an electron spin, $g_L=2.0023$~\cite{Kittel2005}. The spins of the various particles under consideration are reported in Table~\ref{Table:mag_prop}. Diamagnetic species have a negative volume magnetic susceptibility. Fixed values are provided by the literature. These are reported in Table~\ref{Table:mag_prop}.

For free electrons, the Pauli spin paramagnetism and Landau diamagnetism theories are considered. Specifically, the volume magnetic susceptibility of a Fermi gas consisting of free electrons, with either spin up, $\ket{\uparrow}$, parallel to the magnetic field or spin down, $\ket{\downarrow}$, antiparallel to the magnetic field, is given by the Pauli paramagnetism and Landau diamagnetism, respectively. For high-temperature kinetics ($\varepsilon_F/(k_B T) \ll 1$), the total magnetic susceptibility is given according to Stoner by~\cite{Stoner1935}:

For $\varepsilon_F/(k_B T) \gg 1$:
\begin{equation}
    \chi_m^e = \dfrac{\mu_0 \mu_B^2 N}{\varepsilon_F}\left[1-\dfrac{\pi^2}{12}\left(\dfrac{k_B T}{\varepsilon_F}\right)^2 - \dfrac{1}{10} \left(\dfrac{\mu_0 \mu_B H}{\varepsilon_F}\right)^2\right],
\end{equation}

For $\varepsilon_F/(k_B T) \ll 1$:
\begin{equation}
    \chi_m^e = \dfrac{2}{3}\dfrac{\mu_0 \mu_B^2 N}{k_B T}\left[1-\dfrac{1}{3}\left(\dfrac{2}{\pi}\right)^{1/2}\left(\dfrac{\varepsilon_F}{k_B T}\right)^{3/2} - \dfrac{7}{15} \left(\dfrac{\mu_0 \mu_B H}{k_B T}\right)^2\right],
    \label{eq:chi_m_elec}
\end{equation}
where $\varepsilon_F$ is the Fermi energy. For a three-dimensional, non-relativistic system of a non-interacting ensemble of electrons, the Fermi energy is given by~\cite{Kittel2005}:
\begin{equation}
    \varepsilon_F = \dfrac{\hbar^2}{2m_e}\left(3\pi^2N\right)^{2/3}.
\end{equation}

Examining the electron distribution inside a 1D methane-air reacting flow, it is evident that the second equation is appropriate for the high-temperature kinetics investigated in this study (since the ratio $\varepsilon_F/(k_B T) \ll 1$ is 7 orders of magnitude smaller than 1). The electrons' magnetic susceptibility is computed before the calculation of the static fields (in the EMI-static solver) or before the propagation of the electromagnetic waves (in the EMI-FDTD solver). It is also computed again before the computation of the electromagnetic forces. The molar mass of the electron is set to $5.485799\times 10^{-4}$~kg/kmol~\cite{Farnham1995}.

The total volume magnetic susceptibility of the mixture is given according to Wiedemann’s additivity law~\cite{Kuchel2003}: 
\begin{equation}
\begin{aligned}
    \chi_{m,mix} = \dfrac{\sum_s V^s \chi^s_m}{V},
    \label{eq:chi_m_mix}
\end{aligned}
\end{equation}
where from the ideal gas law, the partial volume is equal to $V^s = M_{m}Y_s V/M_s$.

\subsection{Boundary conditions for the reacting flow}
\label{subsubsec:reac_flow_BC}

The boundary conditions for the inflow, outflow, and periodic interfaces in SENGA follow the Navier–Stokes characteristic boundary condition (NSCBC) formulation~\cite{Poinsot1992,Sutherland2003}. Nonetheless, specific boundary conditions for the ions, as well as for the polar, polarizable, ferromagnetic, and diamagnetic species, are needed since their motion is affected by external electric and magnetic fields. For instance, in the case of electric fields, cations move towards the cathode and anions move towards the anode. Also, ferromagnetic fields are attracted by a positive magnetic field gradient. As a result, the inlet and outlet BC may differ for each species and each boundary cell face. In addition, the BC fields can be time-dependent. Note also that the BC is not only dependent on the force but also on the inlet velocity. Specifically, if the inlet velocity is sufficiently high and the electromagnetic fields are sufficiently low, then the motion of molecules is driven by the velocity of the fluid. Hence, the appropriate formulation of BCs should take into consideration spatial and temporal changes in electromagnetic fields, species distribution, and characteristics of the flow. Therefore, for each species and each boundary cell face, the following conditions apply: 
\begin{itemize}
    \item At the outlet B.C.: \\
    if $V^s_{\textrm{drift},i}<0$ and $|V^s_{\textrm{drift},i}>u_i+V^s_{\textrm{Fick},i}|$ then $Y_s=0$.
    \item At the inlet B.C.: \\
if $V^s_{\textrm{drift},i}<0$ and $|V^s_{\textrm{drift},i}|>|u_i+V^s_{\textrm{Fick},i}|$ then $\dfrac{\partial Y_s}{\partial x_i}=0$.
\end{itemize}
From Eq.~\eqref{eq:diffusion_velocity}, $V^s_{\textrm{drift},i}$ corresponds to the drift term of the diffusion velocity, whereas $V^s_{\textrm{Fick},i}$ corresponds to Fick's law diffusion. In all simulations performed in this study, any changes in the mass fraction of species are subtracted from the mass fraction of \ch{N2} to keep the sum of $Y_s$ equal to 1. This formulation of boundary conditions is valid for any local electric or magnetic field values or species' net charge.

\section{Validation, results, and discussion}

\subsection{Validation of the electrostatic field solver}
\label{subsec:validation}

For the sake of validation, a freely propagating premixed laminar flame of stoichiometric methane and air is simulated. The inlet flow had a pressure of 1~bar, a temperature of 300~K, and a velocity of 0.39~m/s. For an accurate discretization of the flame, a total of 1000 nodes were used. The results from EMI-SENGA are compared with Cantera. In these simulations, electrostatic interactions among ionic species are considered. In this study, the reduced chemistry mechanism developed by Di Renzo and Cuenot is used~\cite{DiRenzo2022}. The mechanism includes 26 species and 134 reactions. Two species are cations (\ch{HCO+} and \ch{H2O+}), one species is an anion (\ch{O2-}), and one species corresponds to the electrons. Ionic species are involved in 10 out of 134 reactions. 

\begin{figure}[!tb]
    \centering
    \graphicspath{{Figures/Comparison/ion_EF/}}
    \includegraphics[width=0.33\textwidth]{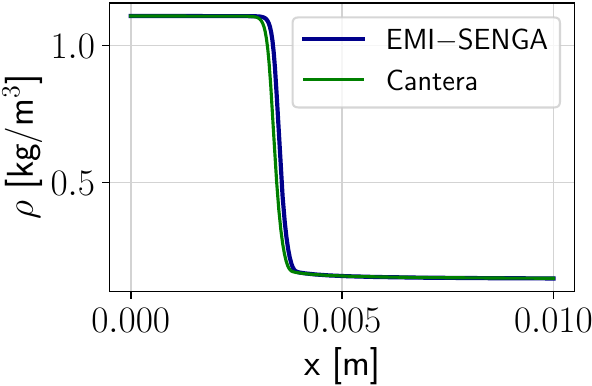}%
    \includegraphics[width=0.315\textwidth]{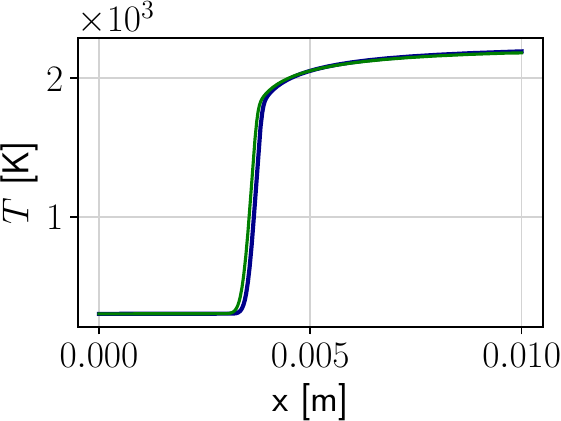}%
    \includegraphics[width=0.33\textwidth]{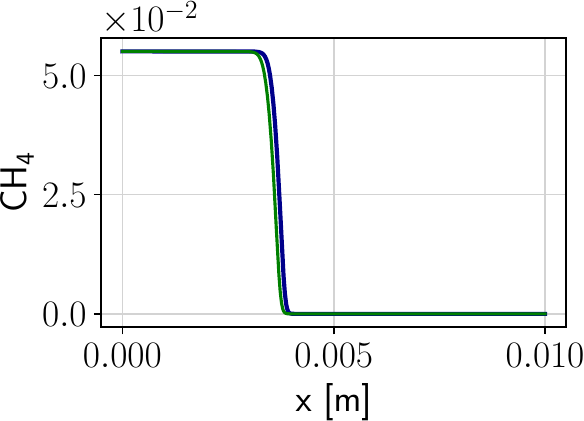}
    \includegraphics[width=0.33\textwidth]{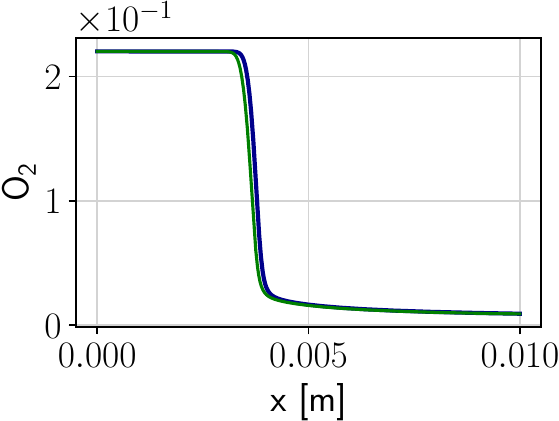}%
    \includegraphics[width=0.33\textwidth]{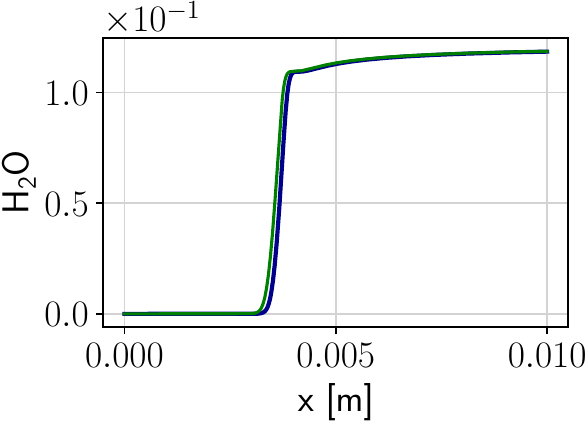}%
    \includegraphics[width=0.33\textwidth]{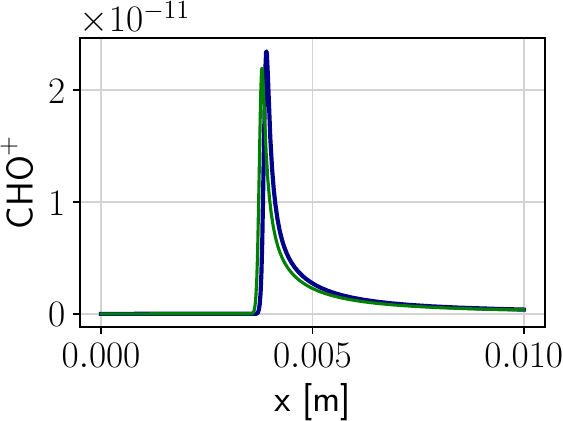}
    \includegraphics[width=0.33\textwidth]{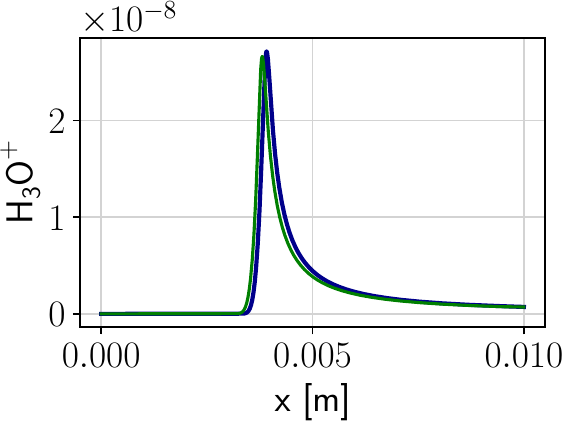}%
    \includegraphics[width=0.33\textwidth]{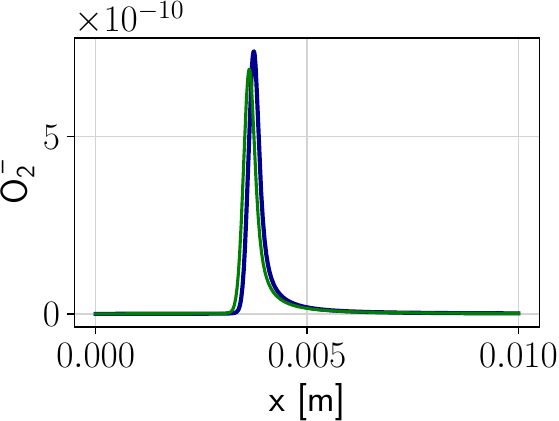}%
    \includegraphics[width=0.33\textwidth]{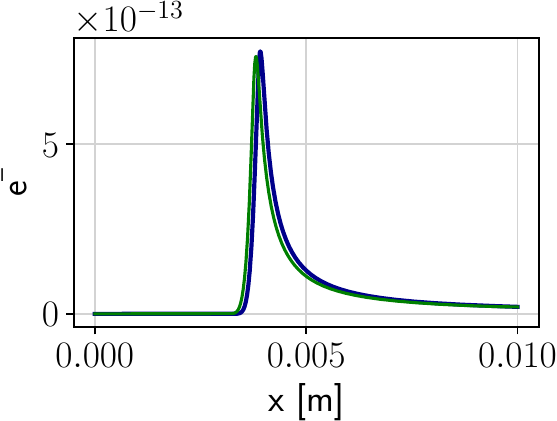}
    \caption{Comparison between Cantera and EMI-SENGA for selected quantities obtained with the ion transport model and with electrostatic interactions activated. A more detailed comparison is given in Section~\ref{SIsubsec:validation}.}
    \label{fig:comparison_ion_EF_selec}
\end{figure}

The comparison between EMI-SENGA and Cantera is shown in Fig.~\ref{fig:comparison_ion_EF_selec} for selected quantities. A more detailed comparison is given in Figs.~\ref{fig:comparison_ion_EF_1} and \ref{fig:comparison_ion_EF_2}. For this comparison and for the remainder of this work, the ion transport model is applied as described in Section~\ref{sisubsubsec:ion_transport_model}, together with the polynomials of Eq.~\eqref{eq:CA_mode}. In addition, electrostatic interactions are activated with zero external potential. Note that simulations in EMI-SENGA were initialized with a steady-state solution obtained from Cantera. Then, EMI-SENGA simulations have been advanced for a total simulated time of 7.5~ms using a timestep of 1~ns. At the end of the simulation, all fields, i.e., temperature, density, energy, pressure, velocity, and species mass fractions, had an absolute difference between the final and the tenth previous timestep of at least 6 orders of magnitude smaller than the maximum final value of the field. In addition, the values of the aforementioned fields at the final timestep and the previous 10~\textmu s timestep differed by less than 3 orders of magnitude compared to the maximum values of the final timestep. On the basis of these observations, it is concluded that both the slow and fast chemical and transport processes have reached a satisfactory steady state.

Overall, the EMI-SENGA solution is in agreement with the results of Cantera. As is evident from Figs.~\ref{fig:comparison_ion_EF_1} and \ref{fig:comparison_ion_EF_2}, the main difference between the two codes is that EMI-SENGA predicts a slightly different flame position. Moreover, the mass fractions of the main species (i.e., \ch{CH4}, \ch{O2}, \ch{CO2}, and \ch{H2O}) predicted by the two codes are in perfect agreement, with only small differences observed for the rest of the species. It is worth noting that the mass fractions of the ionic species predicted by the two codes are in very good agreement (see Fig.~\ref{fig:comparison_ion_EF_selec}), which validates the implementation of the electrostatic fields and the formulations of the equations that are solved for the potential, conservation equations, and forces.

\subsection{1D laminar reacting flow under external electrostatic fields}
\label{subsec:emi_1d_electrostatic}

The developed code is used to investigate the effects of external electrostatic fields on a one-dimensional laminar freely propagating premixed flame. An external electric field parallel to the direction of propagation of the flame front is considered. Similar to Section~\ref{subsec:validation}, the flame was discretized on 1000 nodes and had inflow conditions at 1~bar, 300~K, and 0.39~m/s. Solutions were obtained for 22 different externally applied electric potentials. Also, the methane and air mixture was at stoichiometric conditions. The timestep of each simulation varied depending on the strength of the electric potential and ranged from 0.05 to 1~ns. Moreover, the total simulation time varied among the cases. In all cases, similar to Section~\ref{subsec:validation}, all fields have an absolute error between the final and the tenth previous timestep of at least 6 orders of magnitude smaller than the maximum final value of the field.

The potential is applied on the left boundary of the domain, while the right one is grounded. Both positive and negative potential values are studied. For the ionic species, the left BC described in Section~\ref{subsubsec:reac_flow_BC} is applied. For the right boundary, the subsonic non-reflecting outflow NSCBC is applied for all species. Therefore, at the outflow, the boundary condition of this simulation is not affected by the electric field. This is appropriate for the studied premixed configuration since it is assumed that ionic species exist outside the domain, and thus their mass fraction should not be set to zero at the boundary.

\begin{figure}[t]
    \centering
    \includegraphics[width=0.5\linewidth]{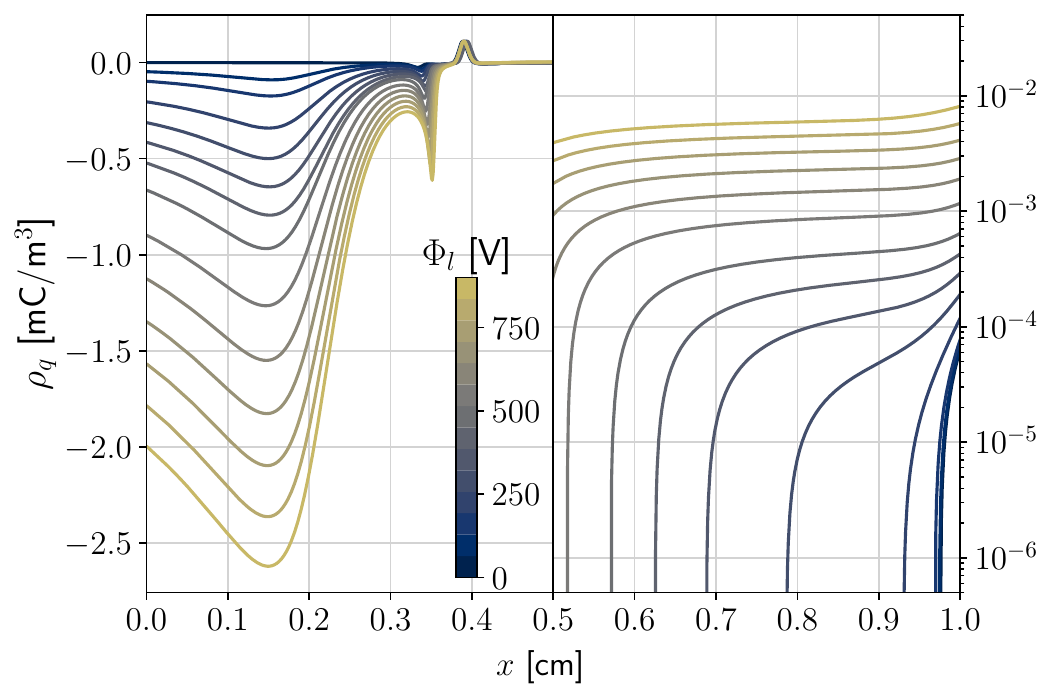}%
    \includegraphics[width=0.5\linewidth]{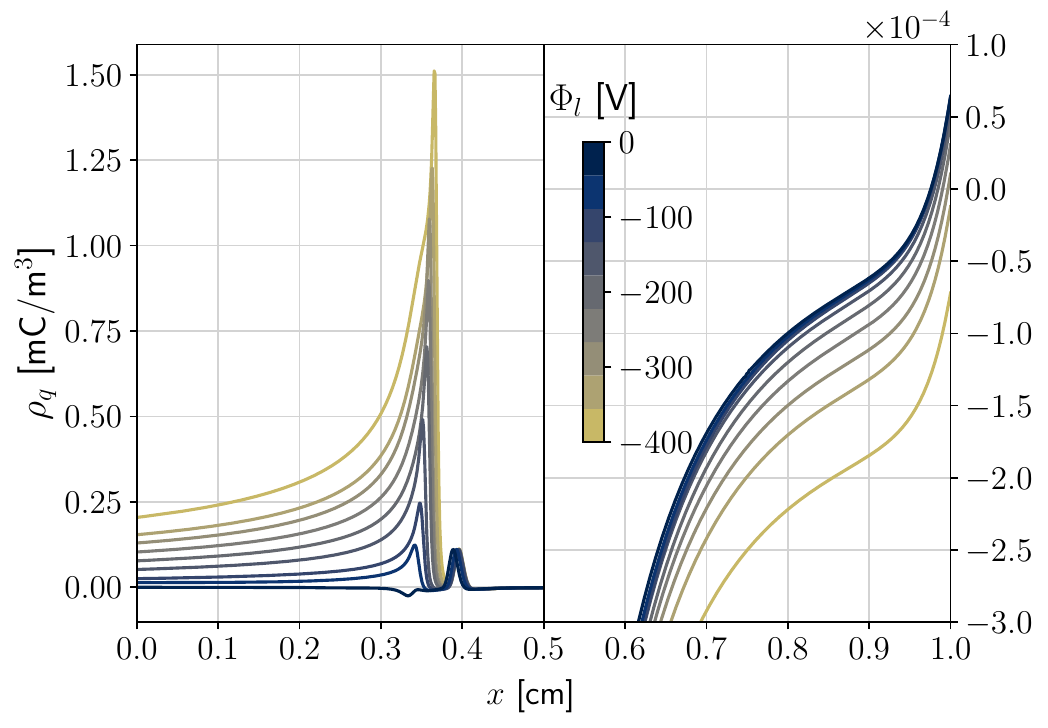}
    \caption{Charge density for a left electrode with a positive (left) and negative (right) voltage. The right electrode is grounded.}
    \label{fig:qden_field}
\end{figure}

The charge density under various external electric potentials is shown in Fig.~\ref{fig:qden_field}. The Lorentz force leads to an accumulation of cations and depletion of anions near the anode and vice versa near the cathode. This behaviour is consistent with previous studies~\cite{Belhi2013,Speelman2015,DiRenzo2018}. Let us consider the positive potential values of the left electrode. On the left side of the reacting zone and near the left boundary of the domain, there is a high accumulation of negatively charged species. Investigation of the distribution of ionic species shows that the profile of the charge density on the left side of the reacting zone matches the profile of the mass fraction of the \ch{O2-} molecules. Di Renzo et al.~\cite{DiRenzo2018} observed a similar increase in \ch{O2-} near the oxidizer stream in their counterflow configuration, which was also the ion with the highest concentration. Two possible phenomena could contribute to the accumulation of \ch{O2-} molecules near the boundary. On the one hand, oxygen anions are generated in the reacting zone and then exit the domain from the left side under the influence of the Lorentz force. On the other hand, the highly mobile electrons travel upstream of the inflow nozzle and ionize the \ch{O2} molecules. The two reactions of the chemical mechanism that can lead to oxygen ionization via collisions with electrons are \ch{e- + 2 O2 <=> O2 + O2-} and \ch{e- + N2 + O2 <=> N2 + O2-}. Considering that both the mass fractions of \ch{O2-} molecules and the charge density have a peak near the inlet boundary but a low value near the reacting zone, the second phenomenon can be deemed as the main reason that governs the generation of \ch{O2-} molecules. Note, also, that the magnitude of the charge density is directly proportional to the strength of the electric potential difference. It should be noted that the charge density on the right side of the reacting zone is significantly smaller than the one on the left side. Even under a strong 900~V potential, the charge density is 2 orders of magnitude lower. This is probably a consequence of the low mobility of the cation species that are quickly consumed before they reach the outlet boundary.

When a negative potential is applied at the left electrode, the shape of the charge distribution is different. In particular, there is a high accumulation of cations near the reacting zone, while there is a negligible concentration of ionic species exiting the domain. Examining the chemical mechanism~\cite{DiRenzo2022}, it can be seen that there are no reactions involving cations and reactants (either fuel or oxidizer). This can explain the absence of generated ionic species near the left boundary compared to the positive electric potentials. Overall, for all values of the electric potential, an increase in the potential magnitude leads to more ionic species exiting the domain from both boundaries. It should be noted that a shift in the flame position compared to the case with no potential difference was observed in all cases with an external electric force. However, this shift was not consistent among the various cases, probably because the simulation time varied among the cases. Therefore, definitive conclusions about the location of the reacting zone cannot be made and are left for future studies.

The electrostatic field, which is the result of the internal charge density and the external potential, is higher for positive electric potentials compared to negative ones. For instance, at 400~V and -400~V, the electric field reaches maximum values of 0.31~MV/m and 0.17~MV/m, respectively. The internal electric field for the different investigated cases is plotted in Figure~\ref{fig:elfx_field}. Moreover, the charge density leads to electric field gradients in the reacting zone. When a positive potential is applied at the left electrode, the shape of the electric field in the reacting zone does not deviate considerably from the case with zero external potential. On the contrary, the electric field profile significantly changes when a negative potential is applied, even for weak values of the potential. The electric field gradients in the reacting zone can introduce a polarization force for polar or field-polarized species.

\begin{figure}[t]
    \centering
    \includegraphics[width=0.5\linewidth]{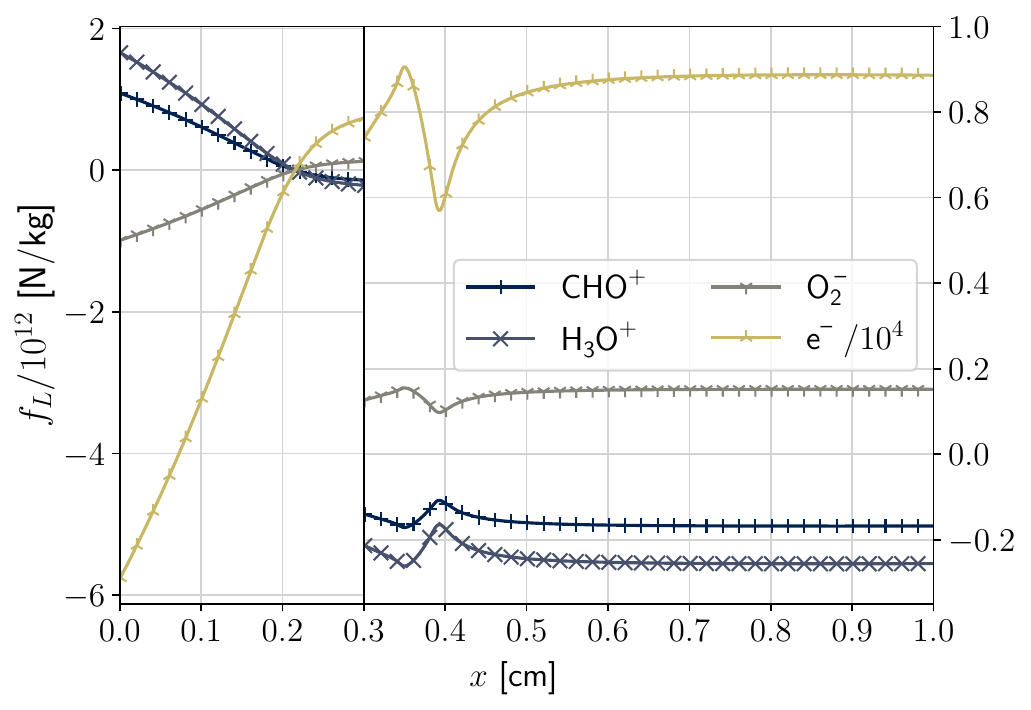}%
    \includegraphics[width=0.5\linewidth]{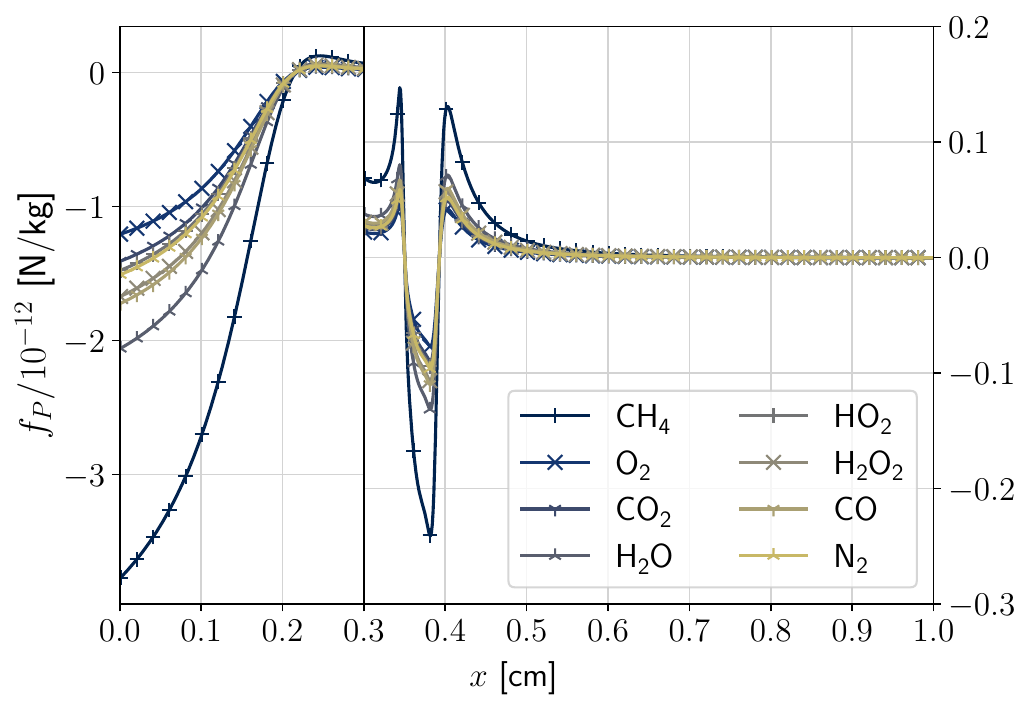}%
    \caption{Lorentz force on charged species (left) and Polarization force on various neutral species (right) for the case with $\Phi_l=500$~V.}
    \label{fig:lorentz_polar}
\end{figure}

The Lorentz force per unit mass (acceleration induced by electric forces) acting on the ionic species is shown in Fig.~\ref{fig:lorentz_polar}. Under the same electric field, electrons experience the highest acceleration due to their smaller mass, followed by the acceleration of \ch{O2-} molecules. Additionally, it is noted that the direction of the forces changes at $x=2$~mm, due to the difference in the sign of the electric field reported in Fig.~\ref{fig:elfx_field}. Figures~\ref{fig:lorentz_polar} and~\ref{fig:polarization_field_2} show the polarization forces of neutral molecules. The polarization forces near the right boundary are negligible for all studied species since there are no electric field gradients. On the contrary, forces are evident in the reacting zone and near the left boundary. The acceleration of neutral species due to the polarization forces is 24 orders of magnitude smaller than the Lorentz forces acting on the ionic species. However, since the mass fractions of ionic molecules can be lower than 10 orders of magnitude compared to neutral molecules (see Figs.~\ref{fig:comparison_ion_EF_1} and \ref{fig:comparison_ion_EF_2}) and since the polarization forces act mainly near the inflow boundary where most ionic species have low concentration, the polarization force may affect the reacting flow. The investigation of these secondary effects is left for future study. It is also worth noting that the developed formulation and code allow for the description of the interactions between highly polarizable or magnetizable particles (e.g., nanomaterials) and electromagnetic fields in reacting flows (where strong polarization and magnetization forces are expected). This is another interesting area for future developments.

\subsection{1D laminar reacting flow under uniform magnetostatic fields}
\label{subsec:1D_mag}

The effects of magnetostatic fields on 1D laminar flames are investigated herein. The configuration of the flame is similar to the one of Section~\ref{subsec:emi_1d_electrostatic}. In this case, simulations are performed with a 68-step chemical mechanism for methane that does not include any ionic species~\cite{Warnatz2006}, in order to focus on the effects of magnetic fields on neutral species. Note that if ionic species were present, the magnetic field would induce Lorentz forces in the direction perpendicular to the direction of flame propagation, leading to ions' motions outside the 1D domain. The external magnetic potential is applied on the left and right boundaries, similarly to the electrostatic potential cases discussed in Section~\ref{subsec:emi_1d_electrostatic}, since the formulation implemented here (see Section~\ref{subsubsec:electrostatic_magnetostatic_fields}) is based on the solution of the internal magnetic potential. The total simulated time is 1~ms with a timestep of 10~ns.

\begin{figure}[!tb]
    \centering
    \includegraphics[width=0.5\linewidth]{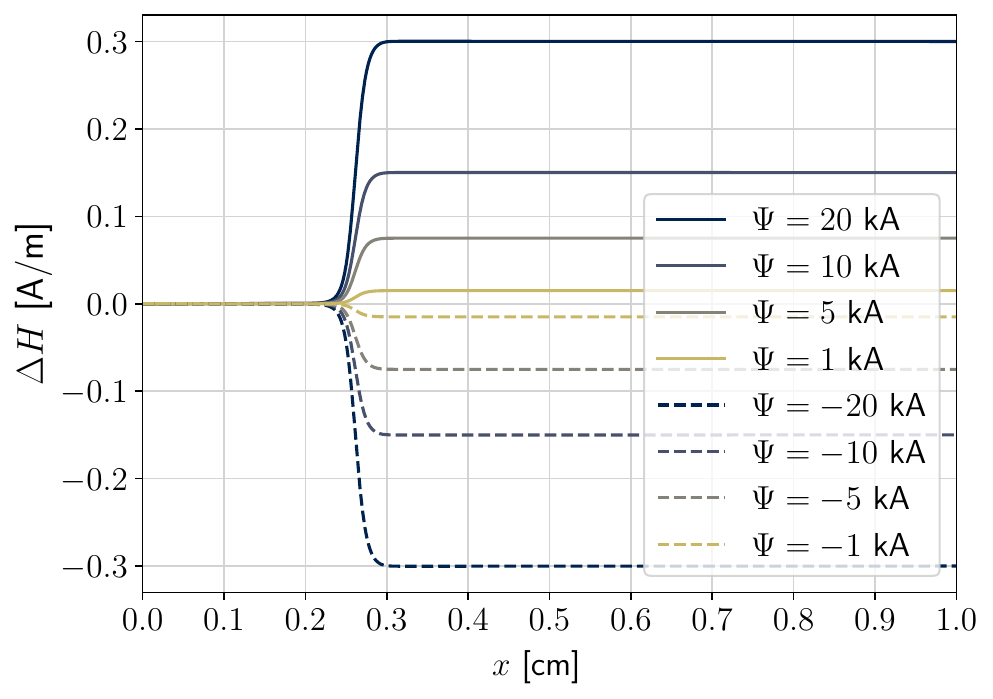}
    \caption{Relative magnetic field with respect to the value at the left boundary.}
    \label{fig:mafx_chim}
\end{figure}

Figure~\ref{fig:mafx_chim} shows the relative magnetic field induced by different external magnetic potentials. The magnetic field changes near the flame front, due to the big decrease in the magnetic susceptibility of the mixture. The magnetic moment of the species affects the magnetic field only slightly. Even under a strong potential of 20~kA, the magnetic field changes only by 0.3~A/m. Moreover, the mixture's magnetic susceptibility is not significantly affected by the external potential (see Fig.~\ref{fig:mafx_chim_si}), which suggests that the species distribution is also not affected. By investigating the profiles of the various quantities, such as temperature, density, and species mass fractions (not shown here), it was found that the magnetic field did not considerably alter them. Note that the flame may require much longer than the simulated time to respond to these small variations in the magnetic field. This should be further investigated in future work.

It is noted that the inability of the reacting flow to induce internal magnetic fields, and thus affect its own behaviour, was experimentally observed in previous studies~\cite{Mayo1967,Mizutani2001}. In 1967, Mayo~\cite{Mayo1967} investigated a small ethylene diffusion flame inside a static magnetic field. It was found that even at a 1~T magnetic field strength, the effects of the magnetic fields on the flame shape were negligible. Only in cases with low fuel inflow velocities was it possible to observe a deflection of the flame. This phenomenon may be attributed to the fact that the magnetic field was relatively uniform, and thus the effects were mainly caused by the Lorentz force acting on charged species produced in the reaction zone. The more recent study by Mizutani et al.~\cite{Mizutani2001} investigated the effects of a 5~T uniform magnetic field on premixed laminar flames. Despite the high strength of the magnetic fields, they found negligible effects on the fast chemical reactions and thus on the flame propagation and temperature.

On the other hand, inhomogeneous magnetic fields with strong gradients can have a considerable impact on the reacting flow. This occurs due to the paramagnetic and diamagnetic properties of species that exist in flames. The effects of inhomogeneous magnetic fields on the flame behaviour were first investigated by Ueno and Harada~\cite{Ueno1987} for methane, propane, and hydrogen flames under magnetic field strengths up to 1.6~T and with gradients up to 220~T/m. In their experiments, they observed that flames bent away from high-intensity magnetic fields. The effects were attributed to the paramagnetic properties of \ch{O2}. The behaviour of flames in an inhomogeneous magnetic field is discussed in Section~\ref{subsec:2d_laminar_magnet}.

\begin{figure}[t]
    \centering
    \includegraphics[width=0.48\linewidth]{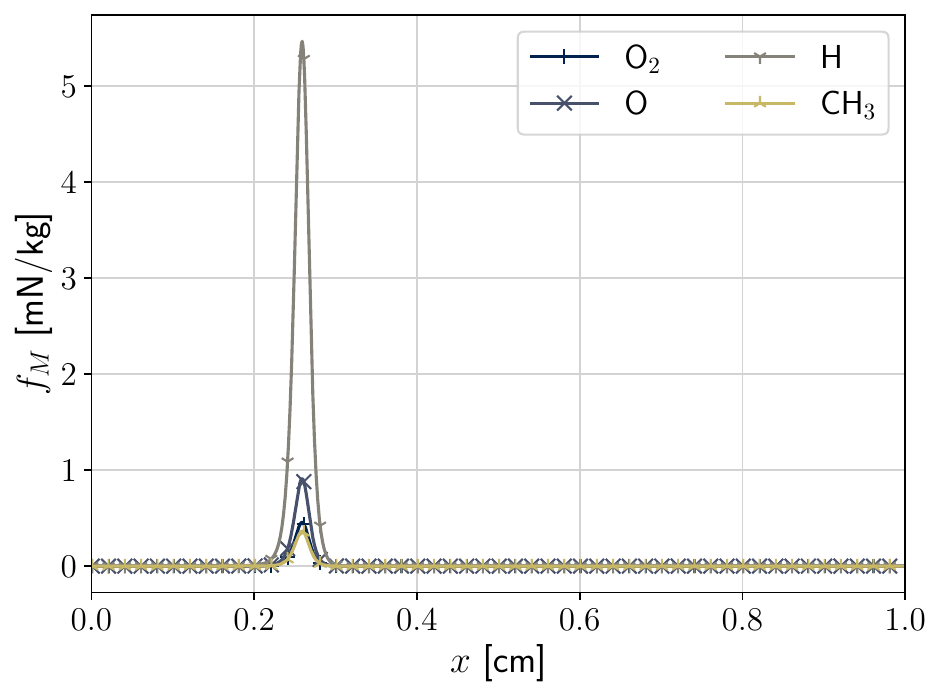}%
    \includegraphics[width=0.5\linewidth]{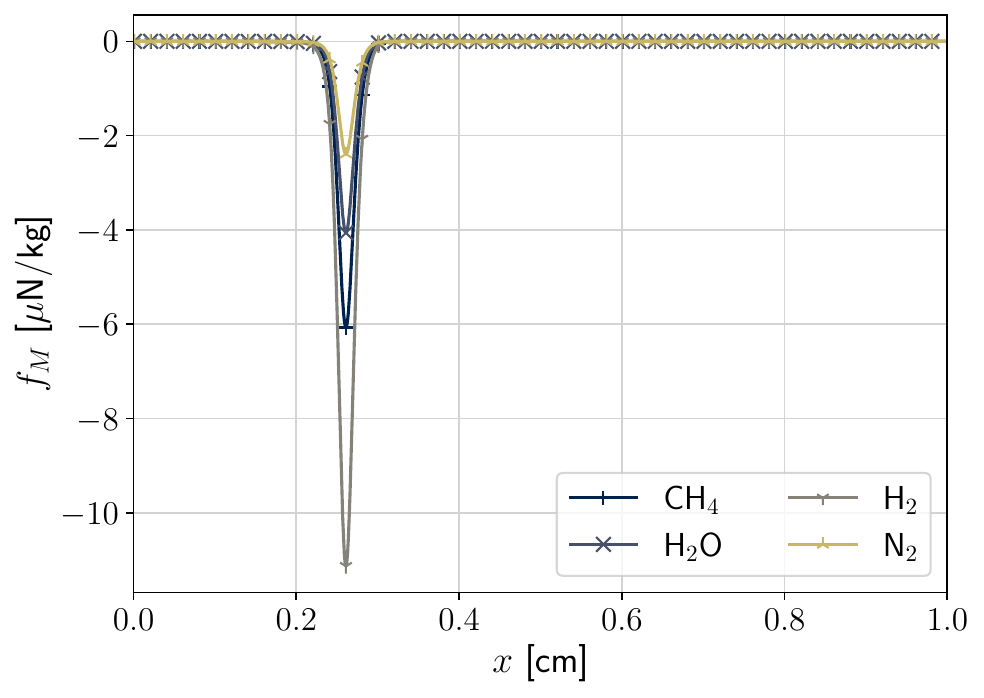}
    \caption{Magnetization force for various paramagnetic (left) and diamagnetic (right) molecules for $\Psi=10$~kA}
    \label{fig:magnetization_param_diam}
\end{figure}

The magnetization forces for selected paramagnetic and diamagnetic species are shown in Fig.~\ref{fig:magnetization_param_diam}. The paramagnetic and diamagnetic molecules have opposing forces. Paramagnetic molecules are attracted to positive magnetic field gradients, whereas diamagnetic species are repelled. Moreover, the magnetization forces on diamagnetic molecules are a couple of orders of magnitude weaker compared to the forces on paramagnetic species, due to the lower magnetic moment of diamagnetic species. It is also interesting to note that the force per unit mass on hydrogen is significantly stronger compared to other species. Therefore, hydrogen's small mass overcomes its small spin (e.g., compared to oxygen atoms with a spin of one), resulting in the strongest magnetization force per unit mass.

\begin{figure}[!tb]
    \centering
    \includegraphics[width=0.5\linewidth]{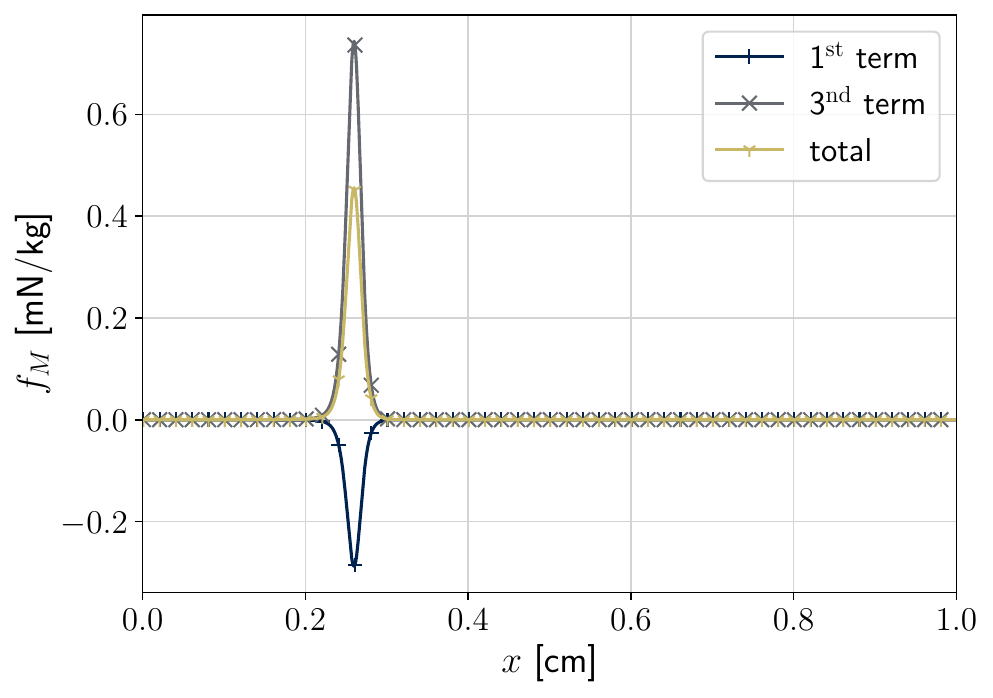}%
    \includegraphics[width=0.5\linewidth]{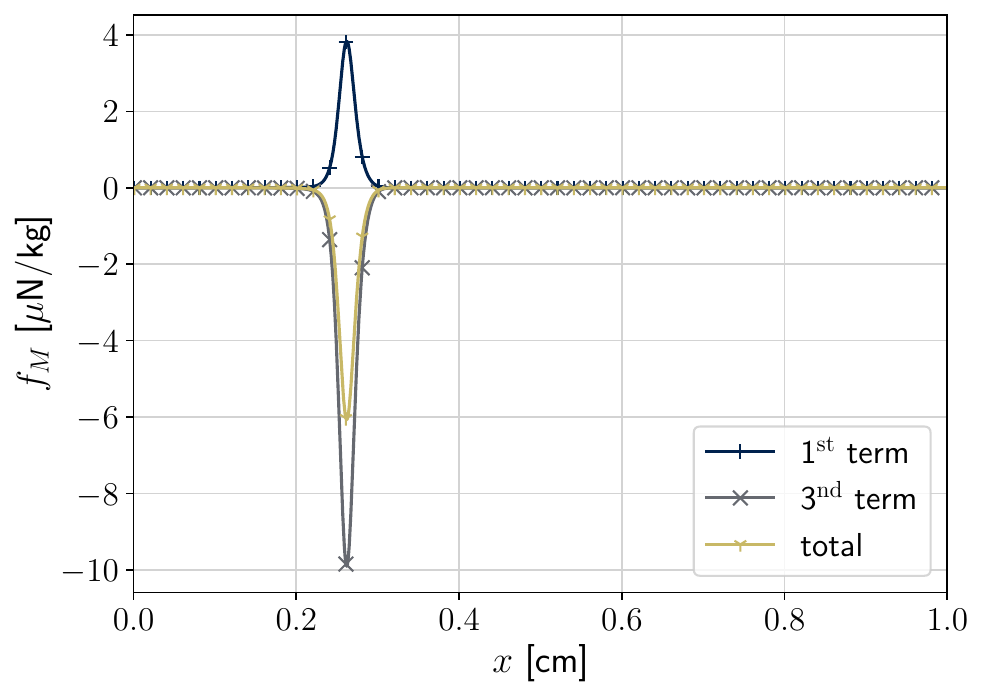}
    \caption{Contributions to the magnetization force for \ch{O2} (left) and \ch{CH4} (right) for $\Psi=10$~kA. The \nth{1} and \nth{3} terms refer to the right-hand side terms of Eq.~\eqref{eq:magnetization_force}. Markers are added to every 20th node of the grid.}
    \label{fig:magnetization_force_terms}
\end{figure}

Finally, the contributions of the gradients of the magnetic susceptibility and magnetic field to the total force of Eq.~\eqref{eq:magnetization_force} are plotted in Fig.~\ref{fig:magnetization_force_terms} for the \ch{O2} (paramagnetic) and \ch{CH4} (diamagnetic) species. It is reminded that for this one-dimensional configuration, the triple vector cross product in Eq.~\eqref{eq:magnetization_force} is zero. Therefore, only the first and third terms are plotted. By examining the distribution, it can be seen that the contribution of the magnetic susceptibility gradient is smaller and opposes the contribution of the magnetic field for both \ch{O2} and \ch{CH4}. As a result, the superposition of the first and third terms on the RHS of Eq.~\eqref{eq:magnetization_force} gives the total force in the direction of the third term. Nonetheless, Fig.~\ref{fig:magnetization_force_terms} also shows that the first term considerably affects the magnitude of the total force. These findings suggest that the gradient of the magnetic susceptibility of Eq.~\eqref{eq:magnetization_force} cannot be neglected for static magnetic fields or other configurations where weak magnetic fields are present, as it may have an important effect on the dynamics of the reacting species.

\subsection{2D laminar reacting flow under inhomogeneous magnetostatic fields}
\label{subsec:2d_laminar_magnet}

The analysis continues with the study of a 2D laminar reacting flow. The same chemical mechanism with only neutral species that was used in Section~\ref{subsec:1D_mag} is applied here. The 2D configuration investigated here is constructed from the 1D configuration of Section~\ref{subsec:1D_mag} by repeating the initial solution along the $y$ direction. The domain has a size of $L_x=1$~cm and $L_y=0.74$~cm and is discretized with $N_x=500$ and $N_y=370$ nodes. 

\begin{figure}[!tb]
    \centering
    \includegraphics[width=0.55\linewidth]{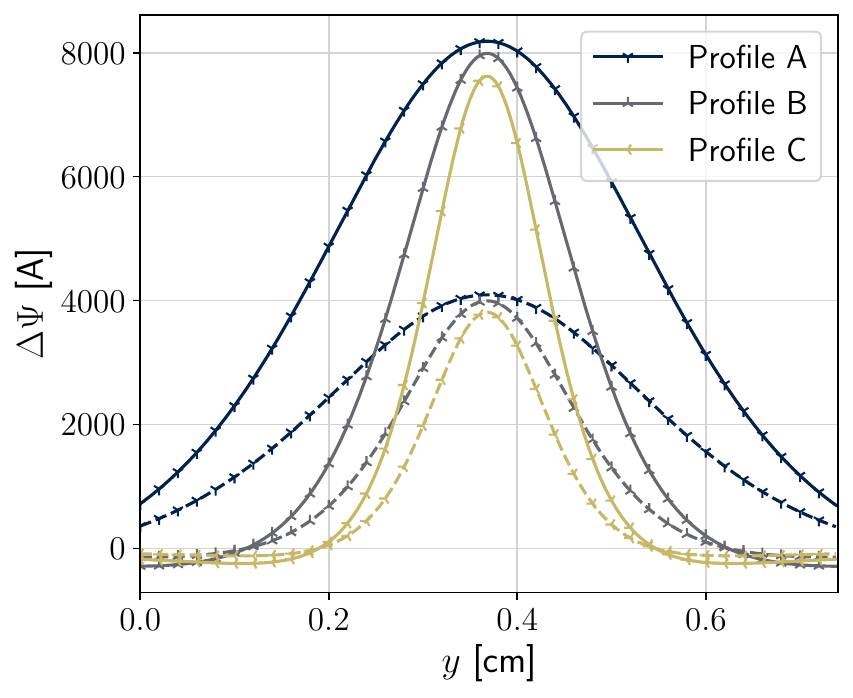}
    \caption{Magnetic potential profiles imposed at the left boundary. The dashed lines were obtained for a magnetization of 1~T. The solid lines were obtained by doubling the potential profiles of the dashed lines.}
    \label{fig:magnetic_potential}
\end{figure}

\begin{figure}[!t]
    \centering
    \includegraphics[width=0.5\linewidth]{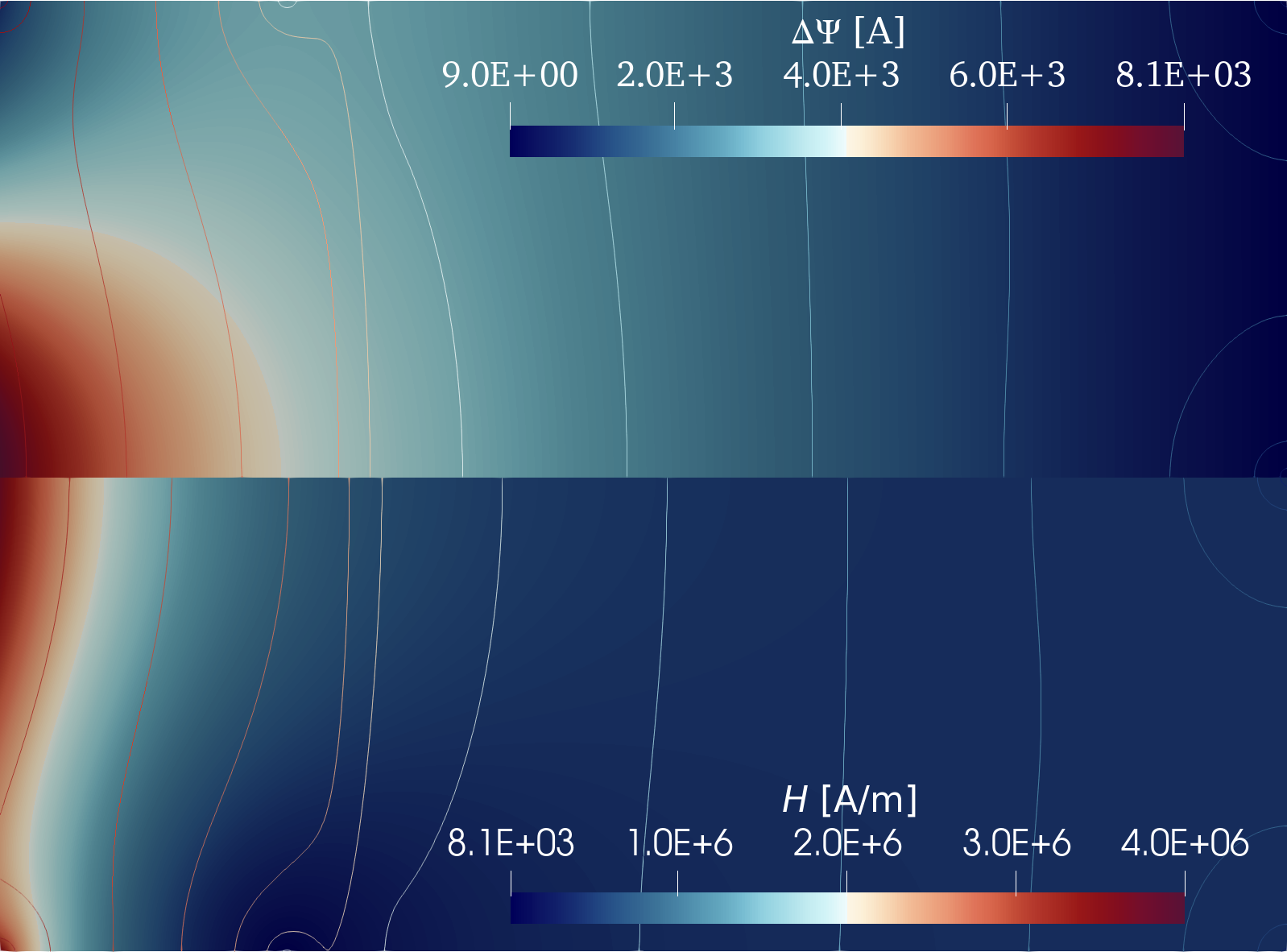}
    \caption{Magnetic potential and magnetic field distribution in the domain for the solid line Profile B of Fig.~\ref{fig:magnetic_potential}.}
    \label{fig:magnetic_field_and_potential}
\end{figure}

The profiles of the external magnetic potentials are shown in Fig.~\ref{fig:magnetic_potential}. They are imposed on the left boundary and are obtained from the magnetic field generated by a cylindrical magnet. The profiles A, B, and C, correspond to a magnetic with diameter $D_m$ of $\dfrac{L_y}{2}$, $\dfrac{L_y}{3}$, and $\dfrac{L_y}{4}$, respectively, and a length $L_m$ of $\dfrac{L_y}{2}$, $\dfrac{L_y}{3}$, and $\dfrac{L_y}{4}$, respectively. The dashed lines of Fig.~\ref{fig:magnetic_potential} were obtained for a magnetization of 1~T. These profiles were obtained using the magpylib code~\cite{magpylib2020}. To investigate the effects of strong magnetic fields, additional profiles were constructed by doubling the previous potential profiles. The magnetic potential and magnetic field distribution are plotted in Fig.~\ref{fig:magnetic_field_and_potential}. It can be seen that the boundary condition directly affects both the longitudinal and lateral distribution of the magnetic field inside the domain. Therefore, 2D and 3D effects can be introduced as a consequence of the BCs. Note also that the magnetic field appears strong in a confined area near the boundary. In the reacting zone (located at approximately 1/3 of the domain length) the magnetic field is significantly smaller than near the boundary.

Previous numerical studies have revealed that magnetostatic fields can have significant effects on reacting flows. Specifically, by simulating hydrogen-oxygen flames, Yamada et al.~\cite{Yamada2002,Yamada2003,Yamada2003a} showed that changes in oxygen distribution due to the magnetic field gradient could affect the reaction process. It was concluded that the spatial distribution of OH radicals changes indirectly due to the diffusion of oxygen molecules and not directly due to the forces induced by the magnetic field gradient. In addition, it was shown that the orientation of the magnetic field can have different effects on the distribution of species.

\begin{figure}[!t]
    \centering
    \includegraphics[width=0.5\linewidth]{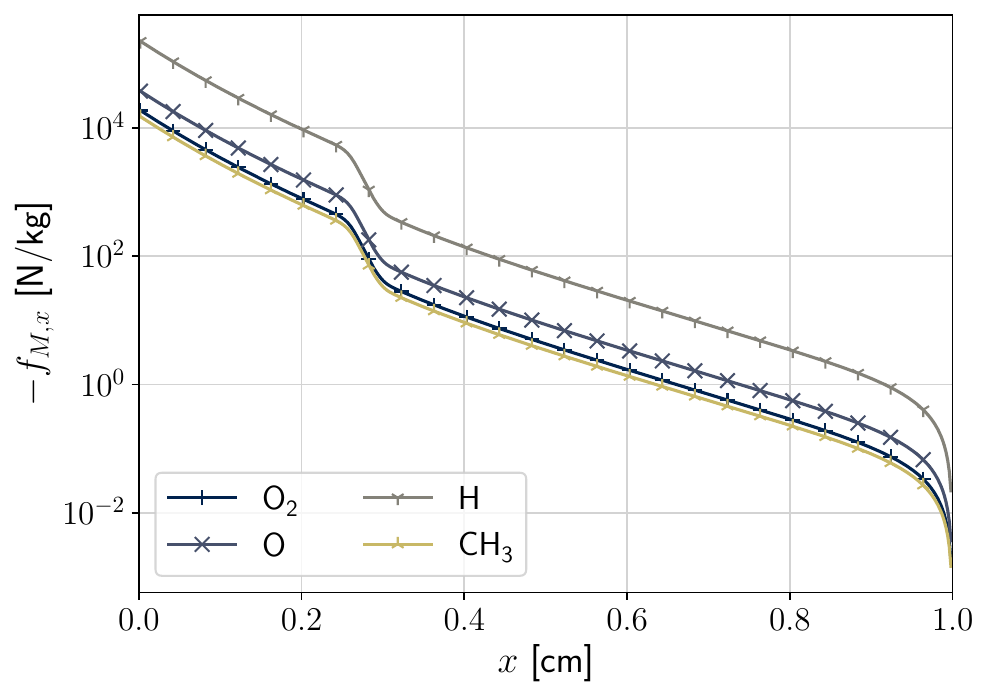}%
    \includegraphics[width=0.5\linewidth]{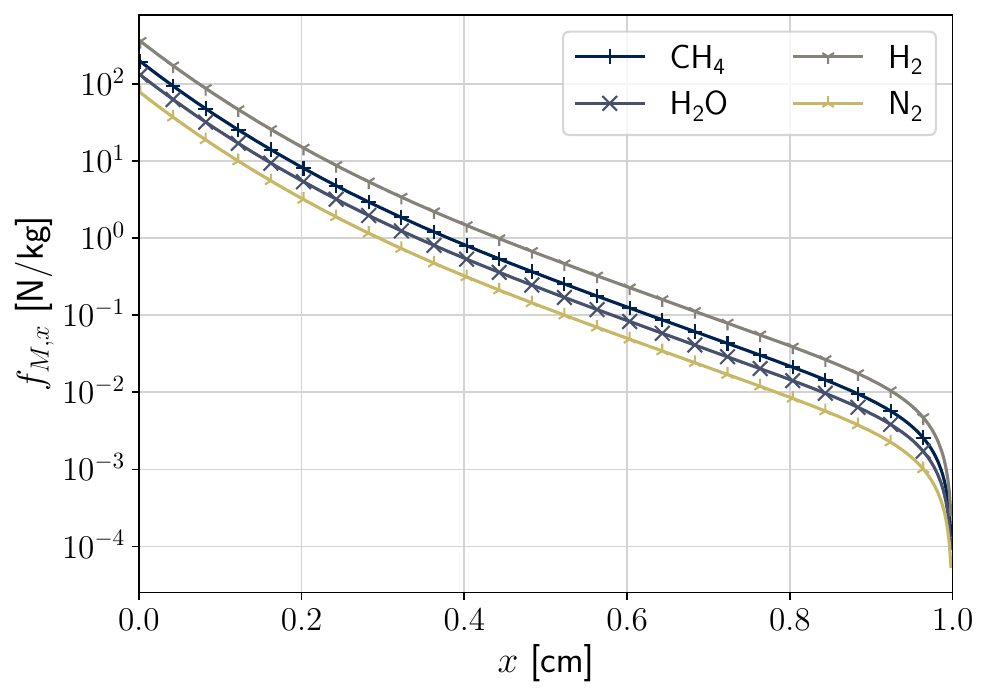}
    \caption{Magnetization force acting on paramagnetic (left) and diamagnetic (right) species for the solid line Profile B of Fig.~\ref{fig:magnetic_potential}.}
    \label{fig:magnetization_force}
\end{figure}

The magnetization forces of paramagnetic and diamagnetic species are shown in Fig.~\ref{fig:magnetization_force}. In addition, the magnetization force of \ch{O2} is plotted in Fig.~\ref{fig:magnetization_force_oxygen}. Note that the effect of the gradient of the magnetic susceptibilities on the magnetization force can be neglected here. The reason is that the magnetization force in the vicinity of the reacting region shown in Fig.~\ref{fig:magnetization_force} is significantly larger than the one shown in Fig.~\ref{fig:magnetization_param_diam}. Moreover, a large discrepancy in the magnitude of the forces can be observed between the left and right $x$ boundaries. The force acting on both the diamagnetic and paramagnetic species near the left boundary is 6 orders of magnitude higher than on the right side. Therefore, in this configuration, the reactants are affected to a higher degree than the products. Note also that for the paramagnetic species only, a large decrease in the forces occurs at $x\approx 3$~mm, i.e., at the reacting zone. This decrease is attributed to the temperature dependence of the magnetic susceptibility of paramagnetic species according to Curie's law, see Eq.~\eqref{eq:Curie_law}.

\begin{figure}[!t]
    \centering
    \includegraphics[width=0.5\linewidth]{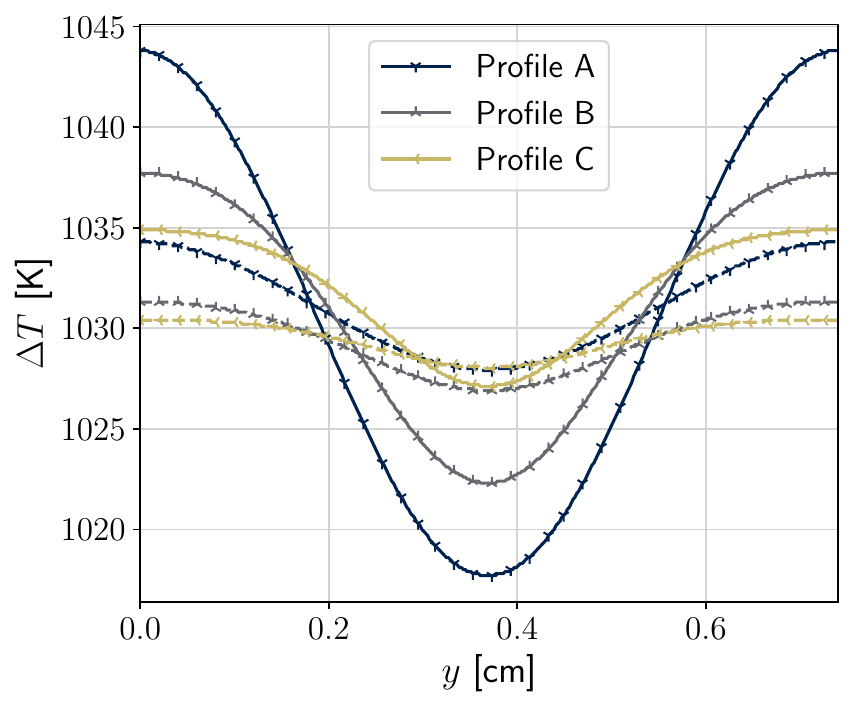}
    \caption{Flame temperature under different magnetic potentials at $x\approx 2.85\times 10^{-3}$.}
    \label{fig:temperature_mag}
\end{figure}

The effects of the various magnetic potential profiles on the reacting flow are investigated by examining the temperature of the reacting region. The temperature across the $y$ direction passing from a specific $x$ point is shown in Fig.~\ref{fig:temperature_mag}. It is evident that the temperature field is directly affected by the external magnetic potential. The magnetization forces affect the species distribution and, in particular, the distribution of paramagnetic species. As shown in Fig.~\ref{fig:magnetization_force_oxygen}, the magnetization forces are not parallel to the $x$ coordinate. As a result, the reacting zone gets curved. A temperature variation of 2 to 26~K can be observed across the $y$ direction for the magnetic potentials shown in Fig.~\ref{fig:magnetic_potential}.

\subsection{Validation of the EMI-FDTD solver}

\begin{figure}[!tb]
    \centering
    \includegraphics[width=0.5\linewidth]{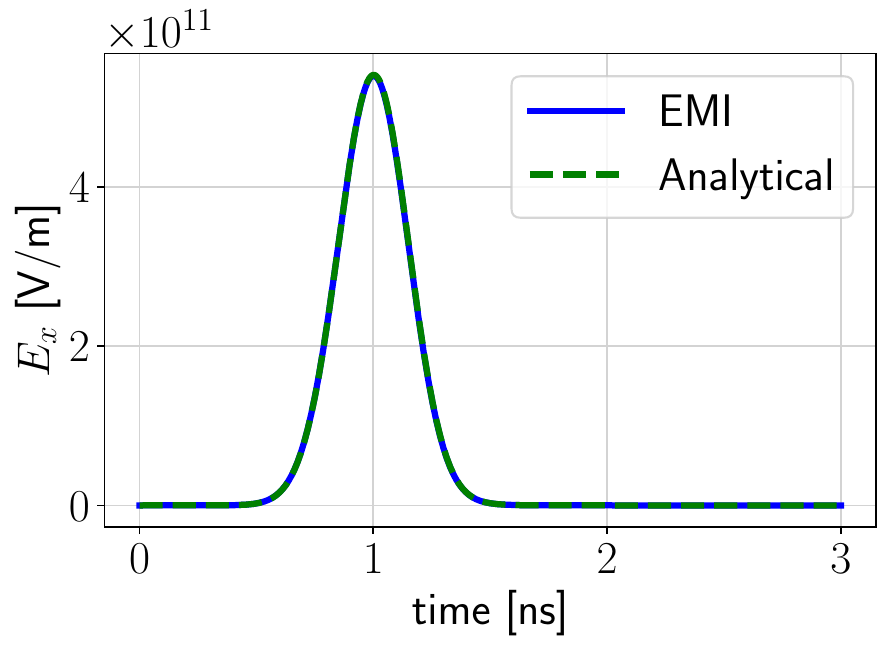}%
    \includegraphics[width=0.5\linewidth]{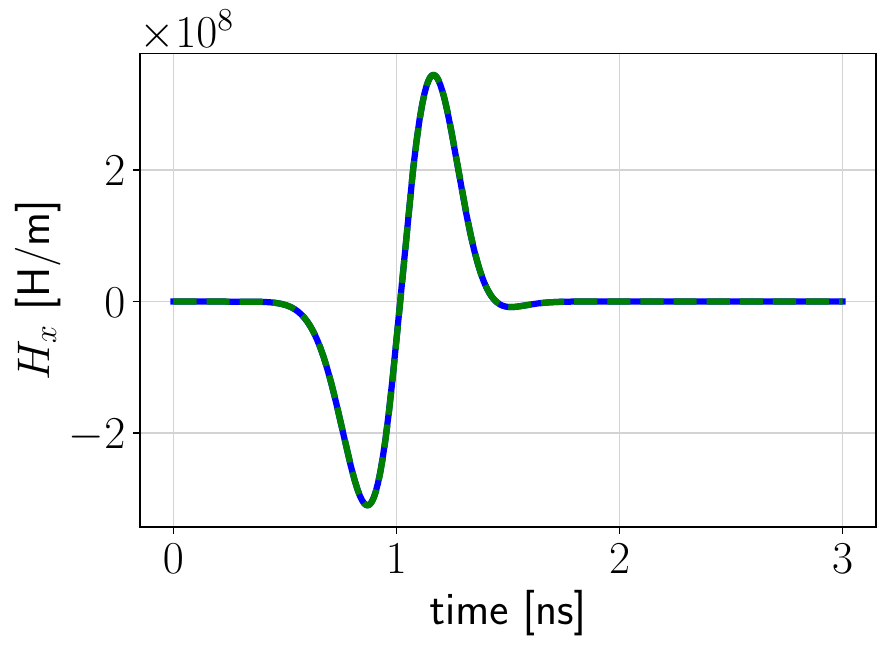}
    \caption{Comparison between EMI-FDTD and the analytical solution of Ref.~\cite{ziolkowski1983} for a Hertzian dipole source with a Gaussian derivative current waveform.}
    \label{fig:fdtd_profiles_ExHx}
\end{figure}

In this section, the implementation of the EMI-FDTD solver is compared with three-dimensional analytical and numerical solutions. A soft Hertzian dipole source oriented in the $z$ direction is applied (see Section~\ref{subsubsec:electromagnetic_waves}). The waveform of the current, $\mathcal{I}$, is the derivative of a Gaussian pulse and is given by:
\begin{equation}
    \mathcal{I}=-4\pi^2 f^2 \left(t-\frac{1}{f}\right) e^{-2\pi^2 f^2 \left(t-\frac{1}{f}\right)^2}.
\end{equation}
The domain is a cube with a length of 0.1~m and grid size $\Delta x =\Delta y = \Delta z = 1$~mm. 10 CPML cells are applied to all boundaries. In addition, the source has a frequency of $f=1$~GHz and is positioned at the node (20, 20, 20), excluding the CPML cells. A probe is placed at the node (35, 35, 35), which monitors the time evolution of the electric and magnetic field components. The analytical solution is obtained from Ref.~\cite{ziolkowski1983}. Results are shown in Fig.~\ref{fig:fdtd_profiles_ExHx} for the $x$ components of the electric and magnetic fields and in Fig.~\ref{fig:fdtd_profiles} for the other components. The numerical and analytical solutions perfectly overlap. Note that the solution of the $H_z$ component is characterized by small fluctuations instead of being equal to zero as predicted by the analytical solution. Nonetheless, this numerical noise is several orders of magnitude smaller than the other components and can be considered negligible. 

\begin{figure}[!t]
    \centering
    \includegraphics[width=\linewidth]{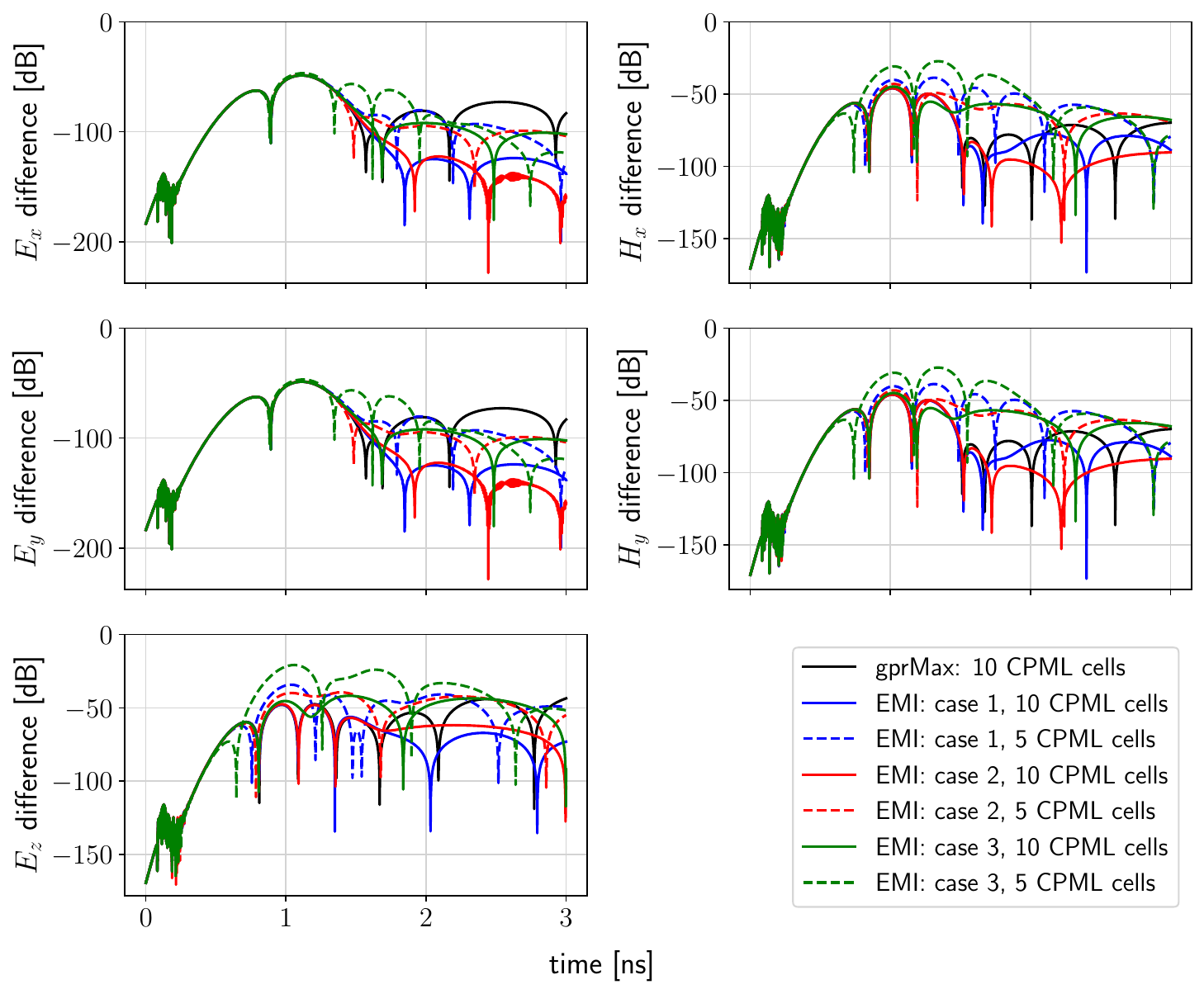}
    \caption{Difference between the numerical solutions obtained with EMI-FDTD and gprMax~\cite{Warren2016} and the analytical solution of Ref.~\cite{ziolkowski1983} for a Hertzian dipole source with a Gaussian derivative current waveform. Results for various CMPL parameters and CPML boundary thicknesses are reported. The cases are detailed in Table~\ref{tab:cpml_parameters_validation}.}
    \label{fig:fdtd_error}
\end{figure}

The accuracy of the EMI-FDTD solution is compared with the widely used gprMax code~\cite{Warren2016} for various CPML parameters and CPML boundary thicknesses. The various cases are reported in Table~\ref{tab:cpml_parameters_validation}. The reader is reminded that the parameters of Table~\ref{tab:cpml_parameters_validation} and the relevant equations that are solved in the CPML regions are detailed in Section~\ref{sisubsec:cpml}. These and similar values have been used in the literature, e.g., see Refs.~\cite{Taflove2005,Elsherbeni2016,Giannopoulos2011}. The relative errors are plotted in Fig.~\ref{fig:fdtd_error}. For most cases, EMI-FDTD performs better than gprMax, as the error at the end of the simulated time is lower with EMI-FDTD. This can mainly be attributed to the different CPML parameters that are used in the two codes. Moreover, it can be observed that large fluctuations exist among the solutions obtained with the different CPML parameters. Therefore, the choice of the parameters for the absorption of reflected waves in reacting flow environments requires further investigation, which should be addressed in the future. Note that, overall, the CPML parameters of Case 2 seem to perform better than the others.

\begin{table}[!b]
    \centering
    \caption{CPML parameters for the various cases shown in Fig.~\ref{fig:fdtd_error}.}
    \label{tab:cpml_parameters_validation}
    \begin{tabularx}{\textwidth}{c|YYYYYY}
        \hline
        Case & $m$ Eq.~\eqref{eq:sigma_x} & $\sigma_{\mathrm{factor}}$ Eq.~\eqref{eq:sigma_x_max} & $\kappa_{max}$ Eq.~\eqref{eq:fdtd_kappa_x} & $\alpha_{max}$  Eqs.~\eqref{eq:alpha_x_e}, \eqref{eq:alpha_x_b} & $\alpha_{min}$ Eqs.~\eqref{eq:alpha_x_e}, \eqref{eq:alpha_x_b} \\
        \hline
        1   & 3 & 0.75 & 15 & 0 & 0.24 \\
        2   & 3 & 1.1 & 11 & 0.05 & 0.05 \\
        3   & 4 & 0.5 & 8  & 0.05 & 0.05 \\
        \hline
    \end{tabularx}
\end{table}

As expected, when thicker CPML boundary regions are used, the electromagnetic waves are better absorbed. Note also that for certain cases, EMI-FDTD with 5 CMPL cells performs better than gprMax with 10 CMPL cells. Overall, EMI-FDTD propagates electromagnetic waves accurately. The accuracy of the software when multiple processors are used in parallel is shown in Fig~\ref{fig:fdtd_error_multi}. The computation is performed with 4 processors in each direction, i.e., a total of 64 processors. This case allows for the assessment of not only the communication between internal subdomains but also of subdomains that hold CMPL regions in the edge, face, and vertex locations of the domain. Results show perfect agreement between the single processor and the multiprocessor solution.

\subsection{Evaluation of the electrostatic assumption in reacting flows}
\label{subsec:evaluation_of_electrostatic_assumption}

In this section, the EMI-FDTD code is used to assess the electrostatic field assumption that is made in previous relevant studies in the literature (see Section~\ref{sec:intro}). The fundamental difference between solving electrostatic or magnetostatic fields and solving electromagnetic waves via the FDTD method is that for static fields, it is assumed that the charges are permanently fixed both in space and time. Therefore, when a static field formulation is applied within the context of reacting flows, a quasi-steady state is assumed, where the effects induced by changes over time in the local charge density are negligible, and thus the magnetic field generated by the electric currents (Ampere's equation) can be neglected. This is an accurate assumption if the concentration times mobility of ionic species and electrons is sufficiently low (i.e., low-intensity currents). 

However, since the produced charged species are mobile, non-negligible currents may be induced. It is already evident from the charge density of Fig.~\ref{fig:qden_field} and the velocity of Fig.~\ref{fig:comparison_ion_EF_1} that even for the intermediate potential strength value, an electric current of the order of 1~mA/m$^2$ will be developed. These currents, albeit weak, may induce electromagnetic fields, which in turn may affect the chemistry rates and thus the formation of ionic species. As a consequence, the charge density is affected and can influence the overall electrodynamics of the reacting flow. The impact of induced electric currents in reacting flows is investigated in the remainder of this section.

It is reminded that the \gls{fdtd} method includes the current density in the numerical formulation and thus the effect of the induced magnetic fields. Specifically, when an external electric field is applied, the FDTD method includes in the numerical formulation the currents given by $\bm{J}=\sigma_e \mathcal{E}$ (see Eq.~\eqref{eq:J_em}), where $\sigma_e$ is given by Eq.~\eqref{eq:sigma_e}. Hence, in the FDTD method, the motion of ionic species produces currents. Overall, the FDTD method provides a more accurate description of electromagnetic interactions in reacting flows. It is also noted that, in the static field approximation, the ions with the highest concentration govern the behaviour of the electric field through Gauss's law. However, with the FDTD method, the species with the highest product of mobility times concentration affect the behaviour of the electric field.

\begin{figure}[!t]
    \centering

    \includegraphics[width=0.5\linewidth]{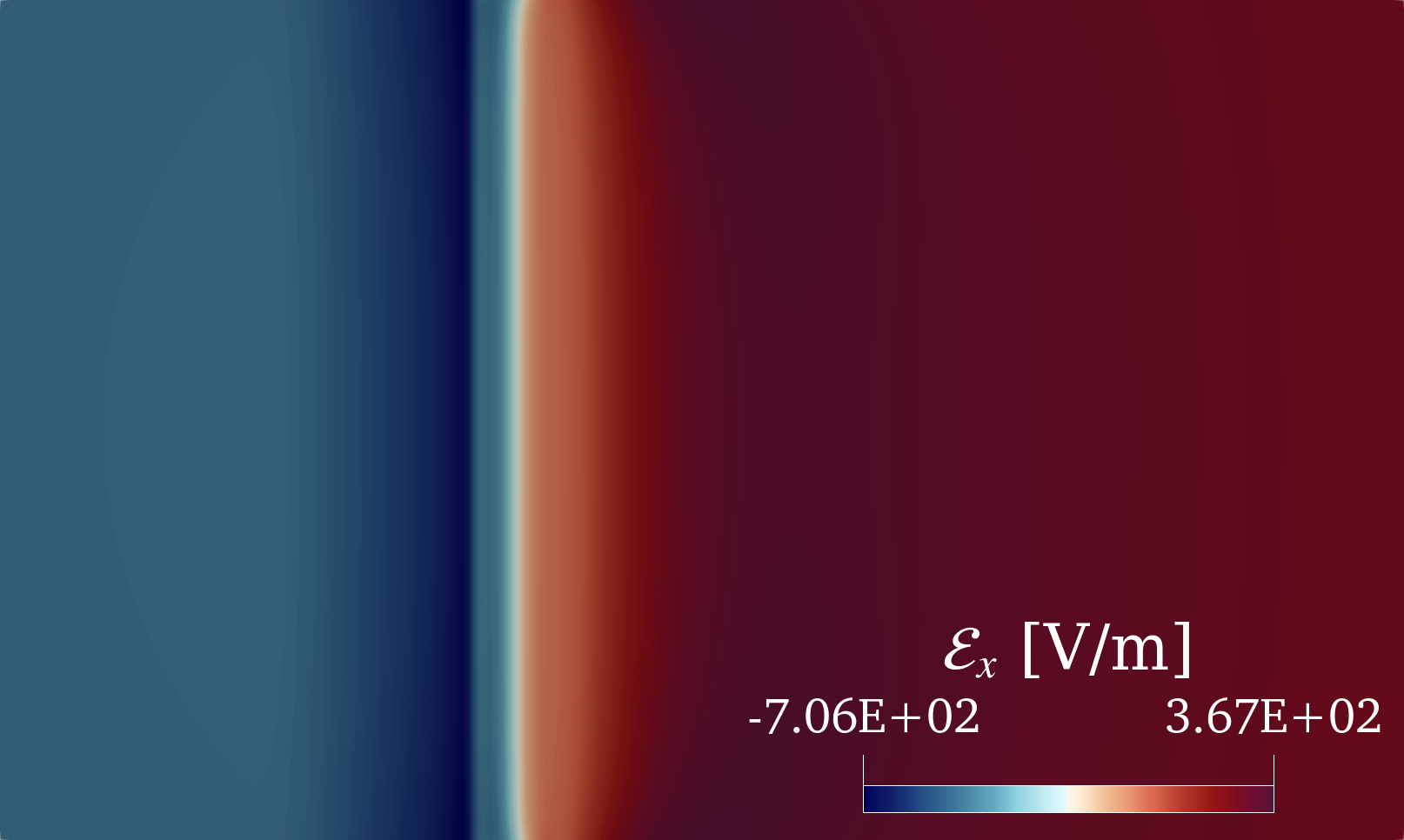}%
    \includegraphics[width=0.5\linewidth]{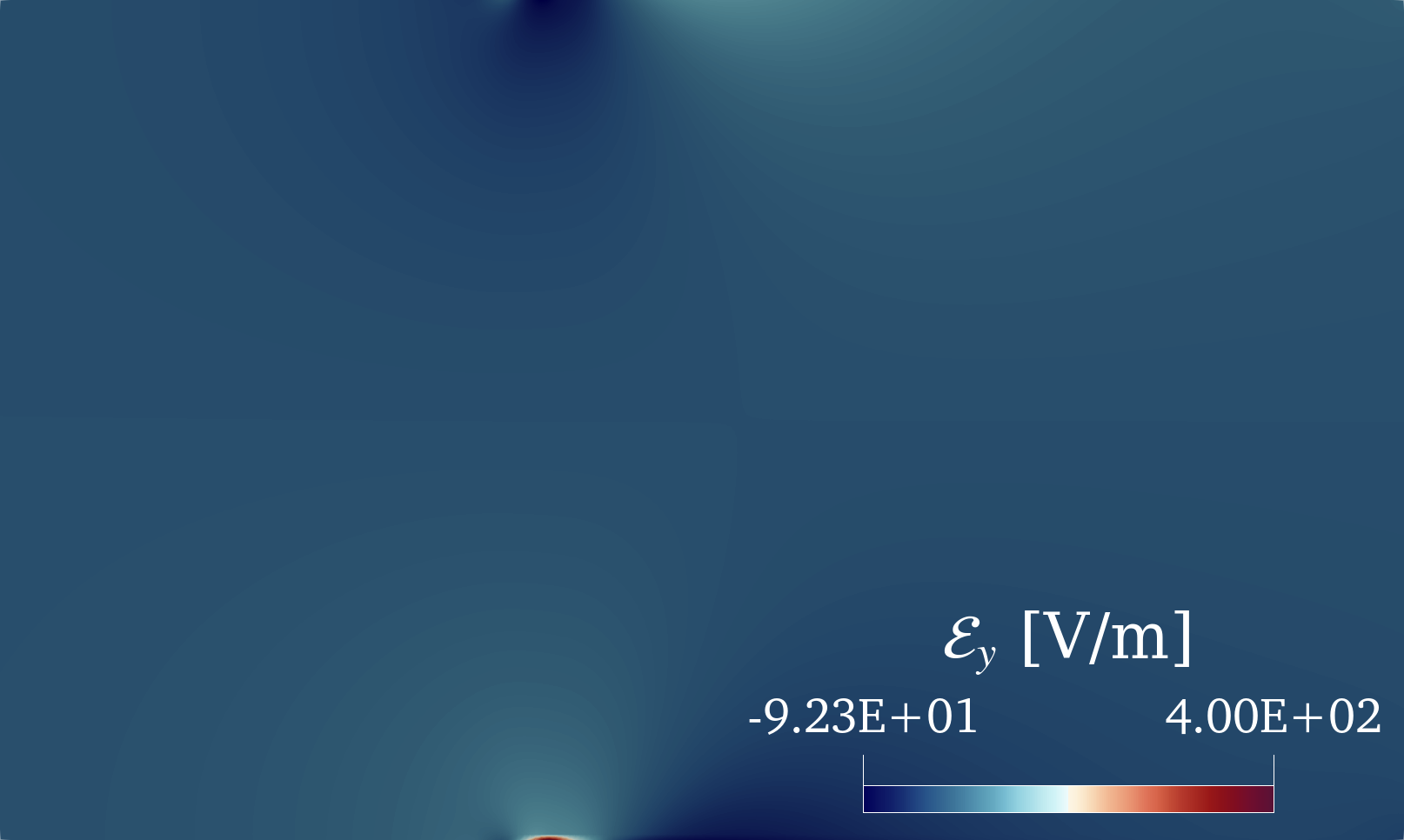}
    
    \includegraphics[width=0.5\linewidth]{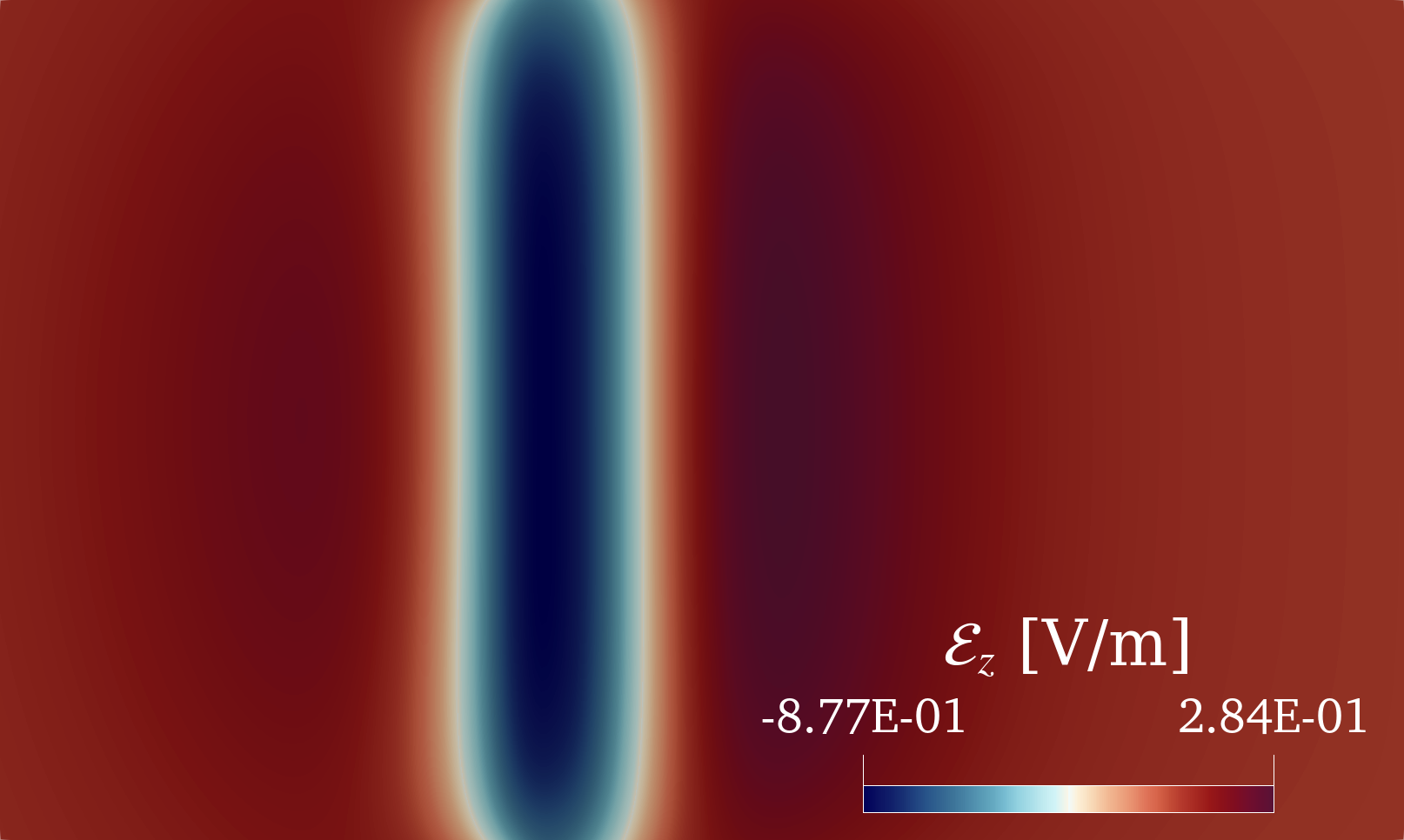}%
    \includegraphics[width=0.5\linewidth]{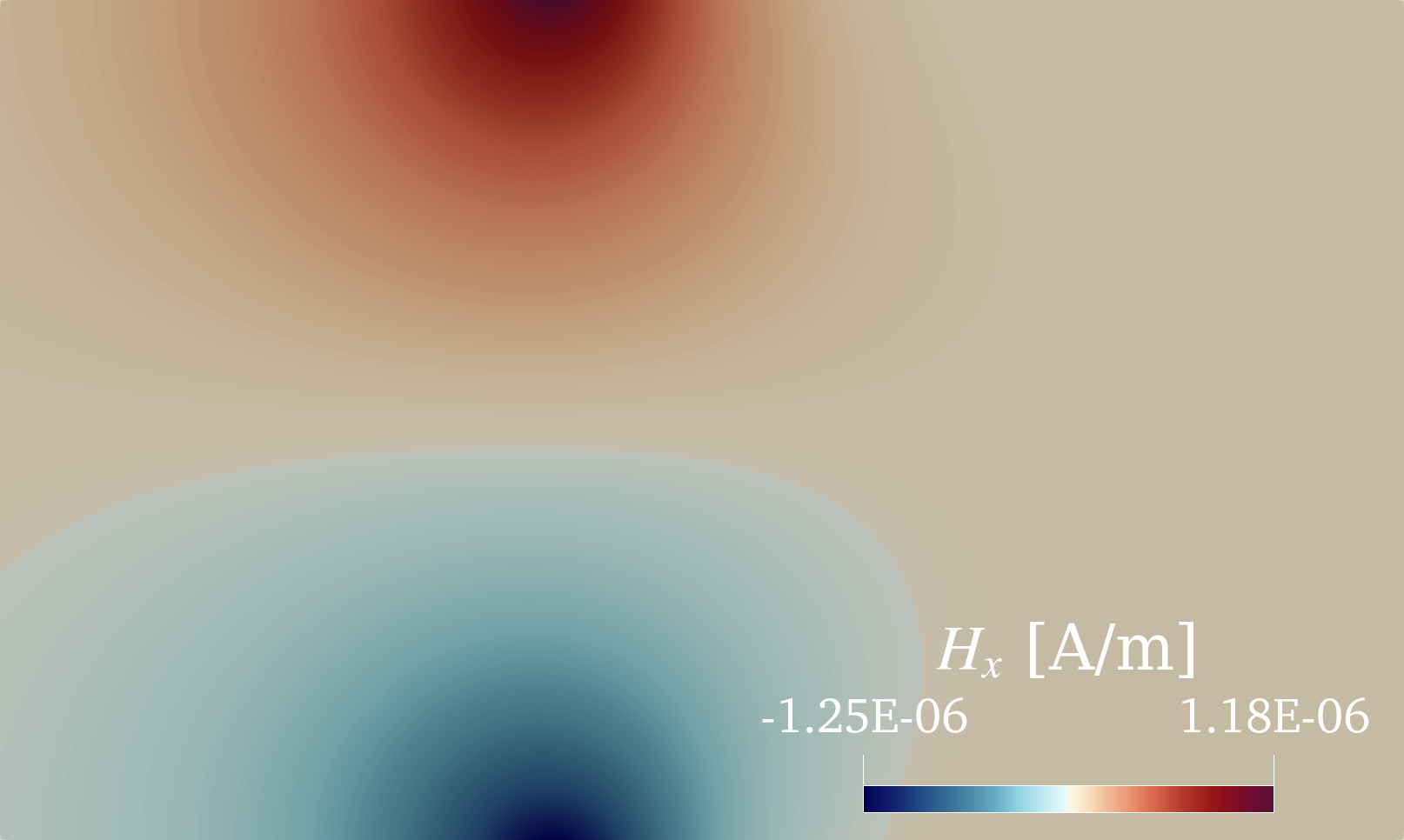}
    
    \includegraphics[width=0.5\linewidth]{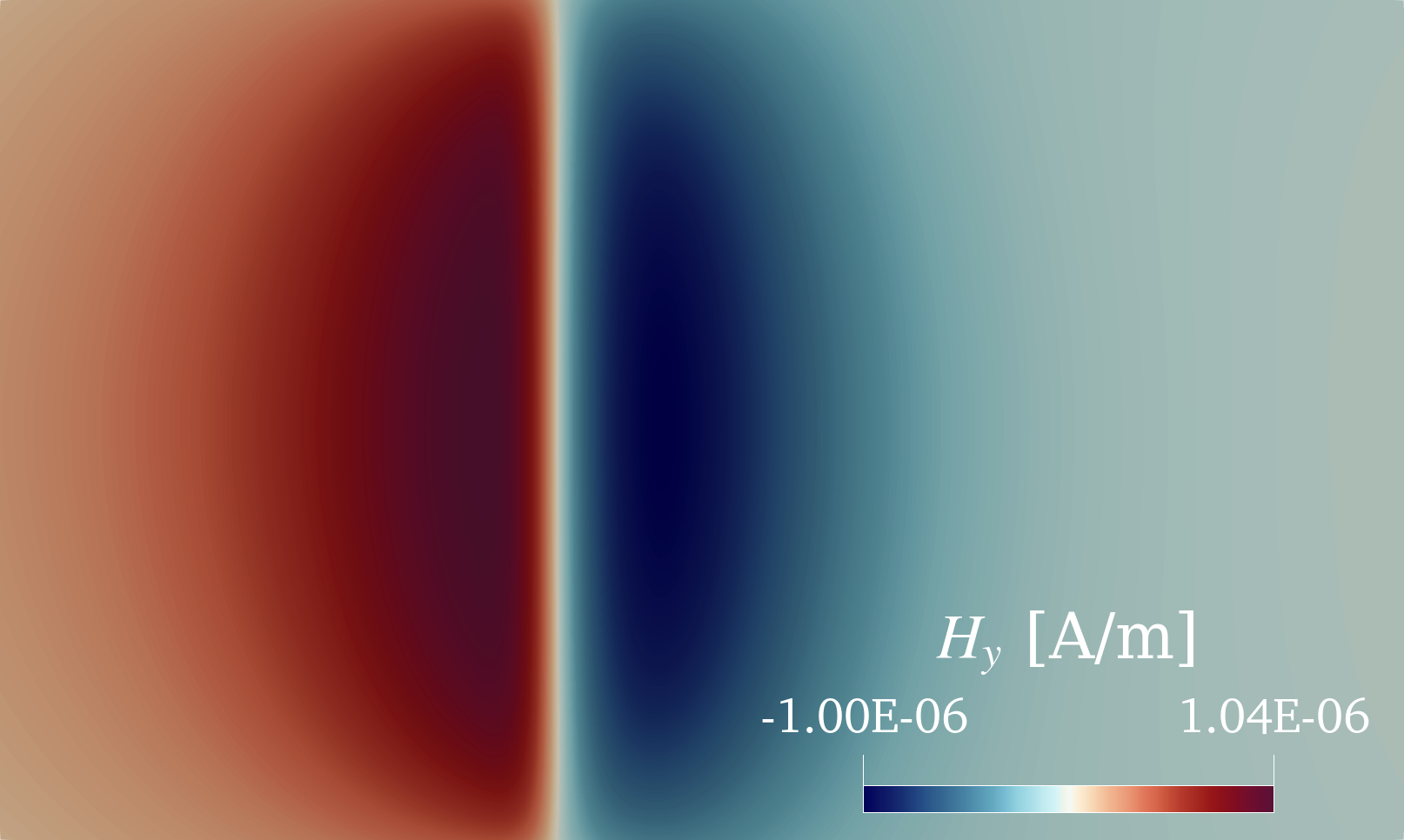}%
    \includegraphics[width=0.5\linewidth]{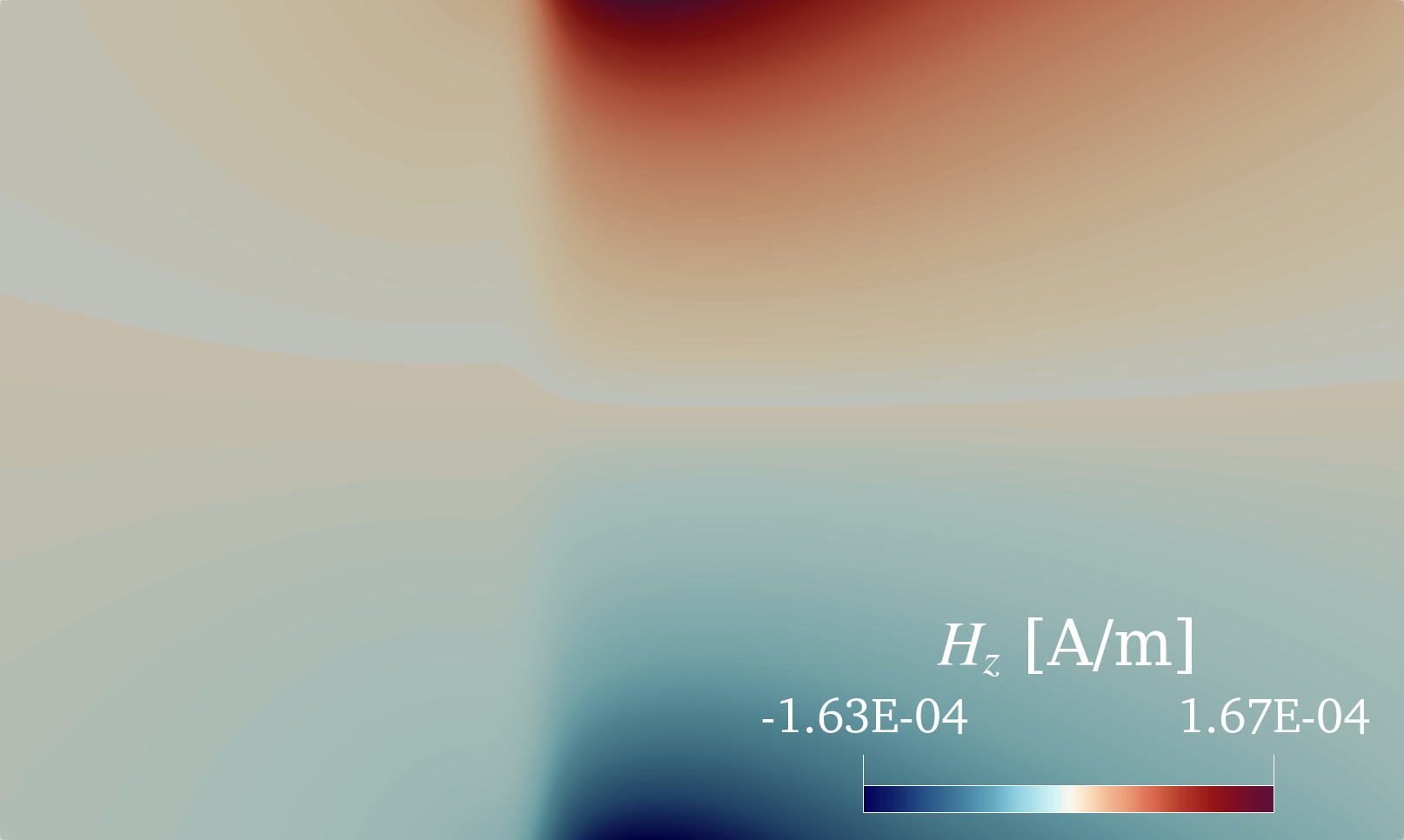}

    \hspace{5pt}%
    \raggedright
    \begin{subfigure}[t]{0.08\linewidth}
        \vspace{-55pt}
        \includegraphics[width=\textwidth]{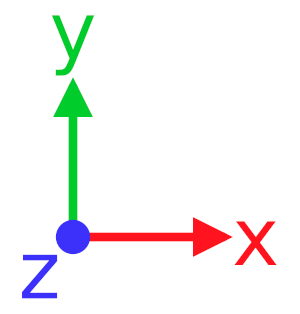}
    \end{subfigure}\hfill

    \caption{Electric and magnetic field components at 10~ns simulated time. The fields are plotted in a slice passing from the middle point of the $z$ coordinate of the domain.}
    \label{fig:FDTD_no_extels}
\end{figure}

To investigate this difference between the two methods, the same flame studied in Section~\ref{subsec:emi_1d_electrostatic} is simulated with the EMI-FDTD code under no external electromagnetic field. Due to the high computational cost of the method, fewer nodes in the $x$ direction are used compared to the electrostatic field simulations of Section~\ref{subsec:emi_1d_electrostatic}. Specifically, the computational domain has a size of 10, 6, and 2~mm, in the $x$, $y$, and $z$ direction, respectively. The size of each cell is equal to 40~\textmu m in all directions. The conservation equations are advanced in time with a timestep of 0.1~ns and the electromagnetic fields with a timestep of 50~fs. CPML boundaries are used with a 5-cell thickness and the CPML parameters of Case 2 of Table~\ref{tab:cpml_parameters_validation}. A steady state solution of a 1D reacting flow, similar to the one presented in Section~\ref{subsec:validation}, but with 250 nodes in the $x$ direction, is imposed as an initial solution for the 3D FDTD simulation. The 1D profiles along the $x$ direction are imposed for all $y$ and $z$ nodes. The existence of a charge distribution requires the initialization of the $\mathcal{E}_x$ profile with the electrostatic solution, as described in Section~\ref{subsubsec:electromagnetic_waves}. Therefore, an initial electrostatic field is also imposed. Simulations are performed for a total simulated time of 10~ns, in order to evaluate the effects of the FDTD method.

\begin{figure}[t]
    \centering
    \includegraphics[width=0.5\textwidth]{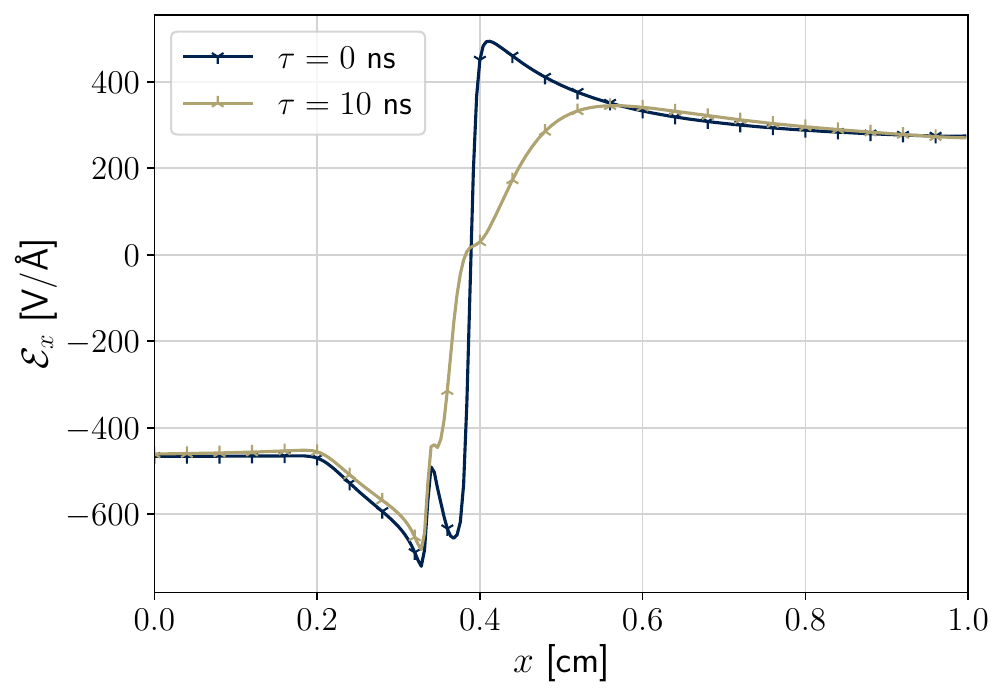}%
    \includegraphics[width=0.5\textwidth]{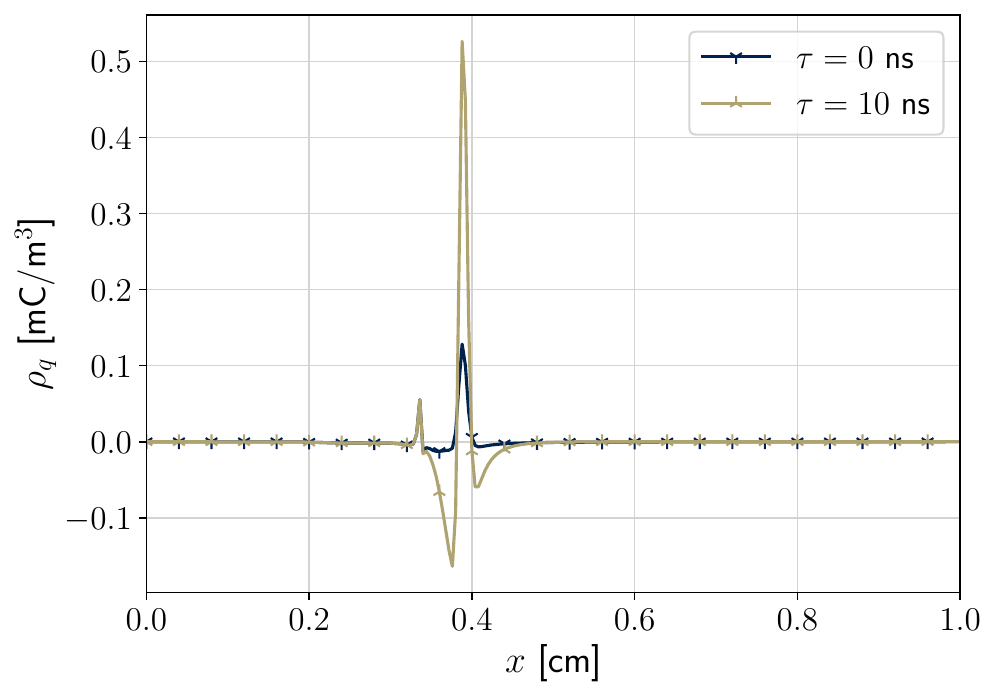}
    \caption{Electric field component (left) and charge density (right) at 10~ns simulated time. Results are plotted for a line passing from the middle $y$ and $z$ coordinate of the domain.}
    \label{fig:no_extels_elfx_qden}
\end{figure}

The electric and magnetic field components at 10~ns simulated time are shown in Fig.~\ref{fig:FDTD_no_extels}. The fields are plotted in a slice passing from the center of the $z$ coordinate. First of all, a change in the $\mathcal{E}_x$ field can be observed. Results are more evident in Fig.~\ref{fig:no_extels_elfx_qden}, which shows the $\mathcal{E}_x$ field and charge density plotted along the centerline in the $x$ direction. Specifically, inside and near the reacting zone, the magnitude of $\mathcal{E}_x$ decreases. A similar but smaller reduction in magnitude can also be seen on the left side of the reacting zone. On the contrary, between approximately $x=0.6$ and $x=0.9$~cm, the electric field slightly increases. This difference is attributed to the induced currents from the mobility of ions and electrons that are considered in the FDTD method. This change in the electric field leads to changes in the mass fraction of species, especially of the charged ones. Subsequently, this change in the charge density affects the magnitude of the induced electromagnetic fields.

Additionally, from Fig.~\ref{fig:FDTD_no_extels}, differences in the electromagnetic field components can be observed near the CPML boundaries. These are caused by the absorption of electromagnetic waves that are incident on the boundaries. It should also be noted that the influence of boundaries (end effects) is not evident in the central part of the reacting flow when the electromagnetic field components with high magnitudes are considered, i.e., the $\mathcal{E}_x$, $\mathcal{E}_y$, and $\mathcal{E}_z$ components. In contrast, the end effects of the boundaries are mainly apparent in the low magnetic field components. It should be noted here that continuing the simulations past the 10~ns simulated time showed a further decrease in the electromagnetic field magnitude. This indicates that a steady state has not been reached. However, the results obtained here are sufficient to highlight the main differences between the FDTD solution and the electrostatic solution. The detailed investigation of the steady-state solution is left for future work.

A few questions arise from the above discussion: (i) Are these changes caused solely by electric currents? (ii) Do the changes in mass fractions significantly affect the electric fields, and in which way? (iii) To what extent do the end effects caused by the CPML boundaries affect the electromagnetic field solution? (iv) How much does the electron's mobility affect the induced currents and the electric field?

To answer the first question, simulations with the FDTD method of the EMI-SENGA code were performed by setting the conductivity of the medium equal to zero. After 10~ns simulated time, the initial $\mathcal{E}_x$ had not changed. Similarly, the rest of the electric field components and the magnetic field components had a zero value (results are not presented here). This suggests that the current density term is mainly responsible for the electromagnetic field effects discussed above. Therefore, this result further supports the above discussion that the differences between static fields and fully coupled electromagnetic fields in reacting flows are attributed to the induced currents and thus to the induced magnetic fields.

To answer the second question, simulations in which only the electromagnetic fields are updated, without solving the conservation equations of the species, momentum, or energy,  were performed. Thus, in these simulations, the initial solution of the reacting flow did not change as the electromagnetic field equations were solved. When the mass fraction of the ionic species is constant, a change in the magnitude of the electric field is observed. This change is almost identical to the one obtained when all conservation equations are solved. This comparison is shown in Fig.~\ref{fig:elfx_comparison_no_e_noy} for the $\mathcal{E}_x$ field. Interestingly, this occurs despite the fact that when the flow field quantities are updated, the charge distribution changes significantly, as shown in Fig.~\ref{fig:no_extels_elfx_qden}. Therefore, these results show that small changes in the concentration of the ionic species may not considerably affect the shape of the electromagnetic waves. In addition, it can be concluded that the electromagnetic fields change mainly because of the induced currents.

To answer the third question, simulations on a larger domain were performed. The cell size was kept constant, whereas the domain size was increased to 2~cm in both the $y$ and $z$ directions. A total of 500 processors are used. This larger configuration provides enough space to further diminish the end effects from the boundaries. The overall shape of the electric field is in good agreement between the bigger and smaller domains. However, between approximately 0.4 and 0.7~nm, the magnitude of the electric field is slightly lower in the bigger domain simulation. The results of this comparison are reported in Fig.~\ref{fig:elfx_comparison_no_e_big} for the $x$-component of the electric field. It should be noted that, as shown in Fig.~\ref{fig:fdtd_error}, the CMPL \glspl{bc} do not perfectly absorb the incoming waves. Thus, reflective waves can reenter the domain and affect the electromagnetic fields inside the domain. These findings suggest that end effects near the CMPL boundaries can affect the magnitude of the electric field. Therefore, it is evident that the size of the domain should be carefully selected, depending on the application under consideration and the medium's characteristics.

Finally, to answer the fourth question, simulations with an electron's mobility of $\upsilon_{e^{-}}=1$~cm$^2$/(Vs) were performed. This mobility value, which is two orders of magnitude lower compared to the value used in the previous simulations, was used by Belhi et al.~\cite{Belhi2010} in their DNS study of premixed flames under external electrostatic fields. The electric field did not significantly change when the low value of the electron's mobility was imposed. Only slight changes near the reacting zone were observed. The simulation results are presented in Fig.~\ref{fig:comparison_low_e_mobility}. Additionally, by examining the conductivity, certain differences between the profiles with high and low electron mobilities can be observed. When a 0.2~m$^2$/(Vs) mobility is used, the conductivity has a profile that is consistent with the mass fraction profile of electrons (see Fig.~\ref{fig:comparison_ion_EF_2}). Therefore, in this case, it is the electrons that determine the conductivity of the mixture. On the contrary, when an electron's mobility of 1~cm$^2$/(Vs) is used, the conductivity has a much lower magnitude. Hence, the electrons do not contribute significantly to the conductivity of the mixture. Specifically, it is the rest of the ions that determine the magnitude of the conductivity. It is reminded that the mobility of the ionic species is determined by their diffusivity. Therefore, the conductivity of the ionic species is directly related to their diffusivities. These findings suggest that the electron's mobility, which is an input value (see Section~\ref{sec:electrical_conductivity}), plays an important role in the conductivity of the mixture and thus directly affects the evolution of the electromagnetic fields. Consequently, this suggests that above a certain magnitude of the mobilities of the ionic species and electrons, the electrostatic assumption ceases to be valid since the induced currents cannot be neglected.

\subsection{Electromagnetic wave sources on reacting flows}

The analysis continues with the application of external electromagnetic sources. As discussed in Section~\ref{subsubsec:electromagnetic_waves}, the FDTD allows the simulation of any form of electromagnetic wave. Therefore, the effects of sources such as continuous sinusoidal waves or discrete pulses can be studied. In this section, the source of Eq.~\eqref{eq:J_source} is applied with the waveform:
\begin{equation}
    \mathcal{I}= C \sin\left(2 \pi f t\right),
\end{equation}
where
\begin{equation}
    C = 
\begin{cases} 
0.25 A_o f t & \text{if $C \leq 1$,} \\
A_o & \text{if $C > 1$.}
\end{cases}
\end{equation}
$A_o$ is the amplitude of the wave. This source was selected from the gprMax software~\cite{Warren2016}, where the amplitude of the sinusoidal wave is ramped during the first cycle, to not introduce a lot of numerical noise in the simulation domain during this initial transient. The computational details reported at the beginning of Section~\ref{subsec:evaluation_of_electrostatic_assumption} are used here. The source is applied along the $z$ direction and to the node at a distance of 40, 75, and 25 nodes in the $x$, $y$, and $z$ direction, respectively. The amplitude of the wave is equal to 0.016 and the frequency is equal to 100~GHz. A high frequency is chosen in order to have a sufficiently low wavelength that will allow for the observation of several crests of the waveform inside the domain. 

\begin{figure}[!t]
    \centering
    \includegraphics[width=0.498\linewidth]{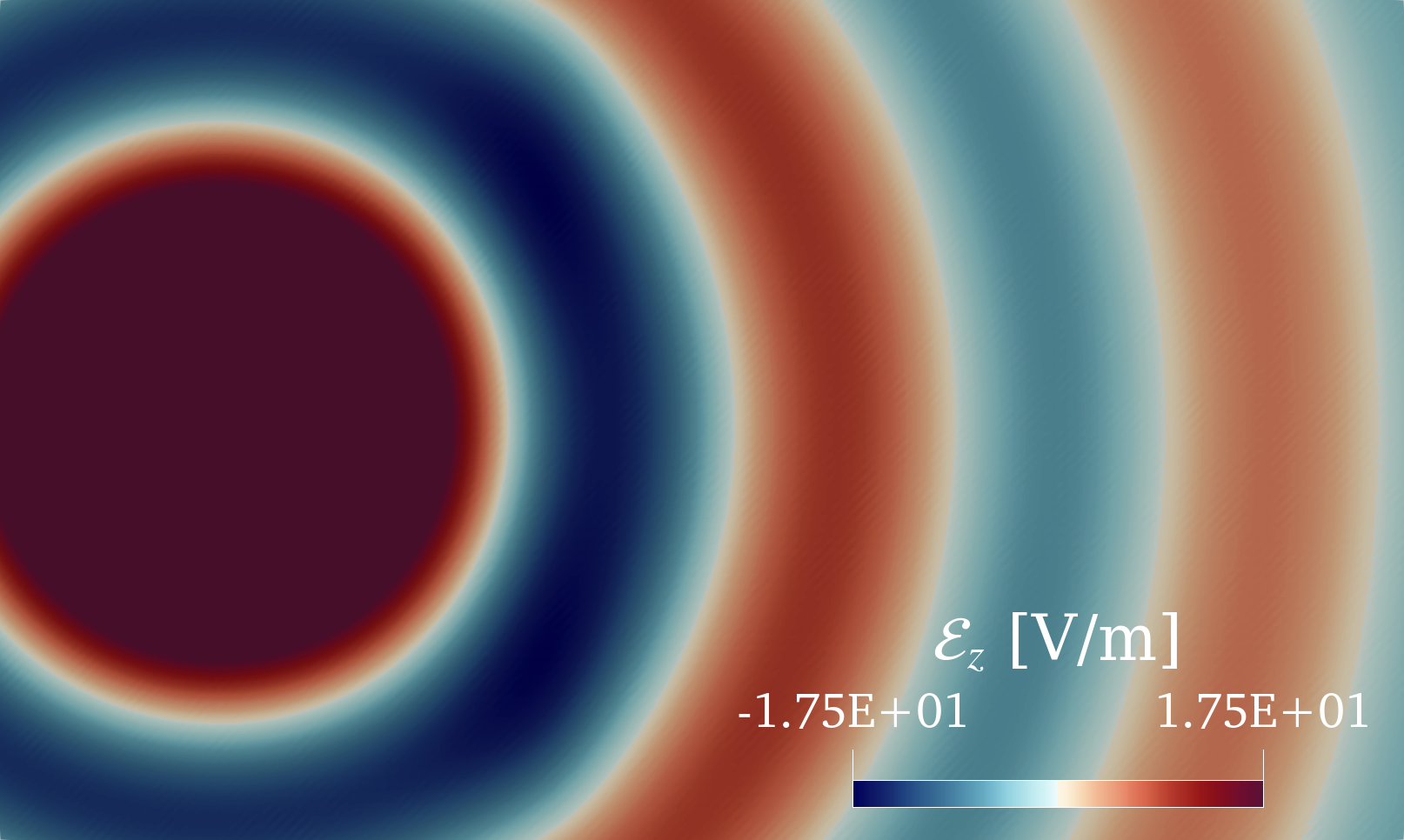}\hfill%
    \includegraphics[width=0.498\linewidth]{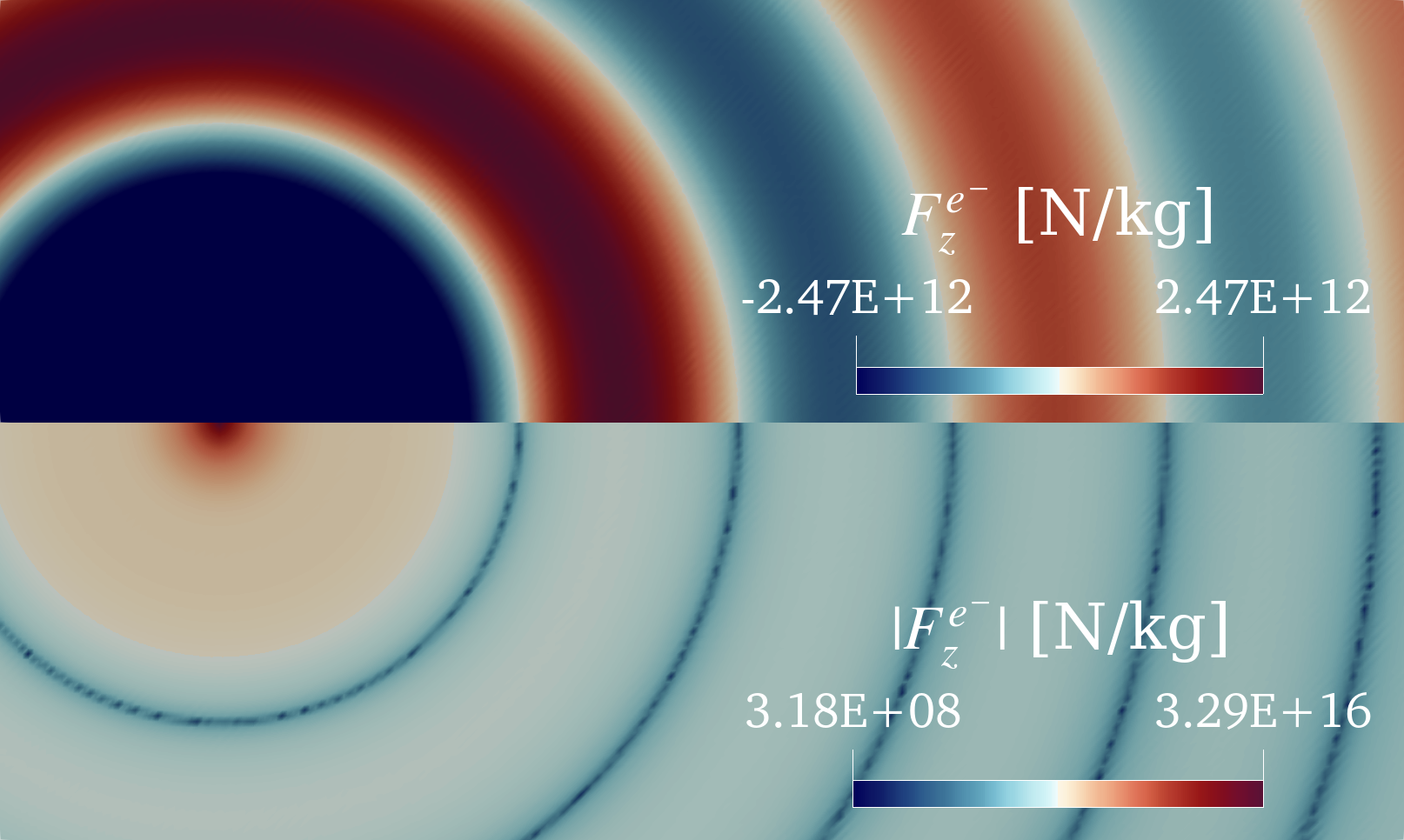}

    \hspace{5pt}%
    \hspace{-165pt}%
    \begin{subfigure}[t]{0.08\linewidth}
        \vspace{-50pt}
        \includegraphics[width=\textwidth]{Figures/FDTD/FDTD_coord.png}
    \end{subfigure}%    
    \hspace{165pt}%
    \begin{subfigure}[t]{0.08\linewidth}
        \vspace{-50pt}
        \includegraphics[width=\textwidth]{Figures/FDTD/FDTD_coord.png}
    \end{subfigure}\hfill

    \caption{$z$-component of the electric field (left) and force on electrons (right) under a sinusoidal wave source at 5~ns simulated time, plotted in a slice passing from the middle point of the $z$ coordinate of the domain. Note that $\mathcal{E}_z \in [-17.5,186990]$, but is rescaled to allow for the observation of the waveform.}
    \label{fig:fdtd_e_force}
\end{figure}

Figure~\ref{fig:fdtd_e_force} shows the $z$-component of the electric field at 5~ns of simulated time. The magnetic field components are given in Fig.~\ref{fig:fdtd_source}. Note that results are rescaled in order for the waveforms to be observable. The reason is that the wave produced by the source has a higher value at and near the source. For instance, at the node where the source is located, the $z$-component of the electric field has an amplitude that is 5 orders of magnitude higher than at an intermediate location along the $x$ direction. At a higher distance from the source, the wave decays. Note also that, in this configuration, the maximum electric field strength is comparable to the strength of the electrostatic fields discussed in Section~\ref{subsec:emi_1d_electrostatic}. Therefore, the electromagnetic waves of this source are expected to affect the reacting flow, similarly to the effects observed in the 1D simulations.

Figure~\ref{fig:fdtd_e_force} also shows the electromagnetic force that the electrons of the reacting flow experience. As expected, the force follows the direction of the sinusoidal electromagnetic wave oscillation. In addition, note that even though the force on electrons is maximum near the source, it is still considerably high in the rest of the domain. Specifically, the magnitude of the $z$ component of the electrons' force is above $10^{10}$~N/kg everywhere and around $10^{12}$~N/kg at the crests and troughs away from the source. The magnitude of the Lorentz force is smaller but comparable to the one presented in Fig.~\ref{fig:lorentz_polar}. Therefore, this approach allows for the investigation of more intricate phenomena such as the interaction of the electrons with the crests, troughs, and zero crossings of the wave. A comprehensive exploration of the effects of electromagnetic wave sources on reacting flows is left for future study. 

Finally, the magnetic susceptibility of electrons, as given by Eq.~\eqref{eq:chi_m_elec}, is investigated. It is possible to introduce electrons with magnetic fields in the current 3D configuration since the Lorentz forces are not constrained in any direction. The peak magnitude of the electrons' magnetic susceptibility is of the order of $4\times 10^{-16}$, which is significantly smaller compared to the one of neutral species (see Fig.~\ref{fig:chim_electron}). Specifically, the oxygen magnetic susceptibility peaks at around $3\times 10^{-7}$. The values for the other neutral species range mainly between $10^{-10}$ and $10^{-12}$, but lower values of approximately $10^{-14}$ exist. The magnetic susceptibility of ionic species can be a few orders of magnitude lower than that of neutrals but similar to the value for the electrons due to their low concentration. Therefore, for small magnetic field gradients, the magnetic susceptibility of the electrons and ions can be neglected. The impact of these small magnetic susceptibilities on reacting flows under sufficiently high magnetic field gradients will be explored in future work.

\section{Summary and conclusions}

A new computational fluid dynamics code, called EMI-SENGA, has been developed for the solution of multi-physics problems involving electromagnetism and reactive flows. The proposed framework can simulate time-dependent or static electromagnetic fields by solving Maxwell's or Gauss's laws, respectively. The variation in space and time of the electric and magnetic properties of the medium, which directly affects the propagation of the electromagnetic fields, is included in the model. In addition, the total electromagnetic force acting on species includes contributions from the Lorentz, polarization, and magnetization forces. The Lorentz force acts on the ionic species, the polarization force on species with an inherent or electrically induced polarization, and the magnetization force on molecules with a net spin.

First, the code has been validated with numerical and analytical solutions from the literature. The results of the EMI-SENGA code were in agreement with the literature data. Then, to demonstrate the capabilities of the framework, EMI-SENGA was used to investigate various laminar premixed flames under external electromagnetic fields. When external electrostatic fields were applied, results showed that the charge density and the related electric field were mainly affected near the inflow of reactants. For positive electric potentials, the electrons travel towards the inflow, where they react with oxygen to form oxygen ions. This demonstrates the capability of the proposed framework to capture the strong coupling between chemistry and diffusion induced by electrostatic interactions, as well as the necessity of appropriate boundary conditions to achieve a realistic behaviour in the vicinity of the boundary region.

Analysis of the magnetostatic fields showed that the correct description of the magnetization force is important for weak magnetic field gradients. Specifically, the gradient of the magnetic susceptibility term is equal in magnitude to the gradient of the magnetic field term and thus cannot be neglected. Strong magnetic fields acting on two-dimensional reacting flows could affect the shape of the flame front. This change is directly proportional to the profile and the magnitude of the boundary magnetic potential. This further highlights the importance of accurate boundary conditions for a reliable prediction of the effect of magnetic fields on the reactive flow.

Furthermore, a comparison between the FDTD solution and the electrostatic solution shows that, for sufficiently high mobilities and concentrations of ionic species and electrons, the induced electric currents significantly affect the electromagnetic fields even in the absence of an external source. This is attributed to the fact that the movement of charged particles, which are generated at the flame front, induces electric currents. Therefore, at high mobilities and concentrations of the charged species, these induced currents cannot be neglected, and the electrostatic field assumption ceases to be valid.

Finally, analysis of the response of a laminar flame to an electromagnetic wave shows that the wave source can affect the distribution of charges in the reacting flow. Specifically, a high-frequency sinusoidal source introduces electromagnetic waves with crests and troughs within the domain. These fluctuations locally influence the electromagnetic forces acting on charged particles. As a consequence, the electromagnetic field effects on ions and electrons become both spatially and temporally dependent.

EMI-SENGA can be used to study the fundamental effects of electromagnetic fields on reacting flows. The numerical formulation maintains a high degree of accuracy for electromagnetic interactions and enables the study of a wide variety of multiphysics phenomena.

\section{Acknowledgments}

This work used the Imperial College Research Computing Service (DOI: 10.14469/hpc/2232) and the ARCHER2 UK National Supercomputing Service (https://www.archer2.ac.uk).

\appendix
\section{Derivation of polarization and magnetization forces}
\label{sec:polarization_magnetization_forces}

The polarization force arises from the intrinsic or field-induced polarization of atoms and molecules, i.e., the distribution of the electron cloud with respect to the nuclei. When a particle with an electric moment is placed in an inhomogeneous electric field, $\mathcal{E}$, it experiences a polarization force equal to~\cite{Griffiths2013}:
\begin{equation}
    \bm{F}_{P}=\left(\bm{p}\cdot \nabla \right) \bm{\mathcal{E}},
    \label{eq:flag_A0_1}
\end{equation}
where $\bm{p}$ is the electric moment. The electric moment is related to the polarization density, $\bm{P}$, by:
\begin{equation}
    \bm{P}=\dfrac{d\bm{p}}{dV}=\dfrac{\sum_i \mathcal{N}_i p_i}{V}=\sum_i N_i p_i,
    \label{eq:flag_A0_2}
\end{equation}
where $\mathcal{N}_i$ is the number of electric dipoles $i$, $V$ is the volume of the mixture, and $N$ is the number density. The number density is given by:
\begin{equation}
\begin{aligned}
     N_i = \rho Y_i \dfrac{N_A}{M_i}.
     \label{eq:flag_A0_3}
\end{aligned}
\end{equation}

Substituting Eq.~\eqref{eq:flag_A0_2} and Eq.~\eqref{eq:electric_flux_density} in Eq.~\eqref{eq:flag_A0_1}, the total polarization force on all the electric dipole moments in a volume $V$ is:
\begin{equation}
    \bm{F}_{P}=V\chi_e\left(\bm{\mathcal{E}}\cdot \nabla\right) \bm{\mathcal{E}},
    \label{eq:flag_A0_f1}
\end{equation}

Therefore, using Eq.~\eqref{eq:flag_A0_f1} with ~\eqref{eq:flag_A0_3}, the polarization force on species $s$ per unit of mass is:
\begin{equation}
    \bm{f}^s_P = \dfrac{\epsilon_0}{\rho Y_s} \chi_e^s\left(\bm{\mathcal{E}}\cdot\nabla\right) \bm{\mathcal{E}}.
\end{equation}

As far as magnetic interactions are concerned, the description of the magnetization force is more complicated. The reason is that, unlike the polarization force, which arises from the electronic structure of the species, the magnetization force arises from the spin of elementary particles. Therefore, the magnetization force that is induced on those elementary magnetic moments depends on how these magnetic moments are modelled~\cite{Boyer1988}. However, it is difficult to define a quantum mechanical phenomenon in the context of classical electrodynamics~\cite{Griffiths2013}. There has been a lot of debate and scrutiny on the form of the magnetization force~\cite{Greene1971,Boyer1988,Aharonov1988,Vaidman1990,Griffiths1992,Spavieri1994,Griffiths2011,Griffiths2013a,Wegrowe2016,Westerberg2022}.

Two approaches have been proposed to model such magnetic dipoles: Ampere's or electric current loop model and Gilbert's or separated magnetic charge model~\cite{Boyer1988,Griffiths2013a}. The electric current loop model assumes that the magnetic dipole arises from the circulation of a charge in an infinitesimal loop. In that case, the magnetization force on a magnetic dipole $\bm{m}$ in a magnetic field $\bm{B}$ is equal to:
\begin{equation}
 \bm{F}_M=\nabla (\bm{m}\cdot\bm{B}).
 \label{eq:flag_A0_EC}
\end{equation}
On the other hand, the separated magnetic charge model assumes that the magnetic dipole is due to magnetic monopoles located at a certain distance. Here, the magnetization force is:
\begin{equation}
    \bm{F}_M=(\bm{m}\cdot \nabla) \bm{B}.
    \label{eq:flag_A0_SM}
\end{equation} 
This latter formulation is equivalent to the polarization force that an electric dipole experiences in an electric field. The two formulations, Eqs.~\eqref{eq:flag_A0_EC} and \eqref{eq:flag_A0_SM}, agree in the case of no spatial dependence of the magnetic moment, no currents, and no time-changing electric fields, due to the identity~\cite{Boyer1988}:
\begin{equation}
    \nabla (\bm{m} \cdot \bm{B}) - (\bm{m} \cdot \nabla) \bm{B} = \bm{m} \times (\nabla \times \bm{B}) + \bm{B} \times (\nabla \times \bm{m}) + (\bm{B} \cdot \nabla) \bm{m}. 
    \label{eq:flag_A0_idm}
\end{equation}
The two formulations are analyzed below.

\subsection{Gilbert's or separated magnetic charge model}

The magnetization density is given by:
\begin{equation}
    \bm{M}=\dfrac{d\bm{m}}{dV}=\dfrac{\sum_i \mathcal{N}_i m_i}{V}=\sum_i N_i m_i.
    \label{eq:flag_A0_1_1}
\end{equation}
Moreover, from Eq.~\eqref{eq:magnetic_flux_density}, the magnetization density becomes equal to:
\begin{equation}
    \bm{M}=\dfrac{1}{\mu_0} \dfrac{\chi_m}{1+\chi_m} \bm{B}.
    \label{eq:flag_A0_1_2}
\end{equation}

Thus, combining Eq.~\eqref{eq:flag_A0_SM}, Eq.~\eqref{eq:flag_A0_1_1}, and Eq.~\eqref{eq:flag_A0_1_2}, the total magnetization force acting on all the magnetic dipoles in a volume $V$ according to the separated magnetic charge model is given by:
\begin{equation}
        \bm{F}_M^s = \dfrac{V}{\mu_0} \dfrac{\chi_m^s}{1+\chi_m^s} \left(\bm{B}\cdot \nabla\right) \bm{B}.
        \label{eq:flag_A1}
\end{equation}
Note that Eq.~\eqref{eq:flag_A1} refers to the magnetization force $\bm{F}_M^s$ of each species $s$ and thus the magnetization and subsequently the magnetic susceptibility of the species $s$ is used.

Equation~\eqref{eq:flag_A1} is similar to the magnetization force that is used in several studies in the literature, such as Refs.~\cite{Yamada2002,Yamada2003,Jiang2020a,Kajimoto2003,Zhang2021}. Note that typically the force is presented by replacing $\left(\bm{B}\cdot \nabla\right) \bm{B}=1/2 (\nabla \bm{B}^2)$, which is derived from the identity shown in Eq.~\eqref{eq:flag_A0_idm} and Ampere's law-- see Eq.~\eqref{eq:Ampere_law}-- if no currents and time-varying electric fields exist. In this study, the magnetic field strength $\bm{H}$ is solved for. Thus, Eq.~\eqref{eq:magnetic_flux_density} is substituted into Eq.~\eqref{eq:flag_A1}, giving:
\begin{equation}
    \begin{aligned}
        \bm{F}_M^s &= \mu_0 \chi_m^s V \left(\bm{H}\cdot \nabla\right) \left[\left(1+\chi_m\right)\bm{H}\right].
        \label{eq:flag_A1_1}
    \end{aligned}
\end{equation}
Note that in Eq.~\eqref{eq:flag_A1_1}, the mixture averaged magnetic susceptibility of Eq.~\eqref{eq:chi_m_mix} appears when substituting the magnetic flux density $\bm{B}$ with the magnetic field $\bm{H}$. Therefore, in Eq.~\eqref{eq:flag_A1_1}, both the species-specific and mixture-averaged magnetic susceptibilities are present.

To retain a high degree of numerical accuracy, Eq.~\eqref{eq:flag_A1_1} must be expanded to the derivatives of each quantity. Eq.~\eqref{eq:flag_A1_1} is equivalent to solving the following equation for a vector $\bm{g}$: 
\begin{equation}
    \bm{g}=\left(\bm{v}\cdot \nabla\right)\left(\alpha \bm{v}\right),
    \label{eq:flag_A1_2}
\end{equation}
where $\bm{v}$ is the vector $\bm{H}$ and $\alpha$ is the scalar field $1+\chi_m$. 

The term $\nabla\left(\alpha \bm{v}\right)$ can be written as:
\begin{equation}
    \begin{aligned}
         \nabla\left(\alpha \bm{v}\right) &= \left(\nabla \alpha\right)\bm{v}^\textrm{T}+\alpha\nabla \bm{v}\\
          &= \nabla \alpha\otimes\bm{v}+\alpha\nabla \bm{v}.
          \label{eq:flag_A1_3}
    \end{aligned}
\end{equation}
Therefore, substituting Eq.~\eqref{eq:flag_A1_3} in the initial Eq.~\eqref{eq:flag_A1_2} leads to:
\begin{equation}
    \begin{aligned}
         \left(\bm{v}\cdot \nabla\right)\left(\alpha \bm{v}\right) &= \nabla\left(\alpha \bm{v}\right)\cdot \bm{v} = \left(\nabla \alpha\otimes\bm{v}\right)\cdot \bm{v}+\alpha \left(\bm{v} \cdot \nabla\right) \bm{v},
    \end{aligned}
\end{equation}
which, using the vector identity $\bm{v}\cdot (\nabla \alpha \otimes \bm{v}) = \bm{v} (\bm{v} \cdot \nabla \alpha)$, is equivalent to:
\begin{equation}
    \begin{aligned}
         \left(\bm{v}\cdot \nabla\right)\left(\alpha \bm{v}\right) &= \bm{v} (\bm{v} \cdot \nabla \alpha) + \alpha \left(\bm{v} \cdot \nabla\right) \bm{v}.
         \label{eq:flag_A1_4}
    \end{aligned}
\end{equation}

Substituting the $\alpha$ and $\bm{v}$ parameters in Eq.~\eqref{eq:flag_A1_4} leads to the total magnetization force with respect to $\bm{H}$ which is equal to:
\begin{equation}
    \begin{aligned}
         \bm{F}_M &= \mu_0\chi_m^s V \bm{H} (\bm{H} \cdot \nabla \chi_m) +\mu_0 V \chi_m^s \left(1+\chi_m\right)\left(\bm{H}\cdot\nabla\right) \bm{H}.
         \label{eq:flag_A1_4_1}
    \end{aligned}
\end{equation}

Therefore, the magnetization force in units of force per mass of each substance $s$ is given by:
\begin{equation}
    \begin{aligned}
    \bm{f}^s_M &= \dfrac{\mu_0}{\rho Y_s} \chi_m^s\bm{H}(\bm{H} \cdot \nabla \chi_m) + \dfrac{\mu_0}{\rho Y_s}\chi_m^s\left(1+\chi_m\right)\left(\bm{H}\cdot\nabla\right) \bm{H}.
    \label{eq:flag_A1_4_2}
    \end{aligned}
\end{equation}

For the one-dimensional case of Section~\ref{subsec:1D_mag}, Eq.~\eqref{eq:flag_A1_4_2} can be simplified further. First, Eq.~\eqref{eq:flag_A1_4} is expanded to: 
\begin{equation}
    \begin{aligned}
         \left(\bm{v}\cdot \nabla\right)\left(\alpha \bm{v}\right) &= \bm{v}\cdot \bm{v} \nabla \alpha+\bm{v} \times (\bm{v} \times \nabla \alpha) + \alpha \left(\bm{v} \cdot \nabla\right) \bm{v}\\
          &= \bm{v}^2 \nabla \alpha + \bm{v} \times (\bm{v} \times \nabla \alpha) + \alpha \left(\bm{v} \cdot \nabla\right) \bm{v},
         \label{eq:flag_A1_4_3}
    \end{aligned}
\end{equation}
using the vector identity $\bm{v} (\bm{v} \cdot \nabla \alpha)=\nabla \alpha (\bm{v} \cdot \bm{v}) + \bm{v} \times (\bm{v} \times \nabla \alpha)$. Therefore, using Eq.~\eqref{eq:flag_A1_4_3}, Eq.~\eqref{eq:flag_A1_4_2} can be written as:
\begin{equation}
    \begin{aligned}
    \bm{f}^s_M &= \dfrac{\mu_0}{\rho Y_s} \chi_m^s\bm{H}^2 \nabla \chi_m + \dfrac{\mu_0}{\rho Y_s} \chi_m^s \bm{H} \times (\bm{H} \times \nabla \chi_m) + \dfrac{\mu_0}{\rho Y_s}\chi_m^s\left(1+\chi_m\right)\left(\bm{H}\cdot\nabla\right) \bm{H}.
    \label{eq:flag_A1_4_4}
    \end{aligned}
\end{equation}

For the one-dimensional case, the triple cross product equals a null vector and thus Eq.~\eqref{eq:flag_A1_4_4} can be simplified along the x Cartesian coordinate to:
\begin{equation}
    \begin{aligned}
    \bm{f}^s_M &= \dfrac{\mu_0}{\rho Y_s} \chi_m^s H_x ^2 \dfrac{\partial\chi_m}{\partial x} + \dfrac{\mu_0}{\rho Y_s}\chi_m^s\left(1+\chi_m\right) H_x  \dfrac{\partial H_x}{\partial x}.
    \end{aligned}
\end{equation}

\subsection{Ampere's or electric current loop model}

Combining Eqs.~\eqref{eq:flag_A0_EC}, Eq.~\eqref{eq:flag_A0_1_1}, and Eq.~\eqref{eq:flag_A0_1_2}, the total magnetization force of all magnetic dipoles of species $s$ in a volume $V$ is given by:
\begin{equation}
        \bm{F}_M^s = \dfrac{V}{\mu_0} \nabla \left( \dfrac{\chi_m^s}{1+\chi_m^s} \bm{B} \cdot \bm{B}\right).
        \label{eq:flag_A1_5}
\end{equation}

Substituting the magnetic flux $\bm{B}$ with the magnetic field $\bm{H}$, using Eq.~\eqref{eq:magnetic_flux_density}, Eq.~\eqref{eq:flag_A1_5} becomes:
\begin{equation}
        \bm{F}_M^s = \mu_0 V \nabla \left[\dfrac{\chi_m^s \left(1+\chi_m\right)^2}{1+\chi_m^s} \bm{H}^2 \right].
        \label{eq:flag_A1_6}
\end{equation}
Note here that in Eq.~\eqref{eq:flag_A1_6}, similarly to Eq.~\eqref{eq:flag_A1_1}, both the mixture averaged and species-specific magnetic susceptibilities are present. 

Eq.~\eqref{eq:flag_A1_6} can be expanded to:
\begin{equation}
        \bm{F}_M = \mu_0 V \bm{H}^2 \left(\dfrac{1+\chi_m}{1+\chi_m^s}\right)^2 \nabla \chi_m^s + \mu_0 V \bm{H}^2 \dfrac{2\chi_m^s (1+\chi_m)}{1+\chi_m^s} \nabla \chi_m + \mu_0 V \dfrac{\chi_m^s \left(1+\chi_m\right)^2}{1+\chi_m^s} \nabla\bm{H}^2.
        \label{eq:flag_A1_7}
\end{equation}

Hence, the magnetization force in units of force per mass of each substance $s$ is given by:
\begin{equation}
        \bm{f}^s_M = \dfrac{\mu_0}{\rho Y_s} \left(\dfrac{1+\chi_m}{1+\chi_m^s}\right)^2 \bm{H}^2 \nabla \chi_m^s + \dfrac{2\mu_0\chi_m^s (1+\chi_m)}{\rho Y_s(1+\chi_m^s)} \bm{H}^2 \nabla \chi_m + \dfrac{\mu_0\chi_m^s \left(1+\chi_m\right)^2}{\rho Y_s(1+\chi_m^s)} \nabla\bm{H}^2.
        \label{eq:flag_A1_8}
\end{equation}

Comparing Eq.~\eqref{eq:flag_A1_8} with Eq.~\eqref{eq:flag_A1_4_4} (which is equivalent to Eq.~\eqref{eq:flag_A1_4_2}, but with similar terms to Eq.~\eqref{eq:flag_A1_8}) the main differences that can be observed are the following: First, an additional term corresponding to the gradient of $\chi_m^s$ appears in Eq.~\eqref{eq:flag_A1_8}. This is because in Eq.~\eqref{eq:flag_A1_6}, unlike Eq.~\eqref{eq:flag_A1_1}, the nabla operator acts on both the magnetic moment and the magnetic field. Second, the triple cross product of Eq.~\eqref{eq:flag_A1_4_2} does not appear in Eq.~\eqref{eq:flag_A1_8}. Finally, using the identity $\left(\bm{H}\cdot \nabla\right) \bm{H}=1/2 (\nabla \bm{H}^2)$, which arises from Eq.~\eqref{eq:flag_A0_idm} and Eq.~\eqref{eq:Ampere_law}, if no currents or time-varying electric fields exist, and assuming that $\chi_m \sim \chi_m^s << 1$, then the third term of the right-hand-side in Eq.~\eqref{eq:flag_A1_8} is twice as big as the corresponding one in Eq.~\eqref{eq:flag_A1_4_2}. Therefore, in the limit of magnetostatics with small constant magnetic susceptibilities, the magnetization force of Ampere's model is double the force of Gilbert's model.

Experimental evidence supports both forms (e.g., see Refs.~\cite{Boyer1988,Aharonov1988,Feng2017}), and both equations are used in the literature (e.g., see Refs.~\cite{Ragsdale1998,Coey2009,Plouffe2014,Benassi2014,Nasiri2021}). Despite the fact that magnetic moments formed from current loops make more physical sense in classical electrodynamics, since, to the best of the author's knowledge, magnetic monopoles do not exist, in this study, simulations are performed following the separated charge model. The reason is that the magnetic force of Eq.~\eqref{eq:flag_A1} resembles the one that was used in several previous studies in the relevant field~\cite{Yamada2002,Yamada2003,Jiang2020a,Kajimoto2003,Zhang2021}. A comparison of the effects of the two forces on the combustion dynamics is left for future work.

\section{Derivation of diffusion and drift velocity under electromagnetic fields}
\label{sec:diffusion_drift_velocity}

According to Williams et al.~\cite{Williams1985} the diffusion velocities $V_k$ of the $N$ species are obtained by solving the system:
\begin{equation}
    \begin{aligned}
    \nabla X_p = \sum_{k=1}^N \dfrac{X_p X_k}{D_{pk}}\left(V_k-V_p\right)+\left(Y_p-X_p\right)\dfrac{\nabla P}{P} + \dfrac{\rho}{P} \sum_{k=1}^N Y_pY_k \left(f_p-f_k\right).
    \label{eq:flag_A2_1}
    \end{aligned}
\end{equation}

The following assumptions are made:
\begin{enumerate}
    \item $D_{pk}=D_p$ for all species $p$. This follows from the Hirschfelder \& Curtiss approximation.
    \item The changes in pressure are negligible, i.e., $\nabla P=0$.
\end{enumerate}

Therefore, Eq.~\eqref{eq:flag_A2_1} can be simplified to:
\begin{equation}
    \begin{aligned}
    D_p\nabla X_p = \sum_{k=1}^N X_p X_k V_k- X_p V_p + \dfrac{\rho D_p}{P} \sum_{k=1}^N Y_pY_k \left(f_p-f_k\right).
    \end{aligned}
\end{equation}
where, for simplicity, the gradient of the average molecular weight of the mixture is neglected. Substituting some of the mole fractions with mass fractions and rearranging the terms leads to the following equation:
\begin{equation}
    \begin{aligned}
    Y_p V_p + D_p\nabla Y_p = Y_p\sum_{k=1}^N X_k V_k + \dfrac{\rho D_p Y_p^2}{P X_p} \sum_{k=1}^N Y_k \left(f_p-f_k\right).
    \label{eq:flag_A2_2}
    \end{aligned}
\end{equation}

Next, Eq.~\eqref{eq:flag_A2_2} is summed over all $p$ species. Moreover, from the mass conservation equation, it holds that $\sum_p Y_p V_p=0$. Therefore, Eq.~\eqref{eq:flag_A2_2} becomes:
\begin{equation}
    \begin{aligned}
    \sum_{k=1}^N X_k V_k = \sum_{p=1}^N D_p \nabla Y_p + \sum_{p=1}^N \left[ \dfrac{\rho D_p Y_p^2}{X_p P} \sum_{k=1}^N Y_k \left(f_p-f_k\right) \right].
    \label{eq:flag_A2_3}      
    \end{aligned}
\end{equation}
Substituting the first term of the right-hand-side of Eq.~\eqref{eq:flag_A2_2} with Eq.~\eqref{eq:flag_A2_3}, gives the following equation:
\begin{equation}
    \begin{aligned}
    V_p=&-\dfrac{D_p}{Y_p} \nabla Y_p + \dfrac{\rho D_p M_{p}}{M_{m} P} \sum_{k=1}^N Y_k\left(f_p-f_k\right) \\
    &+ \sum_{j=1}^N Y_j \left[\dfrac{D_j}{Y_j} \nabla Y_j - \dfrac{\rho D_j M_{j}}{M_{m} P} \sum_{k=1}^N Y_k\left(f_j-f_k\right)\right].
    \label{eq:flag_A2_4}
    \end{aligned}
\end{equation}
Eq.~\eqref{eq:flag_A2_4} corresponds to the equation that is discretized and solved in EMI-SENGA.

 % \bibliographystyle{elsarticle-num} 
 % \bibliography{references}





% \end{thebibliography}
\end{document}

\begin{frontmatter}



\title{A computational approach for the study of electromagnetic interactions in reacting flows}


\author[inst1,inst2]{Efstratios M. Kritikos\corref{cor1}}
\author[inst3]{Stewart Cant}
\author[inst1]{Andrea Giusti}

\affiliation[inst1]{{Department of Mechanical Engineering, Imperial College London, London SW7 2AZ, United Kingdom}}
\affiliation[inst2]{{Department of Applied Physics and Materials Science, California Institute of Technology, Pasadena, 91125, United States}}
\affiliation[inst3]{{Department of Engineering, University of Cambridge, CB2 1PZ, United Kingdom}}

\cortext[cor1]{Corresponding author.\\
E-mail address: emk@caltech.edu.}

\begin{abstract}

This document reports additional information in support of the discussion of the results presented in the main paper. The structure of the Supporting Information is as follows.

\tableofcontents

\end{abstract}
\end{frontmatter}

\section{Methodology}
\subsection{Electrostatic/Magnetostatic fields solver (EMI-static)}
\label{subsec:static_solver}

\begin{figure}[!htbp]
\centering
\includegraphics[width=\linewidth]{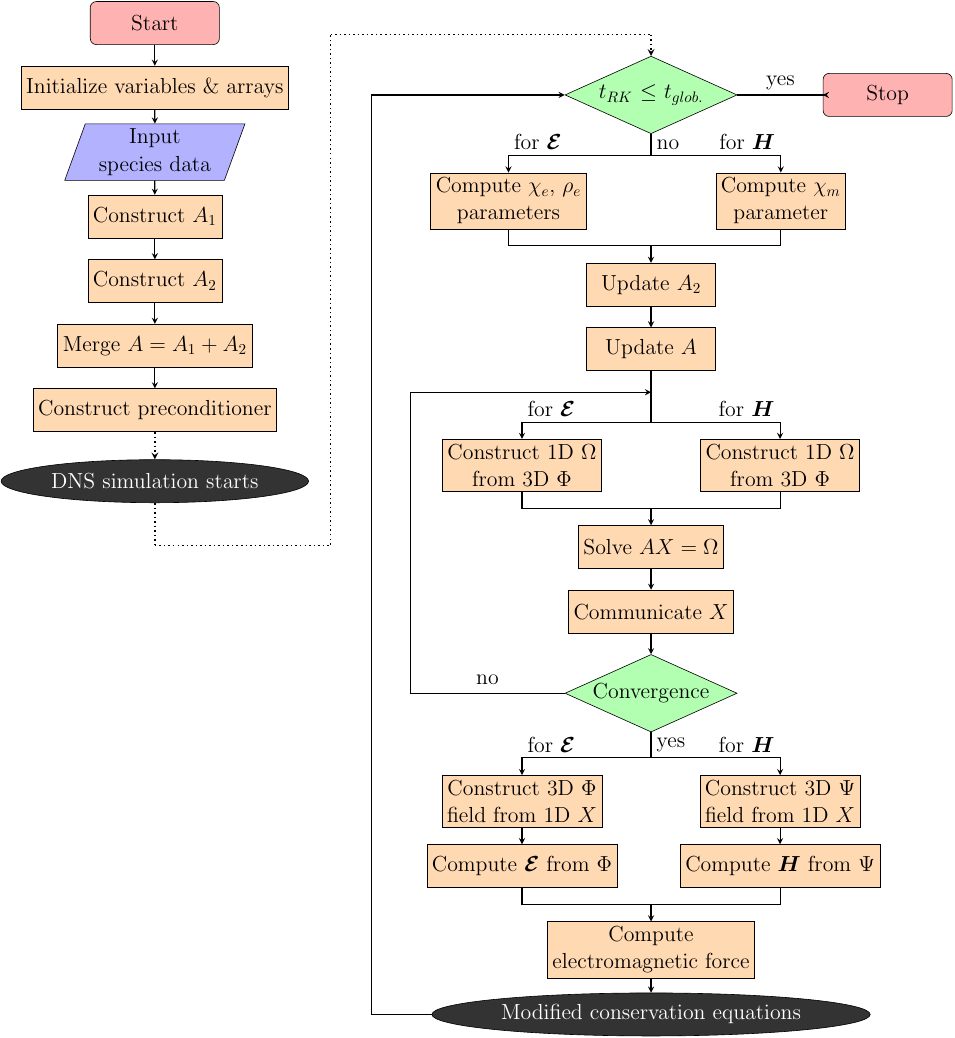}
\caption{Flow chart of the EMI-SENGA implementation for static fields.}
\label{fig:flow_chart_static}
\end{figure}

The developed code is called Electromagnetic Interactions (EMI). EMI is integrated into a modified version of the SENGA~\cite{Jenkins1999,Cant1999} Direct Numerical Simulation (DNS) code. The DNS modifications concern the electromagnetic field effects on the conservation equations, boundary conditions, and molecular transport properties. EMI takes as input the properties of the reacting flow and solves for the propagation of the electromagnetic fields. EMI also computes the electromagnetic forces, which are then introduced into the conservation equations for the flow and the species. SENGA then solves the conservation equations of the reacting flow using the \gls{dns} method. Moreover, SENGA advances the governing equations using a Runge–Kutta algorithm. The flow chart with the structure of EMI-SENGA for static fields is shown in Fig~\ref{fig:flow_chart_static}. Both Eqs.~\eqref{eq:electrostatics} and~\eqref{eq:magnetostatics} of the main paper are solved in a similar manner. The overall process is described below:

\begin{table}[!tb]

\centering
\caption{\nth{1} and \nth{2} order derivative coefficients using central finite difference.}
\label{tbl:coefficiens_discretization}

\begin{tabular}{c|cccccc}
\multicolumn{7}{l}{\nth{1} order derivative}\\
\hline
& \multicolumn{6}{c}{Grid points}\\
Accuracy&$i-5$&$i-4$&$i-3$&$i-2$&$i-1$&$i$\\
\hline
2       &     &     &     &     &-1/2 & 0 \\
4       &     &     &     &1/12 &-2/3 & 0 \\
6       &     &     &-1/60&3/20 &-3/4 & 0 \\
8       &    &1/280&-4/105&1/5  &-4/5 & 0 \\
10  &1/3150&-5/1008&5/126&-5/21&5/3  &-5269/1800\\
\hline
& \multicolumn{6}{c}{Grid points}\\
Accuracy &$i+1$&$i+2$&$i+3$&$i+4$&$i+5$ &\\
\hline
2 & 1/2 &     &     &     &    & \\
4 & 2/3 &-1/12&     &     &     &\\
6 & 3/4 &-3/20&1/60 &     &     &\\
8 & 4/5 &-1/5 &4/105&-1/280&   & \\
10 &-5/3 &5/21 &-5/126& 5/1008   &-1/3150 & \\
\hline
\end{tabular}

\begin{tabular}{c|cccccc}
\multicolumn{7}{l}{}\\
\multicolumn{7}{l}{\nth{2} order derivative}\\
\hline
& \multicolumn{6}{c}{Grid points}\\
Accuracy&$i-5$&$i-4$&$i-3$&$i-2$&$i-1$&$i$\\
\hline
2       &     &     &     &     &1 & -2  \\
4       &     &     &     &-1/12 &4/3 & -5/2 \\
6       &     &     &1/90&-3/20 &3/2 & -49/18 \\
8       &    &-1/560&8/315&-1/5  &8/5 & -205/72 \\
10  &1/3150&-5/1008&5/126&-5/21&5/3  &-5269/1800\\
\hline
& \multicolumn{6}{c}{Grid points}\\
Accuracy&$i+1$&$i+2$&$i+3$&$i+4$&$i+5$ &\\
\hline
2       & 1 &     &     &     &     &\\
4       & 4/3 &-1/12&     &     &    & \\
6       & 3/2 &-3/20&1/90 &     &     &\\
8       & 8/5 &-1/5 &8/315&-1/560&   & \\
10      &5/3 &-5/21 &5/126& -5/1008    &1/3150  &\\
\hline
\end{tabular}
\end{table}

\begin{figure}[t]
    \centering
    \begin{subfigure}[t]{0.245\textwidth}
        \includegraphics[width=\linewidth]{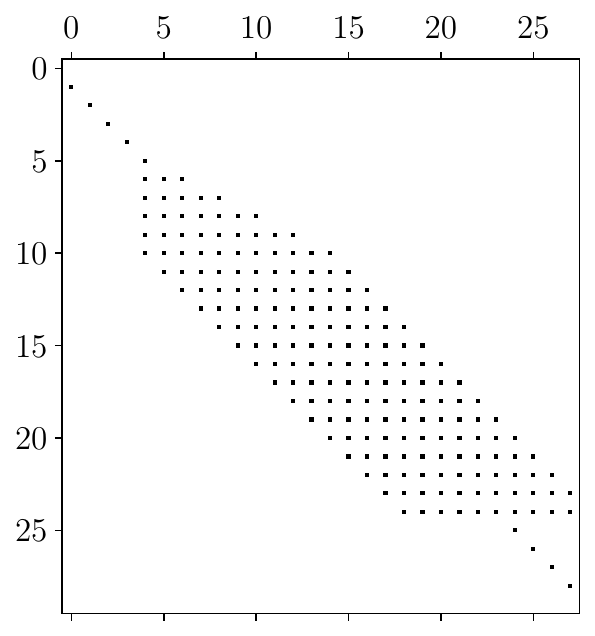}
        \caption{1D simulation,\\ \nth{1} processor}
    \end{subfigure}%
    \begin{subfigure}[t]{0.245\textwidth}
        \includegraphics[width=\linewidth]{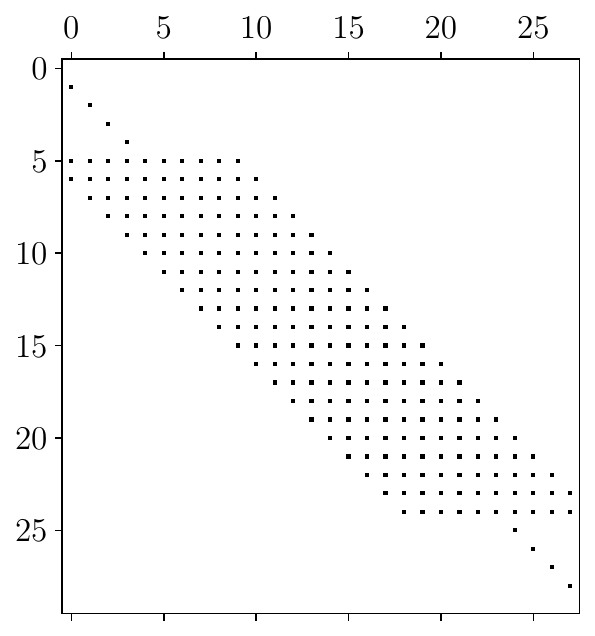}
        \caption{1D simulation,\\ \nth{2} processor}
    \end{subfigure}%
    \begin{subfigure}[t]{0.245\textwidth}
        \includegraphics[width=\linewidth]{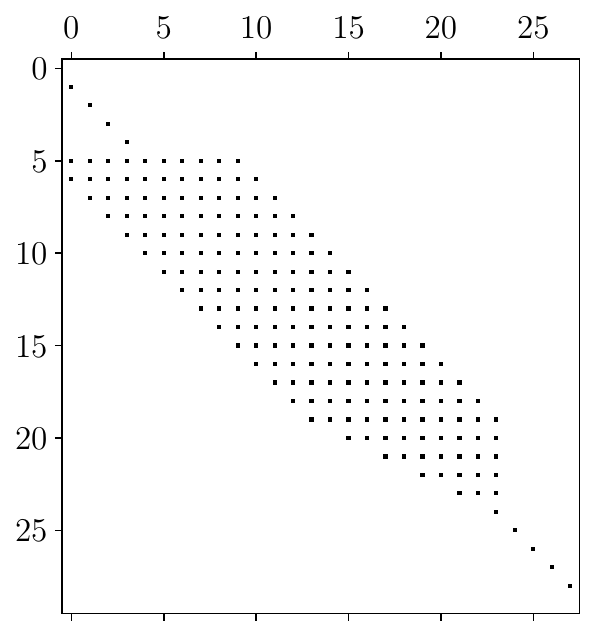}
        \caption{1D simulation,\\ \nth{3} processor}
    \end{subfigure}%
    \begin{subfigure}[t]{0.265\textwidth}
        \includegraphics[width=\linewidth]{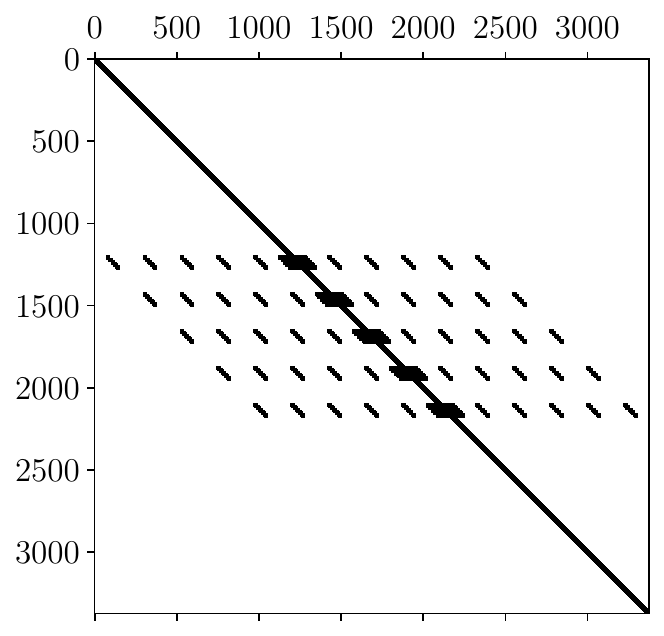}
        \caption{3D simulation,\\ \nth{2} processor}
    \end{subfigure}
    \caption{Population of matrix $A$ for 1D and 3D simulations decomposed into 3 processors along the $x$ direction. For the 1D simulation $N_x=60$, while for the 3D $N_x = 15$, $N_y = 5$, $N_z = 5$.}
    \label{fig:A_population}
\end{figure}

\begin{enumerate}
    \item The \nth{1} and \nth{2} order derivatives are discretized using the finite difference approach and keeping a \nth{10} order accuracy in each $x$, $y$, and $z$ direction.
    \item The accuracy drops from \nth{10} to \nth{2} as the $x$ direction boundaries are approached. This is explained schematically in Fig.~\ref{fig:schematic_electrostatics_magnetostatics} of the main paper. Moreover, Table~\ref{tbl:coefficiens_discretization} shows the coefficients of each term of the discretized equations. Note that it is assumed that the Dirichlet boundary conditions are applied in the $x$ direction. This means that in the $y$ and $z$ directions, the \nth{10}-order accuracy holds everywhere.
    \item The resulting system is of the form of $A X = \Omega$, where matrix $A$ holds the left-hand side parameters of the discretized \nth{1} and \nth{2} derivatives, $X$ holds the discretized $\Phi$ or $\Psi$ potential fields, and $\Omega$ holds the right-hand side parameters. For the study of one- or two-dimensional flows, the discretization in $y$ and $z$ directions is not included in the $A$ matrix.
    \item $A=A_1+A_2$. $A_1$ matrix corresponds to the parameters of the second-order derivatives of Eq.~\eqref{eq:electrostatics} and Eq.~\eqref{eq:magnetostatics}. $A_2$ corresponds to the parameters of the first-order derivatives. Therefore, $A_2$ depends on the $\chi_e$ and $\chi_m$ fields. The population of the $A$ matrix for 1D and 3D domains is shown in Fig.~\ref{fig:A_population}.
    \item For the Dirichlet boundary conditions, the values of matrix $A$ are unity, and the values of $\Omega$ are equal to the electric and magnetic boundaries defined by the user.
    \item The system of equations $A X = \Omega$ is solved using either the direct LU decomposition method or the iterative \gls{bicgstab} of the LAPACK library~\cite{LAPACK}. Due to the exceedingly high computational cost of the former method, the simulations of this study are performed using the \gls{bicgstab} method.
    \item The preconditioner matrix used for the \gls{bicgstab} solver is equal to $A_1^{-1}$. The inverse of the $A_1$ matrix is computed at the beginning of the simulation using routines of the LAPACK library~\cite{LAPACK}. As long as $A_2$ does not change significantly, i.e., as long as the gradients of electric and magnetic susceptibilities are low, it is expected that convergence can be achieved with this preconditioner. 
\end{enumerate}    

The parallelization strategy of the EMI code for static fields is as follows:
\begin{enumerate}
    \item To retain a fully parallel solver, each core holds and solves a separate $A X = \Omega$ system, using either direct or iterative methods. 
    \item If the core is an internal core, then its ghost cells hold the values from the adjacent cores. These ghost cells are treated as Dirichlet boundary conditions. In other words, for the ghost cells, the values of matrix $A$ are unity, and the values of $\Omega$ are equal to the $X$ values from the adjacent cells of the neighbouring processor.
    \item If the core is a boundary core, then it holds the Dirichlet boundary conditions. In this case, the residual ghost cells do not hold any information, and they are not part of the solution.
    \item After the $A X = \Omega$ system is solved, the ghost cells of each processor are updated with values from the other processors, and the system is solved again until convergence is reached. The process is repeated until the specified absolute and relative tolerances are satisfied.
\end{enumerate}   

Note also that, as can be seen from Fig~\ref{fig:flow_chart_static}, the static fields are solved inside the Runge-Kutta solver of SENGA. This coupling is expected to increase the stability and robustness of the code.

\subsection{Electromagnetic waves solver (EMI-FDTD)}
\label{sisubsec:fdtd_solver}

\begin{figure}
    \centering
    \includegraphics[width=\linewidth]{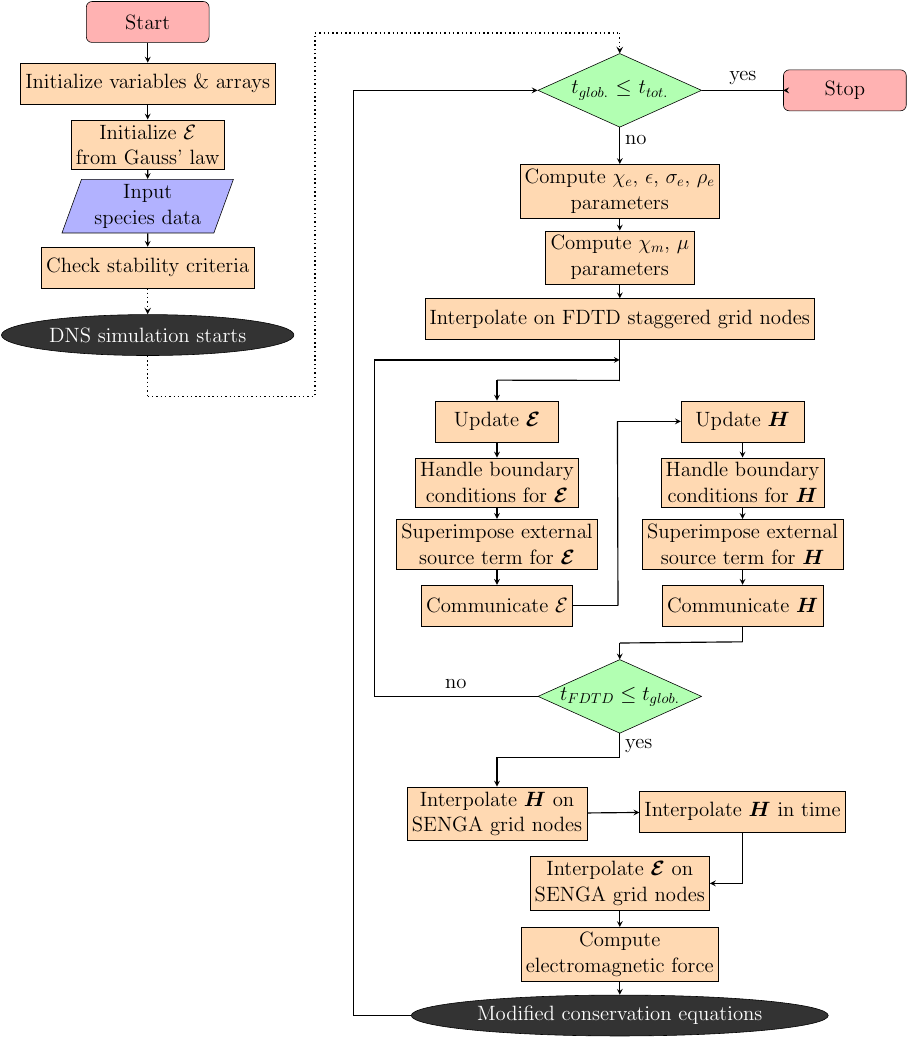}
    \caption{Flow chart of the FDTD method implemented in EMI-SENGA.}
    \label{fig:flow_chart_fdtd}
\end{figure}

A schematic diagram of the structure of the EMI-FDTD code is shown in Fig.~\ref{fig:flow_chart_fdtd}. As discussed in Section~\ref{subsubsec:electromagnetic_waves} of the paper, the \gls{fdtd} solver of EMI code is formulated on a grid different from the SENGA grid. The \gls{fdtd} grid also holds the additional boundary cells. The structure of the code is as follows:
\begin{enumerate}
    \item The medium properties (see Section~\ref{subsec:medium_properties}) are interpolated on the Yee cell positions. Specifically, $\epsilon$ and $\sigma^e$ are interpolated on the positions of the electric field components, while $\mu$ is interpolated on the positions of the magnetic field components.
    \item Inside the main loop of each SENGA timestep, the nested loop of the EMI-FDTD code propagates the electromagnetic waves with a timestep of $\Delta t_{FDTD}$. The electric field components are updated first. Then the electric field sources and boundary conditions are superimposed on the new electric field solution. At this point, the final solution of the electric field is transferred to the other processors. The parallelization strategy will be discussed below in this section. 
    \item Afterwards, the magnetic fields are updated in a similar manner to the electric fields. The boundary conditions are applied, and the ghost cells between processors are communicated.
    \item At the end of electromagnetic field propagation, the fields are interpolated in space and time to be consistent with the SENGA code. Next, the electromagnetic forces are computed.
\end{enumerate}

The parallelization of the EMI-\gls{fdtd} solver follows the domain decomposition method of SENGA. Therefore, the computational domain is decomposed into subdomains, which are solved by each processor. Ghost cells are necessary to communicate information between processors. Since SENGA uses a \nth{10}-order accuracy, at least 5 ghost cells in each direction are required. However, in the EMI-FDTD method, each processor needs to communicate only one row of ghost cells, since the \gls{fdtd} method is of \nth{2}-order accuracy. The \gls{cpml} regions are assigned to the boundary CPU processors. It should be noted here that, since SENGA requires at least 5 ghost cells in each direction, if the thickness of the \gls{cpml} regions is equal to or less than 5 cells, there will not be any additional memory load to the processors.

\subsection{\gls{cpml} formulation}
\label{sisubsec:cpml}

The parameters needed for the \gls{cpml} formulation, namely the $\sigma$, $\alpha$, and $\kappa$ parameters as detailed in Ref.~\cite{Kritikos2023t} and computed in the EMI-FDTD code, are described here. Two widely used approaches for the computation of these parameters are the polynomial grading and geometric grading approaches~\cite{Taflove2005}. A polynomial grading approach is selected in this study for the \gls{cpml} regions, which is commonly found in the literature~\cite{Taflove2005,Taflove2013,Elsherbeni2016}. Therefore,
\begin{itemize}
    \item The $\sigma_w$ term is computed in the $x$ direction by:
    
    \begin{equation}
    \begin{aligned}
        \sigma_x\left(x\right) = \left(\dfrac{x}{l_{x}}\right)^m \sigma_{x,\max},
        \label{eq:sigma_x}    
    \end{aligned}
    \end{equation}
    where $x$ is the distance of the electric or magnetic field component node under consideration from the interface between the main computational domain and the \gls{cpml} region. In addition, $m$ is the order of the polynomial and $l_{x}$ is the thickness of the \gls{cpml} layer. In a similar manner, the $\sigma_y\left(y\right)$ and $\sigma_z\left(z\right)$ are computed in the $y$ and $z$ direction, respectively. It should be noted that the values that $x$ can take are different for the electric and magnetic field components since they are located at different positions (see Fig.~\ref{fig:Yee_cell}). According to Ref.~\cite{Elsherbeni2016}, and references therein:
    \begin{equation}
        \sigma_{x,\max}=\sigma_{\mathrm{factor}} \sigma_{x,\mathrm{opt}},
        \label{eq:sigma_x_max}
    \end{equation}
    where for the electric field:
    \begin{equation}
    \begin{aligned}
        \sigma^e_{x,\mathrm{opt}}\left(x\right) = \dfrac{m+1}{150 \pi \sqrt{\epsilon_r}\Delta x},
        \label{eq:sigma_opt_e}
    \end{aligned}
    \end{equation}
    and for the magnetic field:
    \begin{equation}
    \begin{aligned}
        \sigma^m_{x,\mathrm{opt}}\left(x\right) = \dfrac{\mu_0}{\epsilon_0} \dfrac{m+1}{150 \pi \sqrt{\epsilon_r}\Delta x}.
        \label{eq:sigma_opt_h}
    \end{aligned}
    \end{equation}
    In Eqs.~\eqref{eq:sigma_opt_e} and \eqref{eq:sigma_opt_h}, $\epsilon_r$ is the relative permittivity of the background material. For this implementation, it is assumed that the background material is a vacuum. Therefore, $\epsilon_r=1$.

    \item The $\kappa_w$ term is computed in the $x$ direction using:
    \begin{equation}
    \begin{aligned}
        \kappa_x\left(x\right) = 1+\left(\kappa_{\max}-1\right)\left(\dfrac{x}{l_x}\right)^m,
        \label{eq:fdtd_kappa_x}
    \end{aligned}
    \end{equation}
     for both the electric and magnetic fields. In the above equation, $\kappa_{\max}$ is an input parameter. The parameter $\kappa_w$ is equal to unity at the interface between the CPML region and the main computational domain, while it becomes maximum, i.e., $\kappa_{\max}$, at the outer side of the CMPML region.
    
    \item The parameter $\alpha_w$ in the $x$ direction is computed for the electric field from:
    \begin{equation}
    \begin{aligned}
        \alpha_x\left(x\right) = \alpha_{\min} + \left(\alpha_{\max}-\alpha_{\min}\right)\left(1-\dfrac{x}{l_x}\right)
        \label{eq:alpha_x_e}
    \end{aligned}
    \end{equation}
    and for the magnetic field:
    \begin{equation}
    \begin{aligned}
        \alpha_x\left(x\right) = \dfrac{\mu_0}{\epsilon_0} \left[\alpha_{\min} + \left(\alpha_{\max}-\alpha_{\min}\right)\left(1-\dfrac{x}{l_x}\right)\right].
        \label{eq:alpha_x_b}
    \end{aligned}
    \end{equation}
\end{itemize}
At the CMPL region and computational domain interface, the $\alpha_w$ parameter becomes maximum, i.e., $\alpha_{\max}$, while it reaches a minimum value, i.e., $\alpha_{\min}$, at the outer CMPL side.

It should be noted that the parameters $\sigma_w$, $\kappa_w$, and $\alpha_w$ mentioned above are assumed to have an offset in their $x$ distance from the main computational domain interface~\cite{Elsherbeni2016}. Specifically, when the parameters are computed on the electric field components' nodes, then a 0.25$\Delta x$ offset is subtracted from the $x$ distance. Moreover, when the nodes of the magnetic field components are considered, then a 0.75$\Delta x$ offset is subtracted from the $x$ distance. The reason for this offset is to make sure that the CPML algorithm updates the same number of electric and magnetic field components~\cite{Elsherbeni2016}. The values subtracted correspond to the distance of the components from the \gls{cpml} interface, due to the fact that the fields are staggered in the FDTD formulation (see Fig.~\ref{fig:Yee_cell}).

\subsection{Properties of species}

\begin{table}[H]
\begin{tabular}{c|cc}
 Species   & $q$ [e] & $\alpha$ [{\AA}$^3$ (cgs units)] \\
 \hline
 \ch{N2}   & 0 & 1.710 \cite{Olney1997} \\
 \ch{H2}   & 0 & 0.787 \cite{Olney1997} \\
 \ch{H}    & 0 & 0.667 \cite{Miller1978} \\
 \ch{O2}   & 0 & 1.562 \cite{Olney1997} \\
 \ch{O}    & 0 & 0.802 \cite{Miller1978} \\
 \ch{H2O}  & 0 & 1.501 \cite{Olney1997} \\
 \ch{OH}   & 0 & 1.121 \cite{Loukhovitski2016} \\
 \ch{H2O2} & 0 & 2.300 \cite{Giguere1983} \\
 \ch{HO2}  & 0 & 1.972 \cite{Hait2018} \\
 \ch{CO}   & 0 & 1.953 \cite{Olney1997} \\
 \ch{CO2}  & 0 & 2.507 \cite{Olney1997} \\
 \ch{CH4}  & 0 & 2.448 \cite{Olney1997} \\
 \ch{CH3}  & 0 & 2.341 \cite{Hait2018} \\
\end{tabular}\hspace{1cm}%
\begin{tabular}{c|cc}
 Species   & $q$ [e] & $\alpha$ [{\AA}$^3$ (cgs units)] \\
 \hline
 \ch{CH2}  & 0 & 2.143 \cite{Hait2018} \\
 \ch{CH2(S)} & 0 & 2.143$^\dagger$ \cite{Hait2018} \\
 \ch{C}      & 0 & 1.760 \cite{Miller1978} \\
 \ch{CH}     & 0 & 1.883 \cite{Loukhovitski2016} \\
 \ch{CH3OH}  & 0 & 3.210 \cite{Gussoni1998} \\
 \ch{CH3O}   & 0 & 3.450 \cite{Hait2018} \\
 \ch{CH2OH}  & 0 & 3.450$^\dagger$ \cite{Hait2018} \\
 \ch{CH2O}   & 0 & 2.800 \cite{Haynes2016} \\
 \ch{HCO}    & 0 & 2.485 \cite{Hait2018} \\
 \ch{CHO+}   & 1 & -- \\
 \ch{H3O+}   & 1 & -- \\
 \ch{O2-}    & -1 & -- \\
 \ch{e-}     & -1 & -- \\
\end{tabular}

$^\ddagger$ Due to the low concentration of ionic species, the polarizabilities of ionic species can be neglected, since they do not contribute significantly to the overall concentration (c.f. Eq. \eqref{eq:Clausius_Massotti}).
\caption{Electrical properties of species.}
\label{Table:elec_prop}
\end{table}

\begin{table}[H]

\begin{tabular}{c|c}
\multicolumn{2}{l}{(a) Paramagnetic materials} \\
 Species   & $\mathcal{S}$ \\
 \hline
 \ch{H}    & 1/2  \\
 \ch{O2}   & 1    \\
 \ch{O}    & 1    \\
 \ch{OH}   & 1/2  \\
 \ch{HO2}  & 1/2  \\
 \ch{CH3}  & 1/2  \\
 \ch{CH2}  &  1   \\
 \ch{CH2(S)} & 1  \\
 \ch{C}      &  1  \\
 \ch{CH}     & 1/2  \\
 \ch{CH3O}   & 1/2  \\
 \ch{CH2OH}  & 1/2  \\
 \ch{HCO}    & 1/2  \\
 \ch{H3O+}   & 1/2  \\
 \ch{O2-}    & 1/2  \\
\end{tabular}\hspace{2cm}%
\begin{tabular}{c|c}
\multicolumn{2}{l}{(b) Diamagnetic materials} \\
 Species   & $\chi_m^{molar}/10^{-6}$ [cm$^3$/mol]$^\dagger$ \cite{Haynes2016} \\
 \hline
 \ch{N2}   & -12.0 \\
 \ch{H2}   & -3.99 \\
 \ch{H2O}  & -13.1 \\
 \ch{H2O2} & -17.3 \\
 \ch{CO}   & -11.8 \\
 \ch{CO2}  & -21.0 \\
 \ch{CH4}  & -17.4 \\
 \ch{CH3OH}  & -21.4 \\
 \ch{CH2O}   & -18.6 \\
 \ch{CHO+}   & NF \\
 & \\
 & \\
 & \\
 & \\
 & \\
\end{tabular}

$^\dagger$ Values are given in the CGS system. Note that to transform to volume magnetic susceptibility, values should be multiplied by $\rho Y^s/M_s$.

\caption{Properties of (a) paramagnetic and (b) diamagnetic species.}
\label{Table:mag_prop}
\end{table}

\subsection{Polynomial fitting method}
\label{sisubsubsec:polynomial_fitting_method}

SENGA implementation follows the CHEMKIN~\cite{Kee1996} equations for the computation of the transport properties. Thus, the computation of viscosity $\eta_\alpha$, thermal conductivity $\lambda_\alpha$ and binary diffusion coefficients $D_{\alpha \beta}$ is based on a polynomial formulation~\cite{Kee1986}:
\begin{equation}
\begin{aligned}
    \ln{\eta_\alpha} &= \sum^J_{j=1} a_{\alpha,j} \left(\ln T\right)^{j-1}, \\
    \ln{\lambda_\alpha} &= \sum^J_{j=1} b_{\alpha,j} \left(\ln T\right)^{j-1}, \\
    \ln{D_{\alpha\beta}^{(P_0)}} &= \sum^J_{j=1} d_{\alpha\beta,j} \left(\ln T\right)^{j-1},
\end{aligned}
\end{equation}
where $a$, $b$, and $d$ are polynomial coefficients. The polynomials are fitted with respect to a reference temperature $T_0$. Note that the binary diffusion coefficients are also given with respect to a reference pressure $p_0$. For pressures other than the reference one, the diffusion coefficients are computed according to:
\begin{equation}
    \ln{D_{\alpha\beta}^{(P)}} = \ln{D_{\alpha\beta}^{(P_0)}} \dfrac{P_0}{P}.
\end{equation}

Note that Cantera~\cite{Goodwin2018} follows a different polynomial fitting method for the viscosity, thermal conductivity, and binary diffusion coefficients compared to SENGA. Specifically, the polynomials take the form:
\begin{equation}
\begin{aligned}
    \sqrt{\left(\dfrac{\eta_\alpha}{\sqrt{T}}\right)} &= \sum^J_{j=1} a_{\alpha,j} \left(\ln T\right)^{j-1}, \\
    \dfrac{\lambda_\alpha}{\sqrt{T}} &= \sum^J_{j=1} b_{\alpha,j} \left(\ln T\right)^{j-1}, \\
    \dfrac{D_{\alpha\beta}^{(P_0)}}{T^{3/2}P} &= \sum^J_{j=1} d_{\alpha\beta,j} \left(\ln T\right)^{j-1}.
    \label{eq:CA_mode}
\end{aligned}
\end{equation}

This method was implemented in EMI-SENGA since considerable differences in the concentrations of ionic species were observed between the two polynomial fitting methods when testing a 1D laminar \ch{CH4}-air flame (results are not presented here for the sake of conciseness).

\subsection{Ion transport model}
\label{sisubsubsec:ion_transport_model}

The transport model for ionic species differs from that of neutral molecules. In this study, the ion transport model implemented in Cantera is also implemented in EMI-SENGA. According to this ion transport model, the transport properties are computed as follows~\cite{Goodwin2018}:
\begin{itemize}
    \item The Stockmayer-(n,6,4) model is used for collisions between charged and neutral species~\cite{Selle1999,Selle2000,Chiflikian1995,Viehland1975,Prager2005}.
    \item For the computation of mixture-average diffusion coefficients and mobilities, only binary transport between neutral-neutral and ion-neutral species is considered. Interactions between ions are neglected.
    \item For the computation of mixture-averaged viscosity and thermal conductivity, only neutral species are included.
    \item For collisions between \ch{O2} and \ch{O2-}, experimental collision data are used. The reason is that the Stockmayer-(n,6,4) model cannot accurately describe collisions involving those two molecules due to resonant charge transfers. The data is taken from Prager~\cite{Prager2005}.
\end{itemize}

\section{Results}

\subsection{Validation of the electrostatic field implementation}
\label{SIsubsec:validation}

\begin{figure}[H]
    \centering
    \graphicspath{{Figures/Comparison/ion_EF/}}
    \includegraphics[width=0.33\textwidth]{graph_dens.pdf}%
    \includegraphics[width=0.33\textwidth]{graph_temp.pdf}%
    \includegraphics[width=0.33\textwidth]{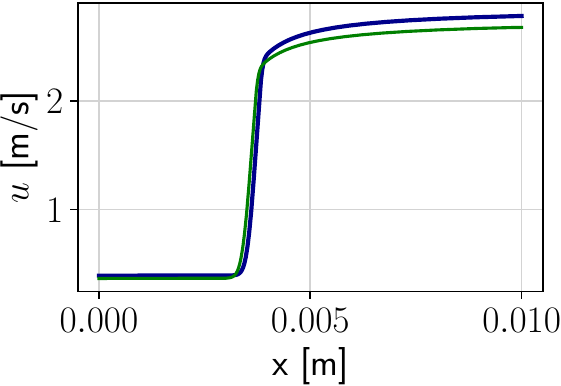}
    \includegraphics[width=0.33\textwidth]{graph_yf01.pdf}%
    \includegraphics[width=0.33\textwidth]{graph_yf02.pdf}%
    \includegraphics[width=0.33\textwidth]{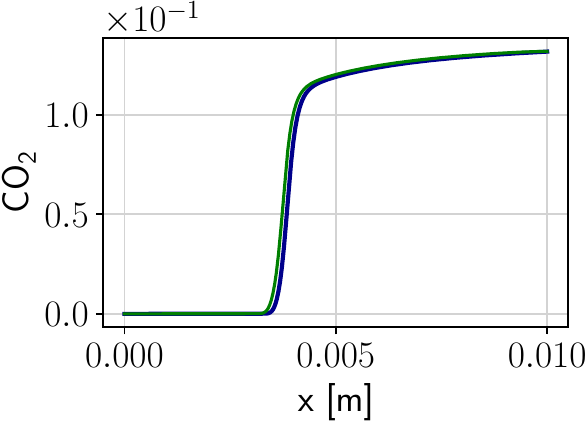}
    \includegraphics[width=0.33\textwidth]{graph_yf04.pdf}%
    \includegraphics[width=0.33\textwidth]{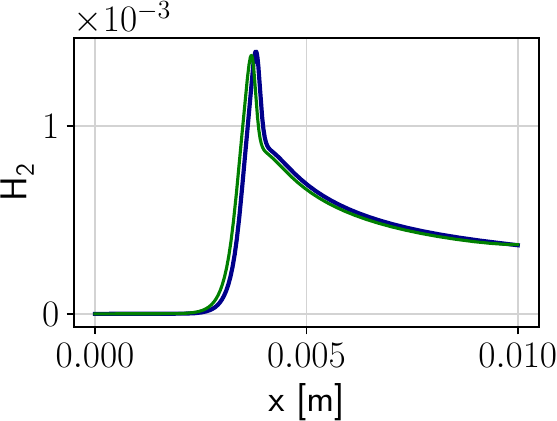}%
    \includegraphics[width=0.33\textwidth]{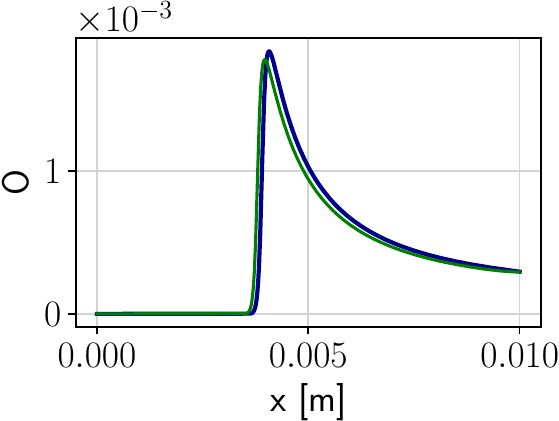}
    \includegraphics[width=0.33\textwidth]{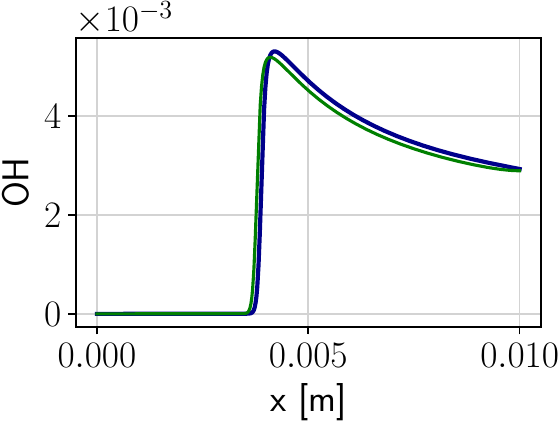}%
    \includegraphics[width=0.33\textwidth]{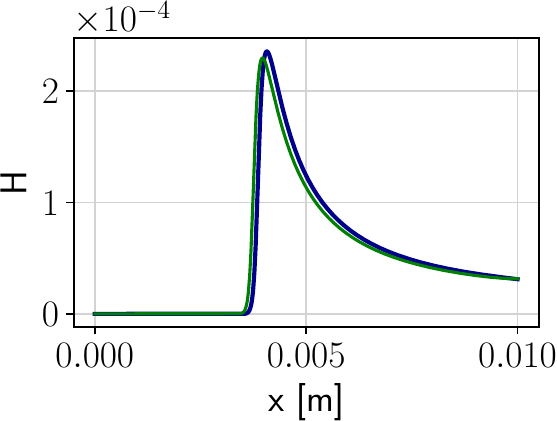}%
    \includegraphics[width=0.33\textwidth]{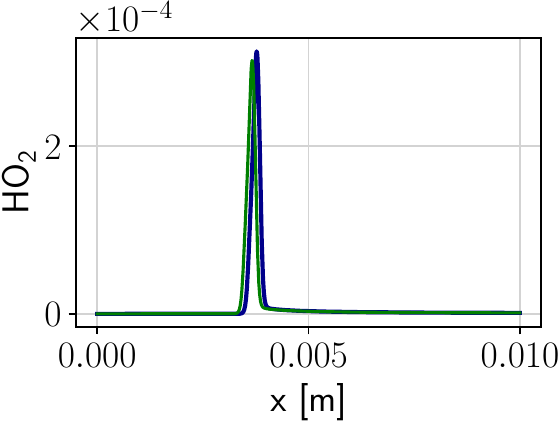}
    \includegraphics[width=0.33\textwidth]{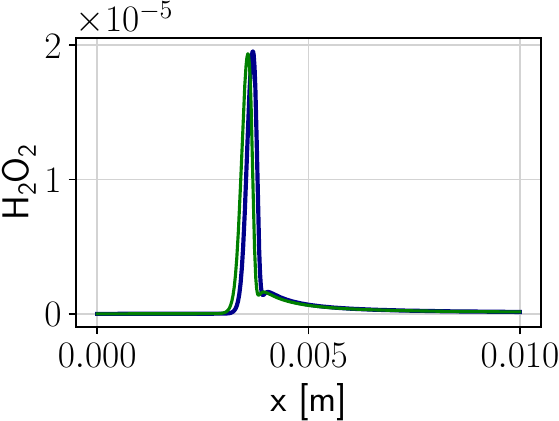}%
    \includegraphics[width=0.33\textwidth]{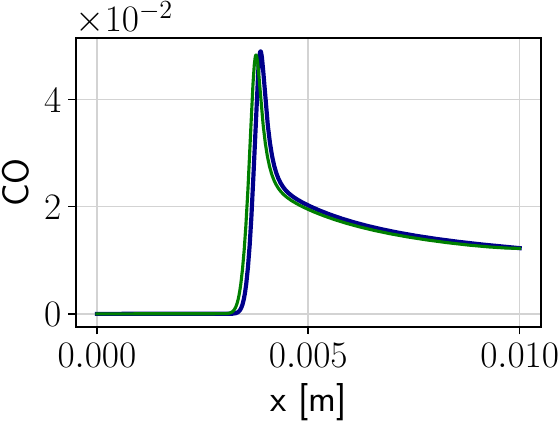}%
    \includegraphics[width=0.33\textwidth]{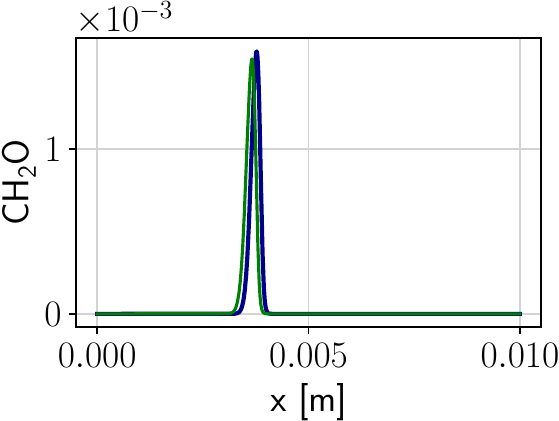}
    \caption{Comparison between Cantera and EMI-SENGA for the ion transport model and with electrostatic interactions activated (continues to Fig.~\ref{fig:comparison_ion_EF_2}).}
    \label{fig:comparison_ion_EF_1}
\end{figure}   

\begin{figure}[H]
    \centering
    \graphicspath{{Figures/Comparison/ion_EF/}}
    \includegraphics[width=0.33\textwidth]{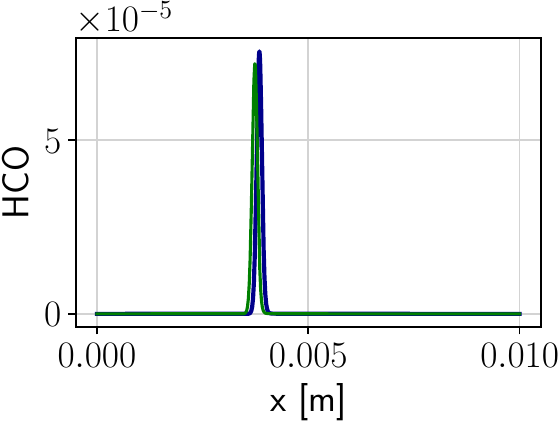}%
    \includegraphics[width=0.33\textwidth]{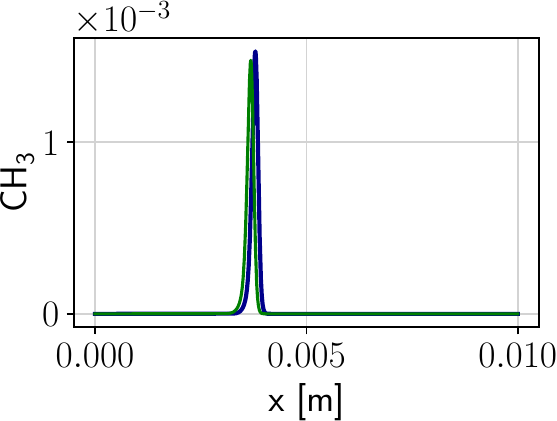}%
    \includegraphics[width=0.33\textwidth]{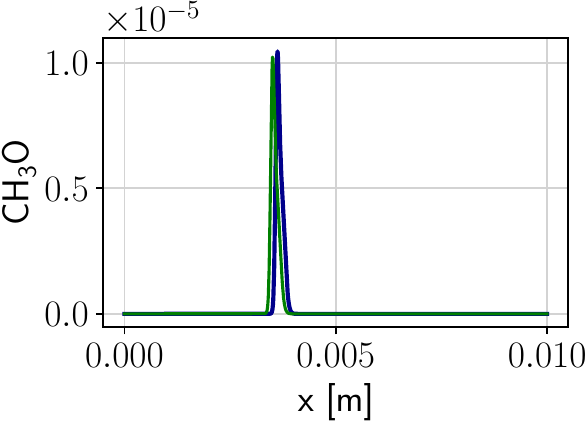}
    \includegraphics[width=0.33\textwidth]{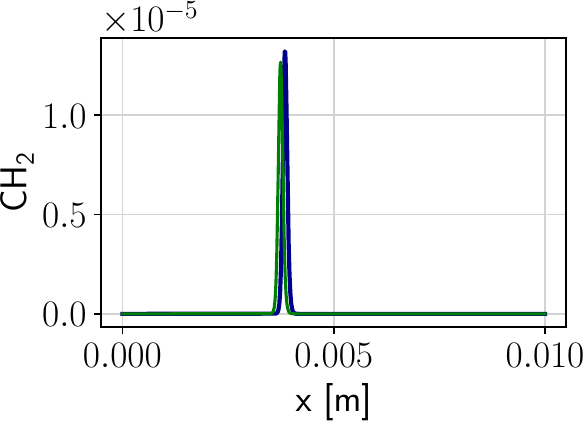}%
    \includegraphics[width=0.33\textwidth]{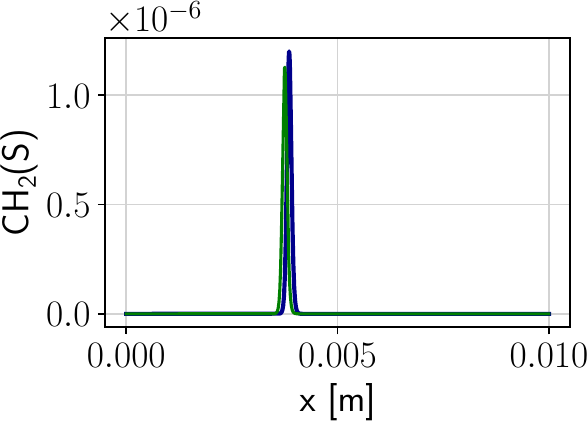}%
    \includegraphics[width=0.33\textwidth]{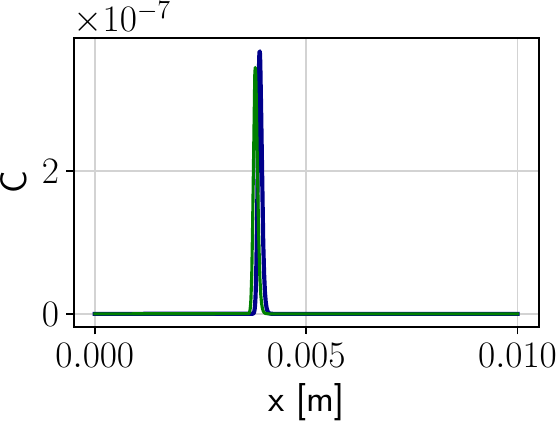}
    \includegraphics[width=0.33\textwidth]{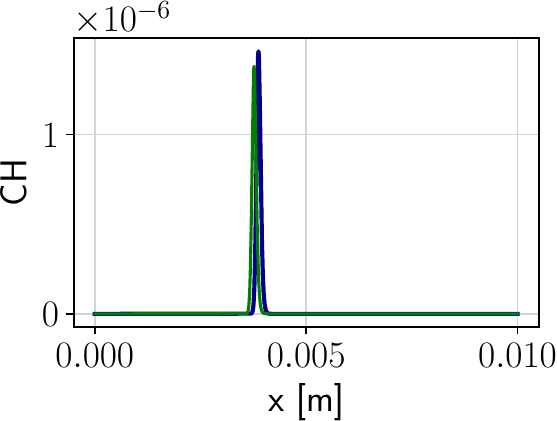}%
    \includegraphics[width=0.33\textwidth]{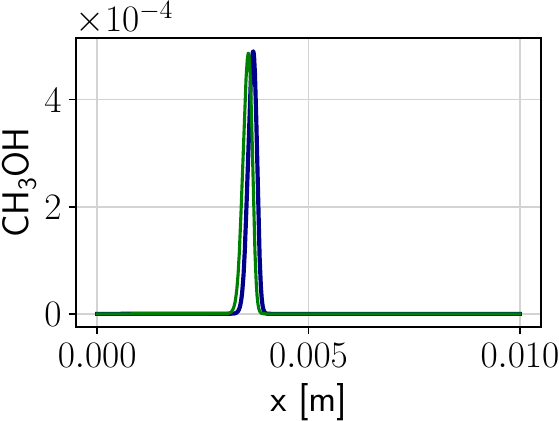}%
    \includegraphics[width=0.33\textwidth]{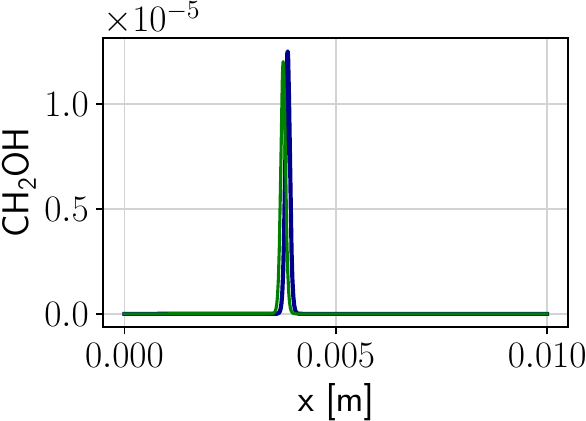}
    \includegraphics[width=0.33\textwidth]{graph_yf22.pdf}%
    \includegraphics[width=0.33\textwidth]{graph_yf23.pdf}%
    \includegraphics[width=0.33\textwidth]{graph_yf24.pdf}
    \includegraphics[width=0.33\textwidth]{graph_yf25.pdf}%
    \includegraphics[width=0.33\textwidth]{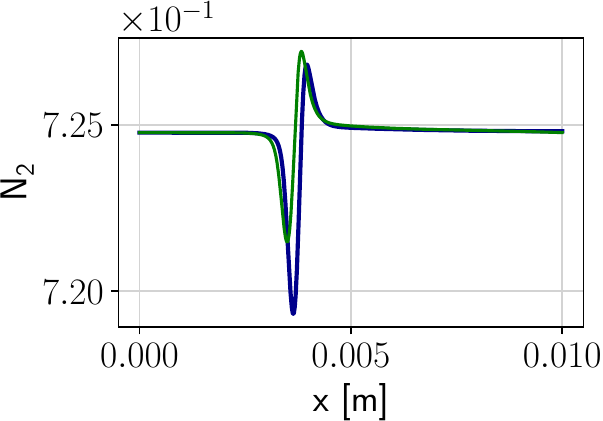}
    \caption{(Continued from Fig.~\ref{fig:comparison_ion_EF_1}) Comparison between Cantera and EMI-SENGA for the ion transport model and with electrostatic interactions activated.}
    \label{fig:comparison_ion_EF_2}
\end{figure}

\subsection{1D laminar reacting flow under electrostatic fields}
\label{sisubsec:1d_laminar_reac_flow_elec_field}

\begin{figure}[H]
    \centering
    \includegraphics[width=0.5\linewidth]{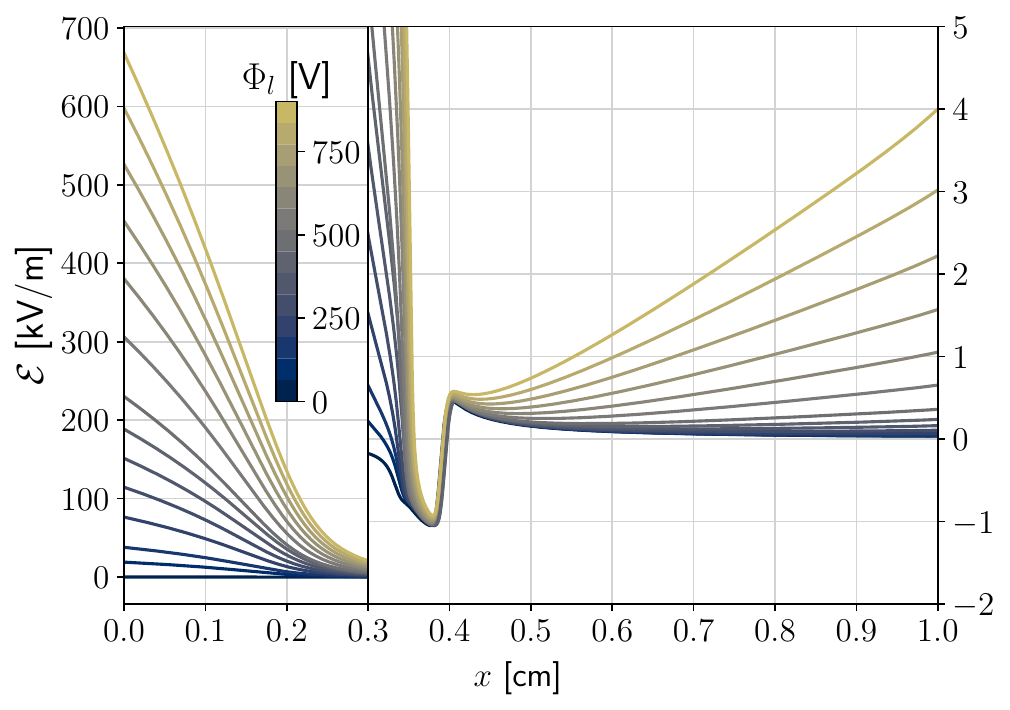}%
    \includegraphics[width=0.5\linewidth]{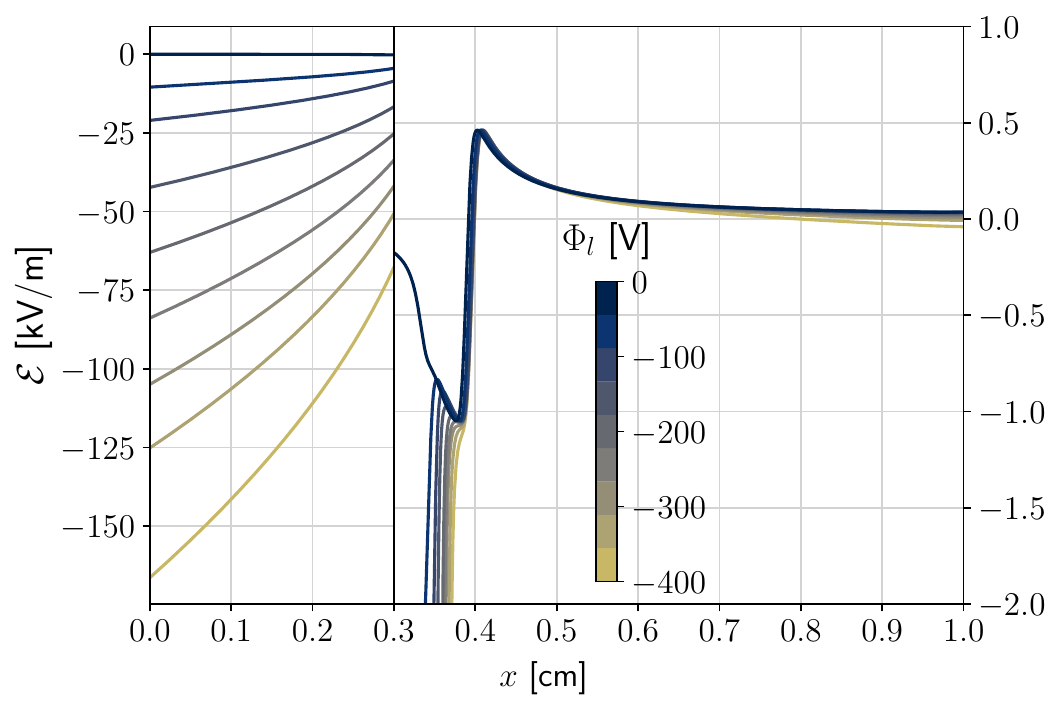}
    \caption{Electrostatic field for a left electrode with a positive (left) and negative (right) voltage. The right electrode is grounded.}
    \label{fig:elfx_field}
\end{figure}

\begin{figure}[H]
    \centering
    \includegraphics[width=0.5\linewidth]{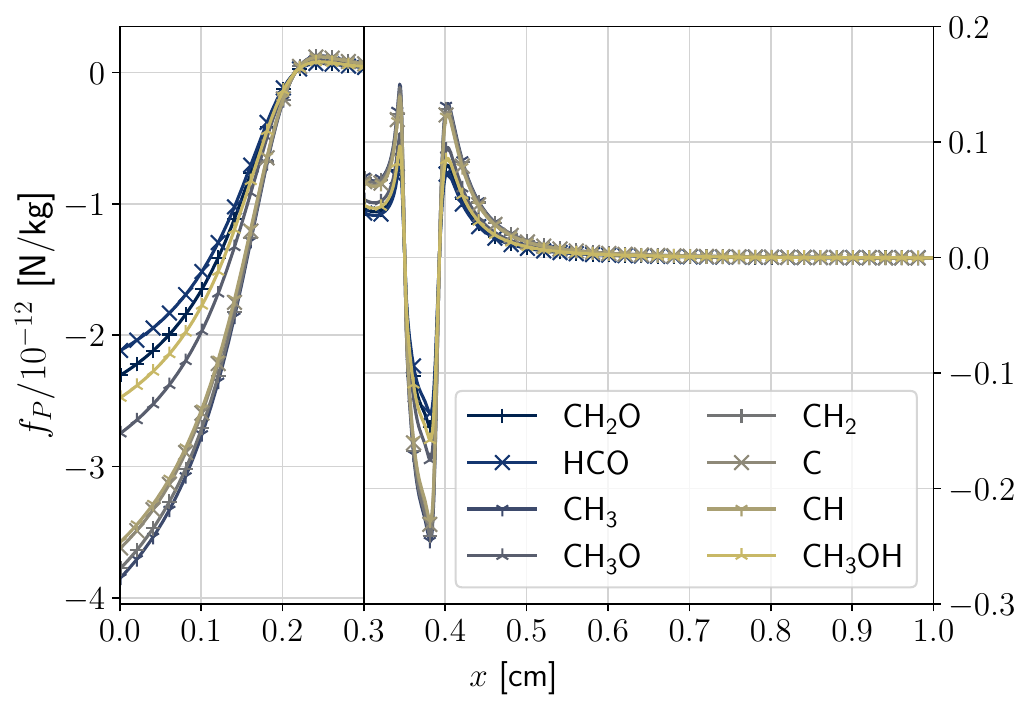}
    \caption{Polarization force on various neutral species for the case with $\Phi_l=500$~V.}
    \label{fig:polarization_field_2}
\end{figure}

\subsection{1D laminar reacting flow under magnetostatic fields}
\label{sisubsec:1D_mag}

\begin{figure}[H]
    \centering
    \includegraphics[width=0.5\linewidth]{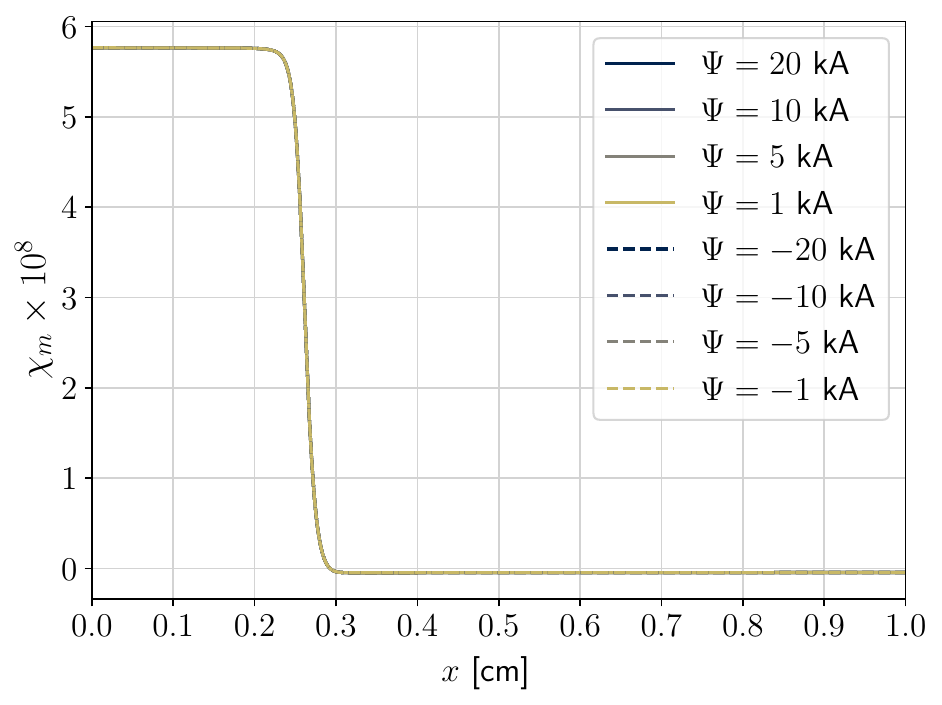}
    \caption{Magnetic susceptibility of the mixture.}
    \label{fig:mafx_chim_si}
\end{figure}

\subsection{2D laminar reacting flow under magnetostatic fields}

\begin{figure}[H]
    \centering
    \includegraphics[width=0.7\linewidth]{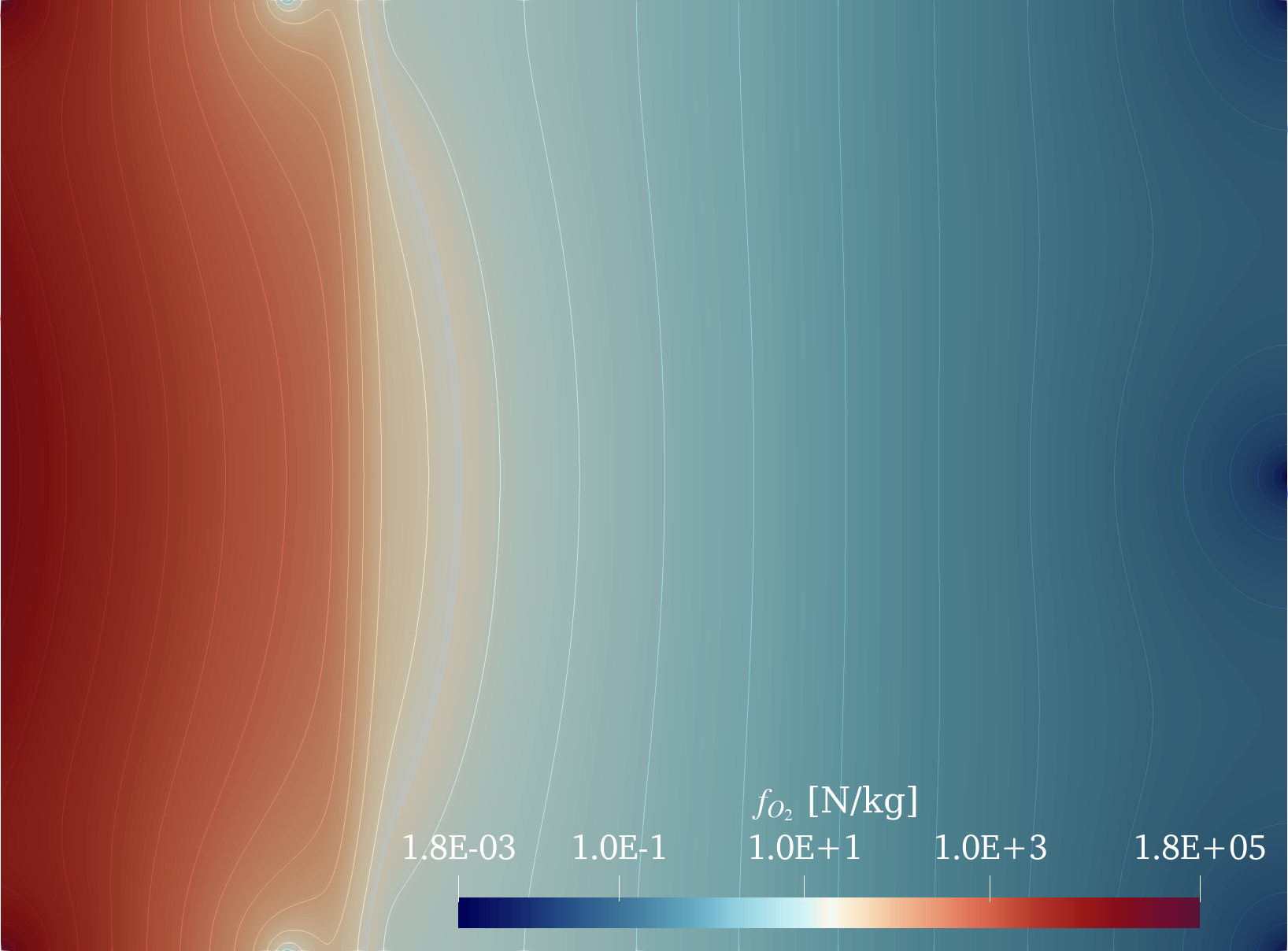}
    \caption{Magnetization force distribution acting on \ch{O2} molecules for the solid line Profile B of Fig.~\ref{fig:magnetic_potential}.}
    \label{fig:magnetization_force_oxygen}
\end{figure}

\subsection{Validation of the EMI-FDTD solver}

\begin{figure}[H]
    \centering
    \includegraphics[width=\linewidth]{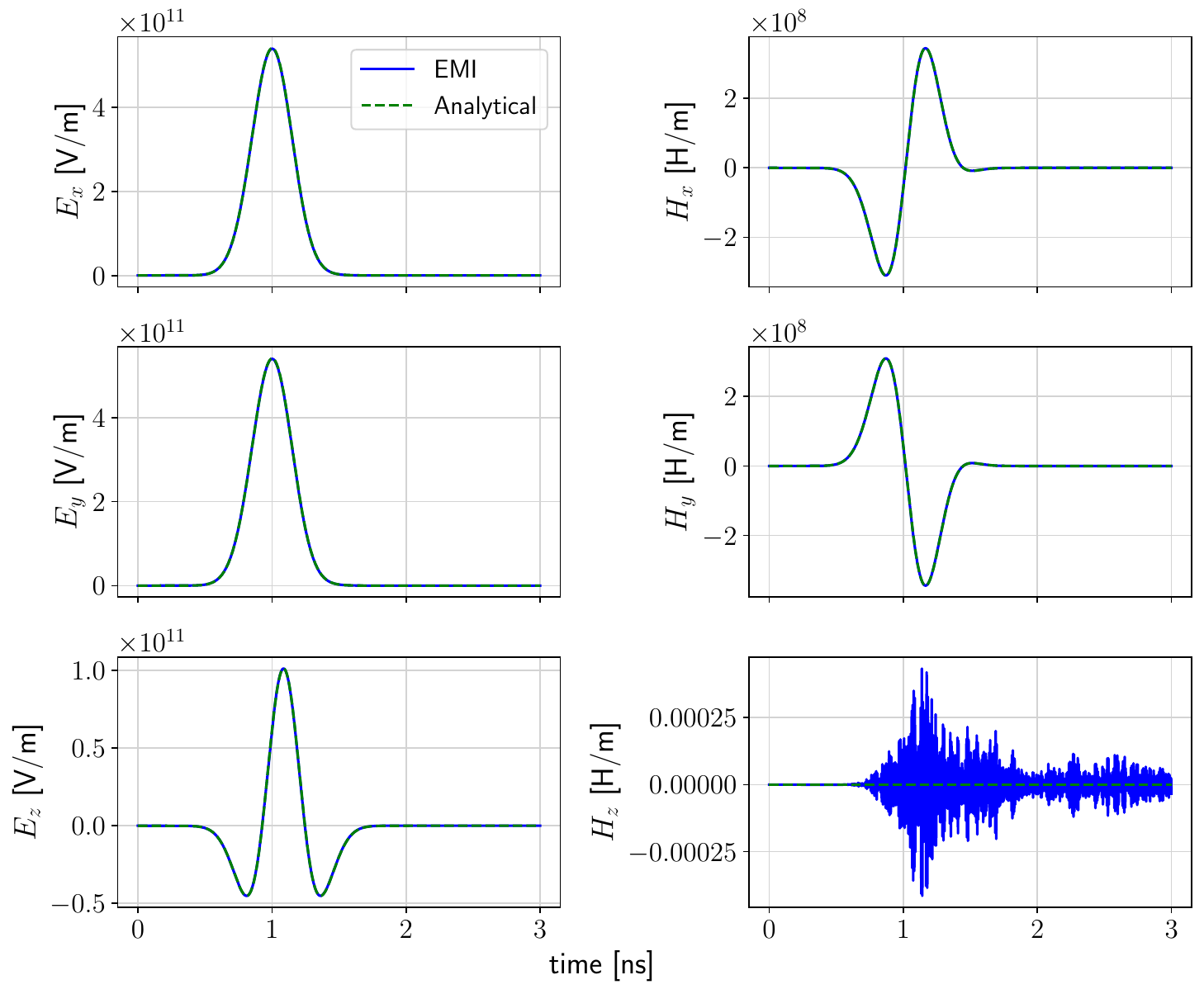}
    \caption{Comparison between EMI-FDTD and the analytical solution of Ref.~\cite{ziolkowski1983} for a Hertzian dipole source with a Gaussian derivative current waveform.}
    \label{fig:fdtd_profiles}
\end{figure}

\begin{figure}[H]
    \centering
    \includegraphics[width=\linewidth]{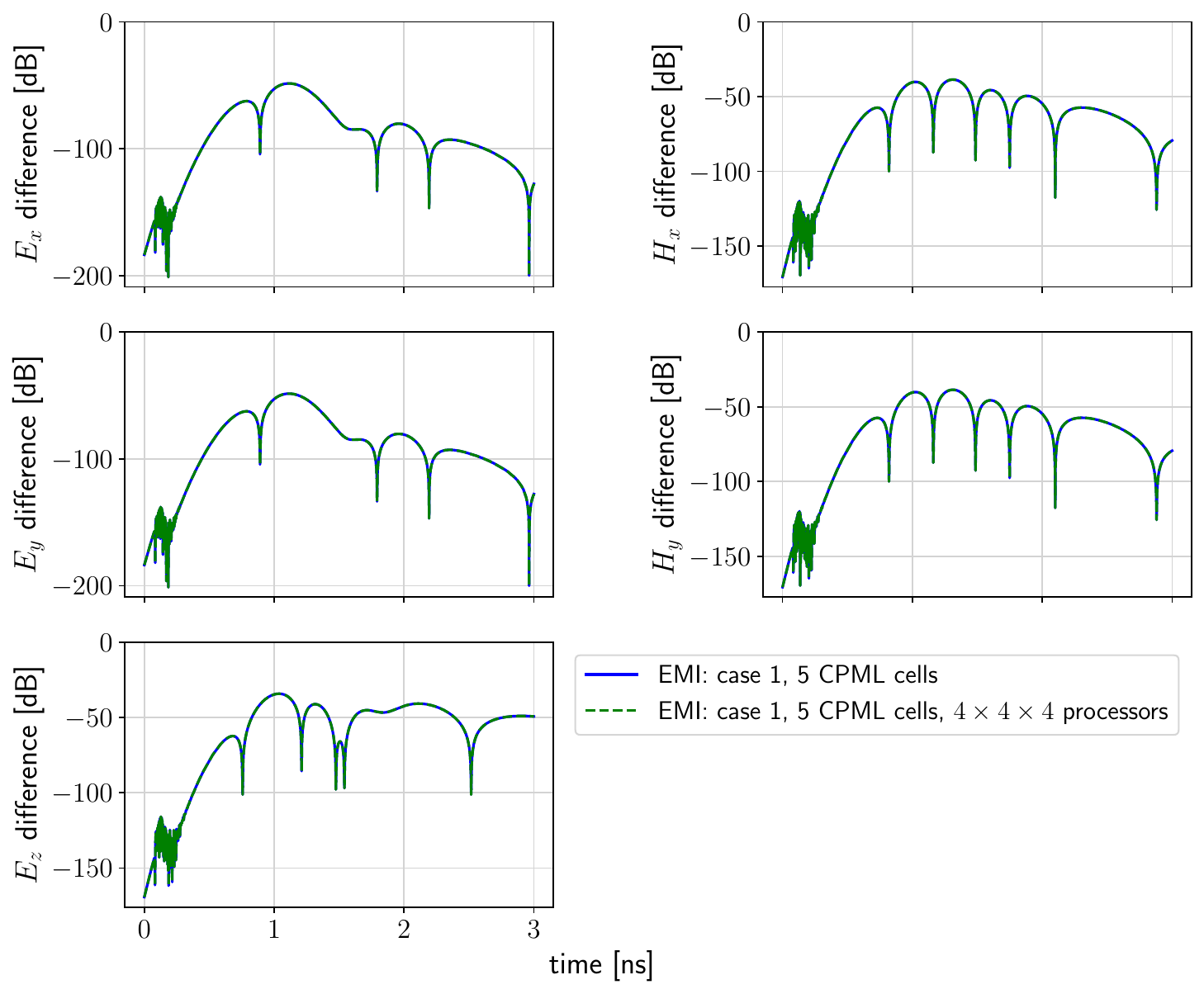}
    \caption{Evaluation of the parallelization strategy of EMI for the configuration described in Fig.~\ref{fig:fdtd_error}.}
    \label{fig:fdtd_error_multi}
\end{figure}

\subsection{Evaluation of the electrostatic assumption in reacting flows}
\label{sisubsec:evaluation_of_electrostatic_assumption}

\begin{figure}[H]
    \centering
    \includegraphics[width=0.55\textwidth]{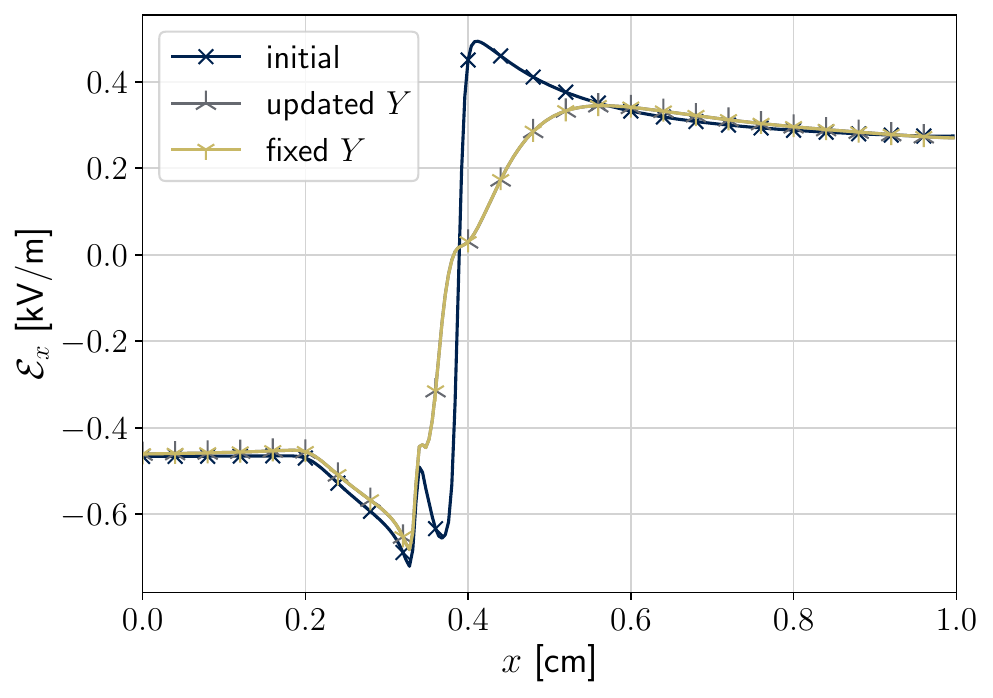}
    \caption{Comparison between the electric field at the beginning of the simulation and at 10~ns simulated time. The ``initial'' case corresponds to the $Y$ profile at the beginning of the simulation; ``updated $Y$'' refers to the final profile, obtained after the advancement of the conservation equations; and ``fixed $Y$'' is the final profile in which case the conservation equations are not solved. Results are plotted for a line passing through the middle $y$ and $z$ coordinates of the domain.}
    \label{fig:elfx_comparison_no_e_noy}
\end{figure}

\begin{figure}[H]
    \centering
    \includegraphics[width=0.55\textwidth]{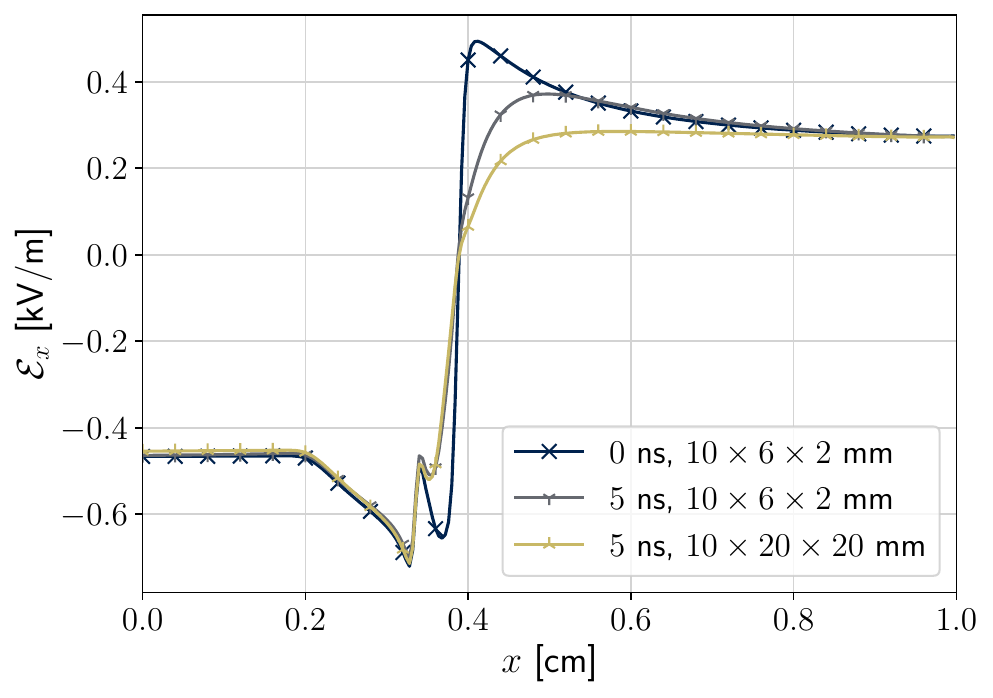}
    \caption{Comparison between the electric field at the beginning of the simulation and at 5~ns simulated time, for the default and for a bigger domain size. Results are plotted for a line passing through the middle $y$ and $z$ coordinates of the domain.}
    \label{fig:elfx_comparison_no_e_big}
\end{figure}

\begin{figure}[H]
    \centering
    \includegraphics[width=0.5\linewidth]{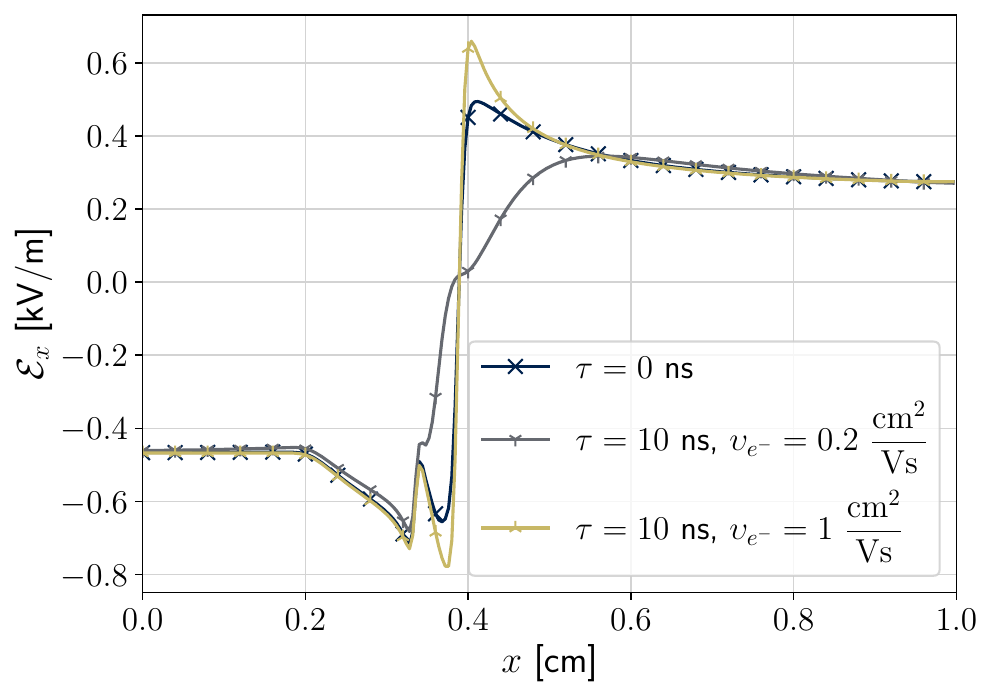}%
    \includegraphics[width=0.49\linewidth]{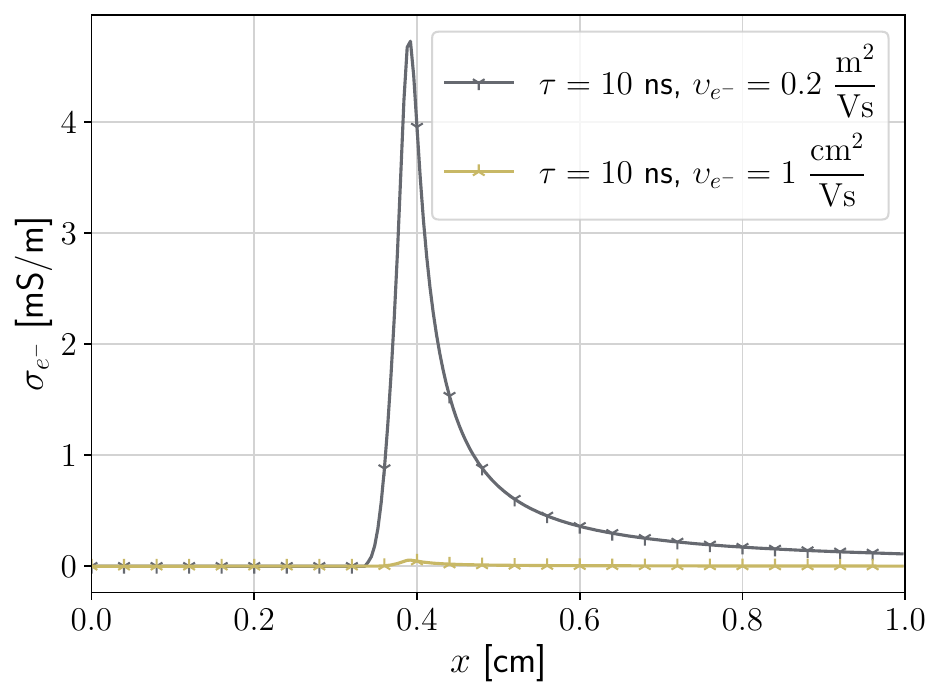}
    \caption{Electric field (left) and conductivity (right) under different electron mobilities at 10~ns simulated time.}
    \label{fig:comparison_low_e_mobility}
\end{figure}

\subsection{Electromagnetic wave sources on reacting flows}

\begin{figure}[H]
    \centering

    \includegraphics[width=0.5\linewidth]{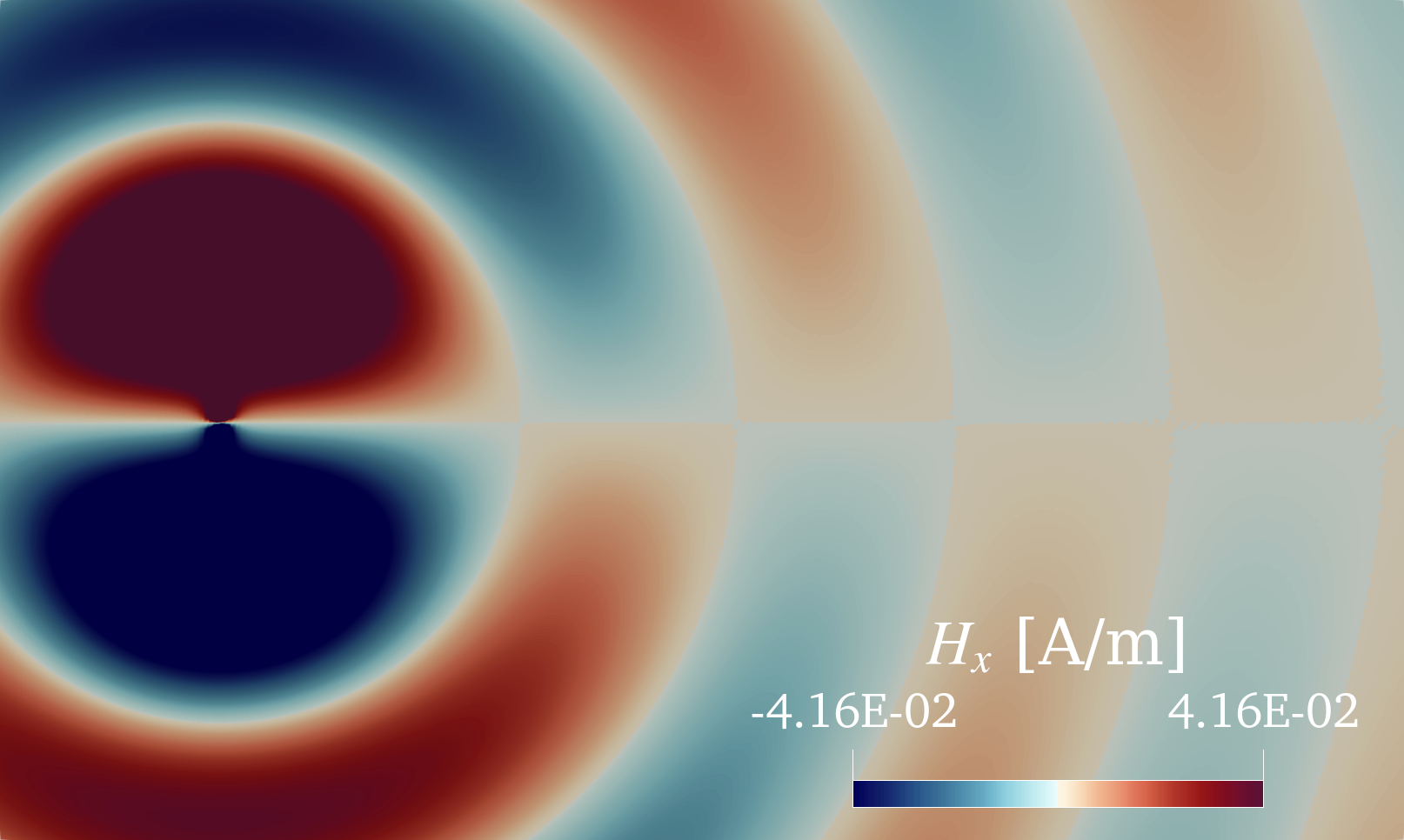}%    
    \includegraphics[width=0.5\linewidth]{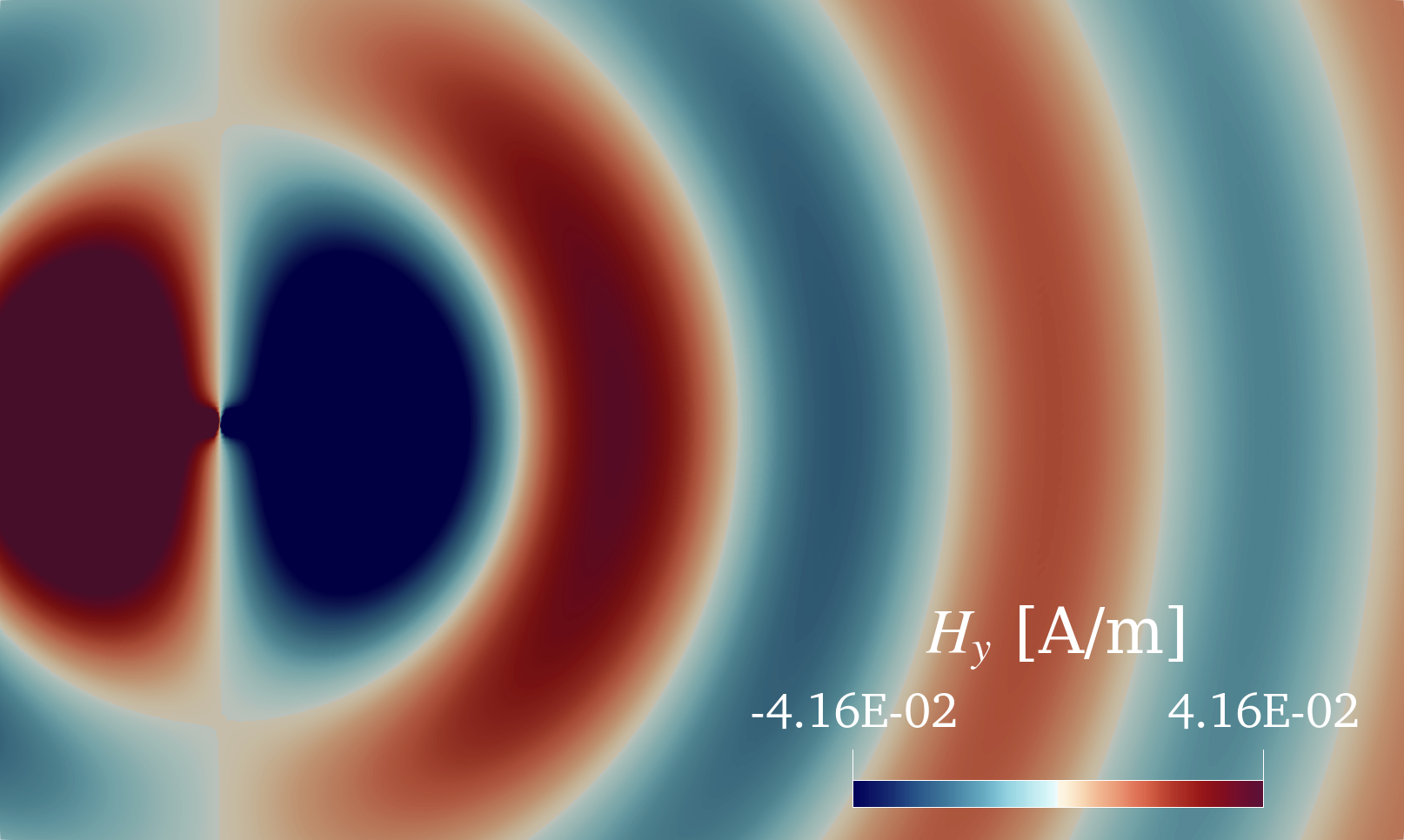}

    \hspace{5pt}%
    \raggedright
    \begin{subfigure}[t]{0.08\linewidth}
        \vspace{-55pt}
        \includegraphics[width=\textwidth]{Figures/FDTD/FDTD_coord.png}
    \end{subfigure}\hfill

    \caption{Selected magnetic field components under a sinusoidal wave source at 5~ns simulated time. The amplitudes of the magnetic field components are an order of magnitude higher than the plotted limits, but they were rescaled to allow for the observation of the waveform. The fields are plotted in a slice passing from the middle point of the $z$ coordinate of the domain.}
    \label{fig:fdtd_source}
\end{figure}

\begin{figure}[H]
    \centering

    \includegraphics[width=0.55\linewidth]{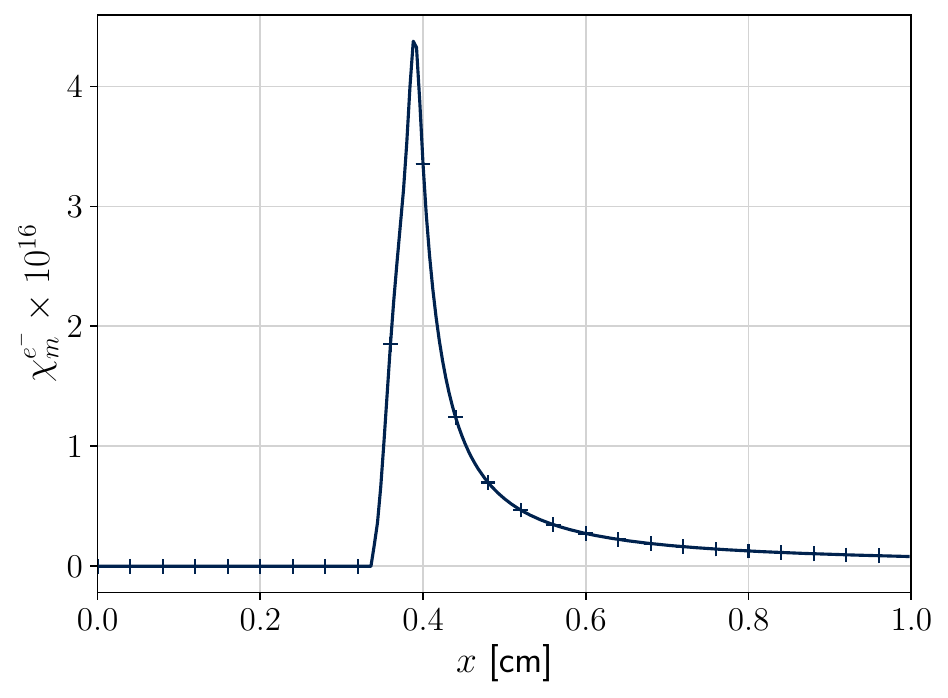}

    \caption{Magnetic susceptibility of electrons along the $x$ direction passing from the middle point of the $y$ and $z$ coordinates.}
    \label{fig:chim_electron}
\end{figure}

 % \bibliographystyle{elsarticle-num} 
 % \bibliography{references}





% \end{thebibliography}